\documentclass[11pt]{article}
\usepackage[utf8]{inputenc}
\usepackage{comment}
\usepackage{graphicx}
\usepackage{wrapfig}
\usepackage{amsmath}
\usepackage{amsfonts}
\usepackage{physics}
\usepackage[dvipsnames]{xcolor}
\definecolor{bg}{rgb}{.9, .9, .9}
\usepackage[frozencache=true]{minted}
\usepackage[justification=centering]{caption}
\usepackage{subcaption}
\usepackage{fancyhdr}
\usepackage[ddmmyyyy]{datetime}
\usepackage{lastpage}
\usepackage{parskip}
\usepackage[table]{xcolor}
\usepackage{hyperref}
\usepackage{multicol}
\usepackage[toc,page]{appendix}
\usepackage[style=numeric, sorting=none]{biblatex}
\usepackage{geometry}
\hypersetup{
    colorlinks = true,
    urlcolor   = blue,
    linkcolor  = dtu_red,
    citecolor  = dtu_red
}

\newcommand{\ASSIGNMENTNAME}{Optimizations and extensions for fair join pattern matching}
\newcommand{\COURSENO}{}

\definecolor{dtu_red}{RGB}{153, 0, 0}

\begin{document}

\begin{center}
\textbf{\LARGE \ASSIGNMENTNAME} \vspace{24mm} \\
\textbf{\Large Spring 2025} \vspace{3mm}\\
\end{center}
\vspace{6mm}
\hrule \vspace{6mm}

\vspace{6mm}
\begin{center}
\textbf{Ioannis Karras}\\ \vspace{2mm}
\end{center}
\vspace{12mm}

\thispagestyle{empty}

\begin{abstract}
Join patterns are an underexplored approach for the programming of concurrent and distributed systems. When applied to the actor model, join patterns offer the novel capability of matching combinations of messages in the mailbox of an actor. Previous work by Philipp Haller et al. in the paper ``Fair Join Pattern Matching for Actors" (ECOOP 2024) explored \emph{join patterns with conditional guards} in an actor-based setting with a specification of \emph{fair and deterministic} matching semantics. Nevertheless, the question of time efficiency in fair join pattern matching has remained underexplored. The stateful tree-based matching algorithm of Haller et al. performs worse than an implementation that adapts the Rete algorithm to the regular version of a join pattern matching benchmark, while outperforming on a variant with \emph{heavy} conditional guards, which take longer to evaluate. Nevertheless, conforming Rete to the problem of join pattern matching requires heavy manual adaptation.

In this thesis, we enhance and optimize the stateful tree-based matching algorithm of Haller et al. to achieve up to tenfold performance improvements on certain benchmarks, approaching the performance of Rete on regular benchmarks while maintaining the advantages of versatility and performance with heavy guards. We also enhance the benchmark suite, adding new features and enhancing its extensibility and user-friendliness. We extend the join pattern implementation with a less ambiguous syntax as well as dynamic pattern switching. Finally, we present a new complex model use case for join patterns, showing their applicability in a microservice web architecture.
\end{abstract}

\newpage
\thispagestyle{empty}
\clearpage
\pagenumbering{arabic}
\tableofcontents

\clearpage
\pagenumbering{arabic}

\newpage

\section{Introduction}

The programming of concurrent and distributed systems notoriously introduces many complexities not found in conventional sequential programming. Many programming paradigms have been developed to tackle these challenges; one such paradigm that has received relatively less attention is the paradigm of \textit{join patterns}. One application of join patterns is in concurrent actor-based message passing systems, where they provide a promising approach with various benefits.

Actors were first introduced in \cite{actors_original}. Informally, the actor model is similar to the model of object-oriented programming (OOP), but with the following key differences:

\begin{itemize}
    \item Instead of objects, the main entity is the \textit{actor}. Like objects, actors store their own encapsulated data. However, while a single thread of execution can affect multiple objects in OOP, each actor instead has its own thread of execution. These threads of execution do not necessarily correspond to platform threads; other execution models such as thread pools may also be used to execute the actors.

    \item The system of communicating data between objects via methods is replaced with an \textit{asynchronous message passing} system. An actor sending a message to another actor does not transfer execution to the other actor; the sender can continue without waiting for a response. Every actor has a \textit{mailbox} that stores all incoming messages.
\end{itemize}

Conventionally, an actor processes messages from its mailbox one at a time, in the order of arrival. If the messages are typed, the actor can have a handler for each message type. The introduction of join patterns allows actors to process \textit{multiple messages} simultaneously and handlers to be associated with specific combinations of message types. Recently, the paper ``Fair Join Pattern Matching for Actors" by Haller et al.~\cite{fair}  explored \textit{join patterns with conditional guards} in an actor-based setting. This is the most expressive conceptualization of join patterns in the literature. While previous implementations of join pattern matching for actors used non-deterministic message matching semantics, Haller et al.~contribute a novel specification of \textit{fair and deterministic} matching semantics. Fair semantics prioritize older messages over newer ones, guaranteeing that any matchable message is eventually consumed. Deterministic semantics ensure that message matching proceeds in a predictable and reliable manner. The authors also present a stateful tree-based algorithm that performs message matching according to these semantics, and prove the correctness of the algorithm with respect to the semantics. They implement this join pattern specification and algorithm in a Scala 3 library called \texttt{JoinActors}. This library dispels the need to create a new programming language supporting join patterns by enabling their definition in a simple and intuitive manner through Scala 3's advanced macro system. Moreover, it includes an original benchmarking suite, with benchmarks tailored to test the nuances of join pattern matching.

Nevertheless, the performance of join pattern matching algorithms has been underexplored. Haller et al. compare their stateful tree-based matching algorithm to an implementation that adapts the Rete algorithm~\cite{rete_original}, using the Evrete library for Java,\footnote{\url{https://www.evrete.org/}} to a join pattern matching benchmark. Evrete outperforms the tree-based algorithm on the regular version of the benchmark, but performs much worse on a variant with so-called \textit{heavy guards}, conditional guards that take a long time to evaluate. Besides, conforming Evrete to the problem of join pattern matching requires a lot of manual adaptation. This conundrum leaves an open question: is there a native algorithm for join pattern matching and an implementation thereof that performs comparably to Evrete in regular benchmarks and also maintains superior performance with heavy guards? In this thesis, we answer this question affirmatively, while building upon the work of Haller et al.

In particular, the main contributions of this thesis are the following:

\begin{enumerate}
    \item We apply various optimizations and new techniques to the stateful tree-based matching algorithm. Some of these improvements are straightforward, while others are more theoretically intriguing, but they all lead to substantial performance improvements. We establish a new baseline that performs better in all cases, as well as various optimizations on top of this baseline which are less generally applicable but provide even better performance in specific cases. Some of our improvements include a more efficient \textit{lazy matching algorithm}, cache-friendly data structures, multithreaded execution, and an algorithm that heuristically uses subexpressions of conditional guards to filter unmatchable messages. These enhancements, described in Section~\ref{sec:optimizations}, constitute the core of our contribution. As a result of these advancements, we achieve performance approximately $10\times$ over the original algorithm on certain benchmarks and approaching that of Evrete on the benchmark where it vastly outperformed the predecessor implementation of Haller et al. Furthermore, we maintain and expand the advantage of the tree-based algorithm on the variant with heavy guards. We argue that these results, along with the fact that Evrete requires elaborate manual adaptation to solve the problem of join pattern matching, render our design the solution of choice on most problem instances.

    \item We comprehensively refactor the benchmark suite of the original implementation to make it more feature-rich, extensible, and user-friendly. Section~\ref{sec:benchmark_suite_improvements} describes this refactoring.

    \item We modify the join pattern macro to support a new, less ambiguous syntax, as described in Section \ref{sec:operator_syntax}.

    \item We introduce support for \textit{dynamic pattern switching} at runtime, enabling a range of expressive use cases. We describe this in Section \ref{sec:dynamic_patterns}.

    \item Lastly, we present a complex, real-world use case of join patterns, demonstrating their advantages within a microservice architecture. We dicuss this case study in Section \ref{sec:payment}.
\end{enumerate}

Our implementation is available in our fork of the \texttt{JoinActors} repository on Github at \url{https://github.com/yaniskas/join-actors}.
\section{Background}

In this section, we provide some background knowledge required to understand our work.

\subsection{Join patterns}

Join patterns are a concurrent programming paradigm originating from a formal model of concurrency called join calculus \cite{join_calculus}. Many implementations of join patterns have been created in various programming languages, with widely varying semantics. Some implementations, like Polyphonic C\# \cite{modern_concurrency_c_sharp}, use them in a \emph{synchronous} concurrency context. Here, we focus on \textit{asynchronous message passing systems}, specifically applied to the actor model. In this context, procedures are driven by messages arriving in the mailbox of an actor.

Join patterns enable a message passing system to match patterns to combinations (i.e., \textit{joins}) of messages in addition to single messages. These combinations may be defined as patterns in a pattern matching construct similar to that found in many programming languages. Additionally, these message passing systems are commonly extended with \emph{guards} on the patterns, which restrict possible matches to those satisfying certain conditions on the message payloads. Such actor-based systems with conditional guards are explored in~\cite{joins_using_extensible_pattern, erlang_with_joins, smart_house_source}.

The predecessor paper by Haller et al.~\cite{fair} gives a formal language specification for join patterns, in which the \emph{join definition} is the main language primitive. A join definition defines a list of join patterns that may be used for matching, where one join pattern specifies a single combination of messages that can be matched. If a join definition combines multiple join patterns, conflicts arise between the patterns whenever more than one pattern can be matched. The main contribution of the predecessor paper is in formalizing a \textit{fair and deterministic} specification for join pattern matching. It imposes an order on possible matches based on the age of the messages matched, and predictably selects matches in a way that prioritizes older messages. This formulation differs from previous implementations, which were non-deterministic.

Formally, a join definition consists of a list of \textit{reaction rules} of the form $J \ \triangleright P$. Here $J$ is a join pattern and $P$ is a process associated with the rule, i.e., the code that is executed when the pattern is matched. A join pattern consists of a sequence of the form $c_1(x) \wedge c_2(y) \wedge ...$ where $c_1$ and $c_2$ are \textit{constructor patterns}, corresponding to message types, and $x$ and $y$ are variables that obtain the values of message payloads when the pattern is matched by incoming messages. For example, if the messages $c_1(m_1)$ and $c_2(m_2)$ arrive in the actor's mailbox, the pattern is matched, $x$ takes the value of $m_1$, and $y$ takes the value of $m_2$. These variables can be referred to in the reaction rule $P$. It is also possible for messages to have no payload or multiple payloads; in each case the shape of the pattern needs to reflect the number of payloads. For instance, $c_3()$ denotes a message with no payloads, and $c_4(a, b, c)$ a message with three payloads. When messages are matched, they are removed from the mailbox and cannot be used in subsequent matches: they are said to be \textit{consumed}.

Haller et al.~\cite{fair} implement their join pattern specification in a Scala library called \texttt{JoinActors}. With this library, a join definition is written using the Scala syntax\footnote{\url{https://docs.scala-lang.org/scala3/book/fun-partial-functions.html}} for \emph{partial functions}, i.e., functions defined only for certain argument values. Using the metaprogramming facilities of Scala 3, the library implements a macro that changes the semantics of partial functions. Namely, the macro transforms a partial function into a definition for a data structure that contains information about the join definition. This data structure is then passed to a matching algorithm, and the matching algorithm is used by an actor in a prototype actor framework.

Listing \ref{lst:factory_example} shows an example of a join definition from~\cite{fair} written in this library. This example model a monitoring program that receives events from devices on a factory shop floor. The join definition starts with the macro call \texttt{receive} on line 2. After this, the join patterns are given using partial function syntax. This consists of a parameter, which is a reference to the actor (\texttt{ActorRef}) (line 2), followed by an arrow, and then a list of the function cases, each defining a single join pattern. The first pattern starts on line 3 with a \texttt{case} statement, followed by a tuple of the constructor patterns that make up the join pattern on lines 3 and 4 (split across two lines to enhance readability).

After the constructor patterns, there is a guard on line 4 starting with the \texttt{if} keyword. A guard states a condition that should be met for the pattern to be matched; if no guard is desired, it can be omitted. In this case, the guard states that the pattern can be matched when the actor has in its mailbox a \texttt{Fault} message and a \texttt{Fix} message both referring to the same fault (\texttt{fid1 == fid2}).

The guard is followed by an arrow \texttt{=>} on line 4 and the code defining the reaction rule on lines 5 and 6. A reaction rule always ends with a returned value that tells the actor what to do next: either \texttt{Continue} to continue matching, or \texttt{Stop(v)} to stop the actor, where \texttt{v} is the value to return once the actor stops. In this case, the reaction rule updates some maintenance statistics and then returns \texttt{Continue} to continue execution.

The \texttt{Actor} class has two type parameters: the type of message received and the type contained within the \texttt{Stop} signal. In the example on line 1, this message type is set to \texttt{Event} and the return type is set to \texttt{Unit}, a standard type in functional programming denoting no information. Since the return type is \texttt{Unit}, the stop signal is parameterized with the unit value \texttt{()}. Finally, after the partial function, the macro receives an algorithm (line 22) as an instance of an enumeration type called \texttt{MatchingAlgorithm}.

\begin{listing}[h!]
\begin{minted}[bgcolor=bg, linenos]{scala}
def monitor() = Actor[Event, Unit] {
  receive { (self: ActorRef[Event]) => {
    case (Fault(_, fid1, _, ts1),
          Fix(_, fid2, ts2)) if fid1 == fid2 =>
      updateMaintenanceStats(ts1, ts2)
      Continue

    case (Fault(mid, fid1, descr, ts1),
          Fault(_, fid2, _, ts2),
          Fix(_, fid3, ts3)) if fid2 == fid3 && ts2 > ts1 + TEN_MIN =>
      updateMaintenanceStats(ts2, ts3)
      log(s"Fault ${fid1} ignored for ${(ts2 - ts1) / ONE_MIN} minutes")
      self ! DelayedFault(mid, fid1, descr, ts1) // For later processing
      Continue

    case (DelayedFault(_, fid1, _, ts1),
          Fix(_, fid2, ts2)) if fid1 == fid2 =>
      updateMaintenanceStats(ts1, ts2)
      Continue

    case Shutdown() => Stop(())
  }}(algorithm)
}
\end{minted}
\caption{The factory shop floor example \cite{fair}}

\label{lst:factory_example}
\end{listing}

One of our contributions is to enable a new operator-based syntax for join pattern definitions. Moreover, our addition of dynamic pattern switching requires explicitly providing type parameters to the \texttt{receive} macro. Therefore, the example would be written differently in the current version of the library; Listing \ref{lst:factory_example_new} shows the first few lines. Lines~3--4 show the new join syntax, where instead of putting the constructor patterns in a tuple, we separate them with a custom operator \texttt{\&:\&}. Line 2 shows the explicit parametrization of \texttt{receive}, where the type parameters previously given to the actor are now given to \texttt{receive} and propagated to the actor by type inference. We discuss our new syntax in Section \ref{sec:operator_syntax}, and the dynamic pattern switching feature in \ref{sec:dynamic_patterns}. To avoid confusion, we use this new syntax in our code snippets henceforth.

\begin{listing}[h!]
\begin{minted}[bgcolor=bg, linenos]{scala}
def monitor() = Actor {
  receive[Event, Unit] { (self: ActorRef[Event]) => {
    case Fault(_, fid1, _, ts1)
         &:& Fix(_, fid2, ts2) if fid1 == fid2 =>
      updateMaintenanceStats(ts1, ts2)
      Continue
    ...
}
\end{minted}
\caption{The first part of the factory shop floor example given in the predecessor paper, written with our new syntax}
\label{lst:factory_example_new}
\end{listing}

The \texttt{JoinActors} library defines two matching algorithms that can be used by actors. The first is a brute-force algorithm implemented in the \texttt{BruteForceMatcher} class, which enumerates all possible message combinations, finds all possible matches, and then searches for the fairest match among these. The second is the aforementioned stateful tree-based algorithm, implemented in the \texttt{StatefulTreeBasedMatcher} class, which improves upon the brute force algorithm by storing partial matches for every message received; we describe this algorithm in detail in the next section. Both matchers have corresponding cases in the \texttt{MatchingAlgorithm} enumeration. The \texttt{receive} macro takes an instance of this enumeration and uses the corresponding class.

The library uses an extensible framework, allowing for new matching algorithms to easily be implemented and plugged in. Both \texttt{BruteForceMatcher} and \texttt{StatefulTreeBasedMatcher} implement the \texttt{Matcher} trait. Scala traits serve a role similar to that of interfaces in Java and other object-oriented programming languages. A new matching algorithm can be added by creating a new \texttt{Matcher} implementation, and then registering it in the \texttt{MatchingAlgorithm} enumeration. In this thesis, we make full use of this extensible structure.

\subsection{The stateful tree-based matching algorithm}
\label{sec:stateful_tree_based_algorithm}

In addition to formalizing the notion of fair matching, Haller et al.~\cite{fair} specify and implement an efficient stateful algorithm for computing matches. The principal data structure used by this stateful matching algorithm is the \textit{matching tree}.\footnote{The predecessor paper refers to this data structure with the terms \textit{mailbox tree}, \textit{M-tree}, and \textit{matching tree}; we opt for the term matching tree.} The paper provides a formal definition of this data structure; here we summarize it in an informal, operational manner.

\begin{figure}[h!]
    \centering
    \includegraphics[width=0.75\linewidth]{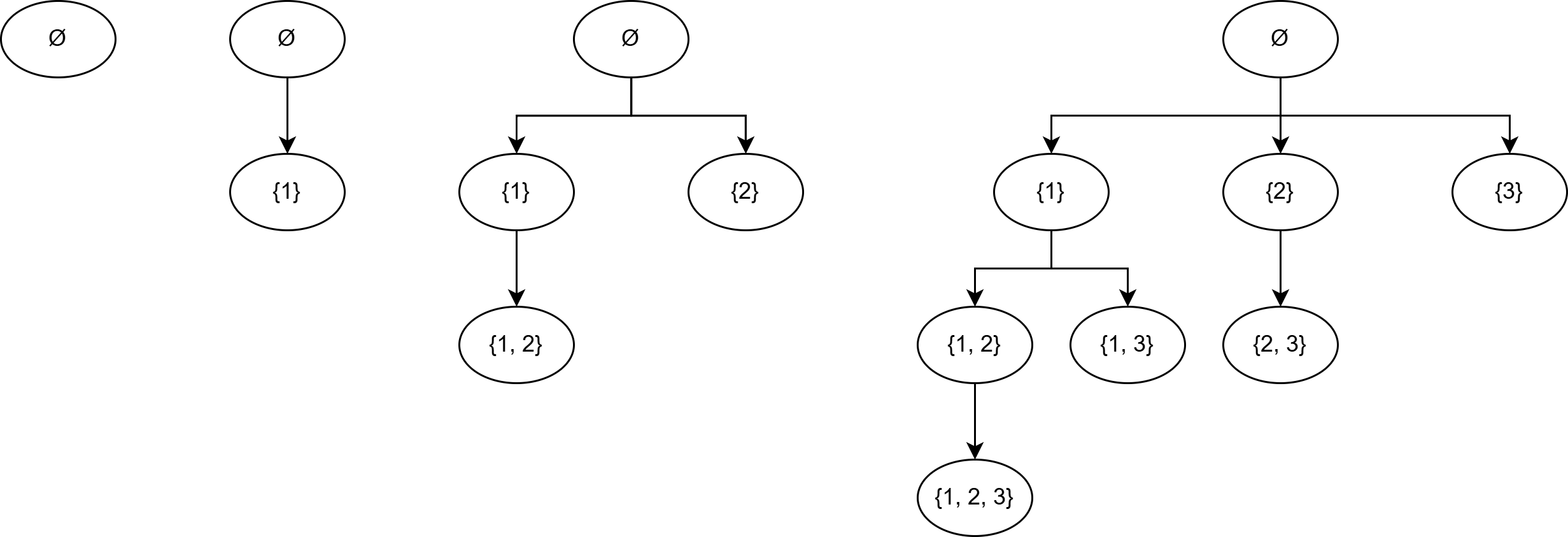}
    \caption{The evolution of an initially empty matching tree as the message indices 1, 2, and 3 are added to it}
    \label{fig:mailbox_tree}
\end{figure}

The nodes of a matching tree hold sets of natural numbers, where each number represents a message index. The tree begins with a root node holding an empty set. When a new message arrives, it is given an index, and this index is added to the matching tree in a process called \textit{ramification}. The message indices start at 1 and increment with every incoming message. When a matching tree without restrictions (as explained below) is ramified with a new message index \textit{i}, every node in the tree gains a child, where this child is a copy of the parent node with the number \textit{i} added to its set. Children are ordered lexicographically with respect to the indices they contain, in ascending order. Figure \ref{fig:mailbox_tree} illustrates the evolution of an initially empty matching tree as the message indices 1, 2, and 3 are added to it.

Every join pattern in a join definition has its own matching tree, whose nodes represent either full or partial matches for the corresponding join pattern. Therefore, the indices inside nodes should correspond to messages that may be matched by the pattern. To that end, since a join pattern consists of a limited number of constructor patterns, each with a specific type, some restrictions are due on the ramification of a matching tree in the context of a join pattern. When a new message of type $T$ arrives and we ramify the tree, a node $n$ is assigned a child only if the number of messages of type $T$ already represented in $n$ is less than the number of constructor patterns of type $T$ in the join pattern. Thus, nodes are extended only until they contain just enough messages of each type to match the join pattern: we refer to such nodes as \textit{complete}.

By themselves, matching trees are just a way of storing message combinations. However, their usefulness derives from the fact that a depth-first, left-to-right traversal of all complete nodes in a matching tree corresponds exactly to the \emph{fairness ordering} of potential message matches. Assuming no message guards, the first complete node returned by a depth-first, left-to-right traversal of the matching tree yields the fairest match. However, if the pattern contains multiple constructor patterns of a particular type, we should also consider the \textit{permutation} of the messages being matched, i.e., which message of that type is assigned to which instance of that constructor type in the pattern. The fairness ranking of permutations is also specified and implemented in~\cite{fair}; we omit the details here.

When guards are involved, one needs to further examine every complete node to determine which permutations satisfy the guard and select the fairest of these. If no such permutation exists, one may move on to the next fairest complete node. If no complete node in the tree yields a permutation that satisfies the guard, then a match cannot be produced by the pattern associated with this tree. Once every matching tree is ramified, the algorithm picks the fairest match that was yielded overall, thus yielding the fairest match that can be made by the join definition.

One last detail that greatly increases the efficiency of the matching tree structure is the fact that it can be \textit{pruned}. Whenever messages are consumed, we can delete each node in the tree that contains at least one index of a message that has been consumed. Since child nodes always contain all the indices of their ancestors, all children of a deleted node can also be deleted. This way, we can delete entire branches of the tree whenever a message is consumed. This pruning procedure is performed on all matching trees after a match is produced.

Although the tree-based matching algorithm depends on the matching tree data structure, this data structure is never created explicitly in the implementation. Instead, the nodes of the matching tree are stored in a Scala \texttt{TreeMap},\footnote{\url{https://scala-lang.org/api/3.x/scala/collection/immutable/TreeMap.html}} which is implemented as a red-black tree. Red-black trees are binary trees while matching trees are not, so this representation does not correspond exactly to the theoretical view of a matching tree. For this representation to work, the \texttt{TreeMap} uses a specially defined ordering on the nodes it stores, which ensures that the iterator of the red-black tree yields nodes in the order they would be yielded by a depth-first, left-to-right traversal of the matching tree, thereby maintaining the principal property of the matching tree.

\subsection{Benchmarks}
\label{sec:benchmarks}

The original implementation \cite{fair} includes a benchmark suite for testing matching algorithms. In the following, we describe the benchmarks from the original suite which we ported to our refactored suite.

\subsubsection{Size}
\label{sec:size}

This is a simple benchmark that does not model any real use case, but simply tests the effect of the size of join patterns on their evaluation speed. It uses six different join definitions containing a pattern that varies in size between the different definitions. The pattern always matches a join between pairwise different message types. For example, the first definition has the pattern \texttt{A()}, the second \texttt{A() \&:\& B()}, the third \texttt{A() \&:\& B() \&:\& C()}, and so on.

There are two variants of this benchmark. In the first, the only messages sent to the actor are the ones needed to match the join pattern, e.g. for the definition with size three this would be a repeating sequence of \texttt{A()}, \texttt{B()} and \texttt{C()}. In the second, random noise messages that do not contribute to matching the pattern are also interleaved. These noise messages cause permanent growth of the matching tree, as they are never consumed.

\subsubsection{Size with guards}
\label{sec:size_with_guards}

This is a version of the Size benchmark where the messages have payloads, and the join pattern has a guard that tests for equality of the payloads. For example, the definition with size three now matches the pattern \texttt{A(x) \&:\& B(y) \&:\& C(z)}, and the guard tests that \texttt{x == y} and \texttt{y == z}. Moreover, in addition to the variant where only matching messages are sent and the variant where noise messages are interleaved, a third variant is provided. In this third variant, some messages are sent that satisfy the pattern in terms of types, but which do not satisfy the guard. These messages not only contribute to permanent growth of the matching tree, but they also cause many failing guard evaluations to be run.

\subsubsection{Complex Smart House}
\label{sec:complex_smart_house}

This benchmark simulates a few functions of a model smart house coordinator, inspired by \cite{smart_house_source}. It uses an actor with a join definition that has four patterns. The first three patterns each match a join on three different statuses from different sensors in the house, and the patterns are equipped with guards testing a complex condition on the payloads. The last pattern matches a shutdown signal to stop the actor. The messages sent consist of a sequence of matching messages intercalated with randomly generated noise messages.

\subsubsection{Simple Smart House}
\label{sec:simple_smart_house}

This benchmark is a simplified version of Complex Smart House with two of its complex join patterns removed, so that there is only one performance-critical join pattern. Moreover, no random messages are sent. This benchmark sends a series of pairs of ``prefix" messages, where each pair contains the first two messages required to match the complex join pattern. After the prefix messages, it sends a message that can complete the match with any of the previously sent pairs.

\subsubsection{Bounded Buffer}
\label{sec:bounded_buffer}

This benchmark implements a scenario where a bounded buffer stores messages generated by producers and provides these messages to consumers, adapted from \cite{savina_suite}. The bounded buffer is implemented as an actor with join patterns, while the producers and consumers are simple threads running a loop.

\subsection{The boundary/break API}
\label{sec:boundary_break_API}

Here we describe a nontrivial feature of the standard library which we use in one of our optimizations. The way we use this is described in Section \ref{sec:lazy_mutable_matcher}.

Many programming languages include a \texttt{break} statement for breaking out of loops, but Scala does not. Nevertheless, there are ways to achieve performant \texttt{break} semantics in Scala, and these are included in the standard library. Prior to Scala 3, a \texttt{break} statement was provided by the \texttt{scala.util.control.Breaks} API.\footnote{\url{https://www.scala-lang.org/api/3.x/scala/util/control/Breaks.html}} However, in Scala 3 a superior alternative was introduced in \texttt{scala.util.boundary}.\footnote{\url{https://www.scala-lang.org/api/3.x/scala/util/boundary$.html}} In particular, this new API provides a more expression-oriented version of the \texttt{break} statement similar to that found in the Rust programming language, where a value can be returned along with the break.

To use this API, one must first define a scope using the \texttt{boundary} statement. Within this scope, it is possible to use the \texttt{break} statement by itself. If it used this way, once the program reaches the \texttt{break}, the program will skip to the next statement after the scope. This is similar to how the \texttt{break} statement works in most C-like programming languages. Listing \ref{lst:break_imperative} shows this kind of simple, imperative use of the API.

\begin{listing}[h!]
\begin{minted}[bgcolor=bg, linenos]{scala}
var i = 0

boundary:
  while true:
    i += 1
    if (i == 4) then break

assert(i == 4)
\end{minted}
\caption{A simple, imperative use of the boundary/break API}
\label{lst:break_imperative}
\end{listing}

However, a \texttt{break} can also be parameterized with a value. This way, when the program reaches the break, instead of simply skipping past the scope, it will make the whole scope assume the value that was given to the \texttt{break}. To use this value, one needs to assign the \texttt{boundary} expression to a variable. Listing \ref{lst:break_with_expression} demonstrates this more complicated use of the API. Here, we assign the variable \texttt{j} to the result of the boundary expression. When the variable \texttt{i} reaches the value 4, the \texttt{break} statement is executed with the parameter 12. Since \texttt{j} is assigned to the boundary expression, \texttt{j} assumes this value of 12.

\begin{listing}[h!]
\begin{minted}[bgcolor=bg, linenos]{scala}
var i = 0

val j = boundary:
  while true:
    i += 1
    if (i == 4) then break(i*3)

assert(j == 12)
\end{minted}
\caption{A more complex, expression-oriented use of the boundary/break API}
\label{lst:break_with_expression}
\end{listing}
\section{Optimizations}
\label{sec:optimizations}

The predecessor paper \cite{fair} focuses on creating a novel specification, while its implementation of the stateful tree-based algorithm is intended as a proof of concept without much focus on performance. Here, we implement many performance optimizations to the algorithm, resulting in substantial experimental performance gains.

We begin in Section \ref{sec:design} by outlining the theoretical design of our optimizations. Then, in Section \ref{sec:implementation}, we discuss our implementation of these optimizations and show the performance each algorithm achieves compared to the previous one. In Section \ref{sec:overall_comparison}, we show the performance improvements in all the benchmarks and present plots comparing all algorithms at once. The Rete algorithm~ \cite{rete_original} solves a similar problem to the problem of fair join pattern matching, and the predecessor paper compares the performance of its implementation to the highly optimized implementation of Rete in the Evrete library for Java. In Section \ref{sec:evrete_comparison}, we compare our most performant new algorithms to Evrete and show that we achieve performance close to that of Evrete. Finally, in Section \ref{sec:failed_optimizations} we talk about some attempted optimizations that did not yield any consistent performance benefit.

\begin{figure}[h!]
    \centering
    \includegraphics[width=0.88\linewidth]{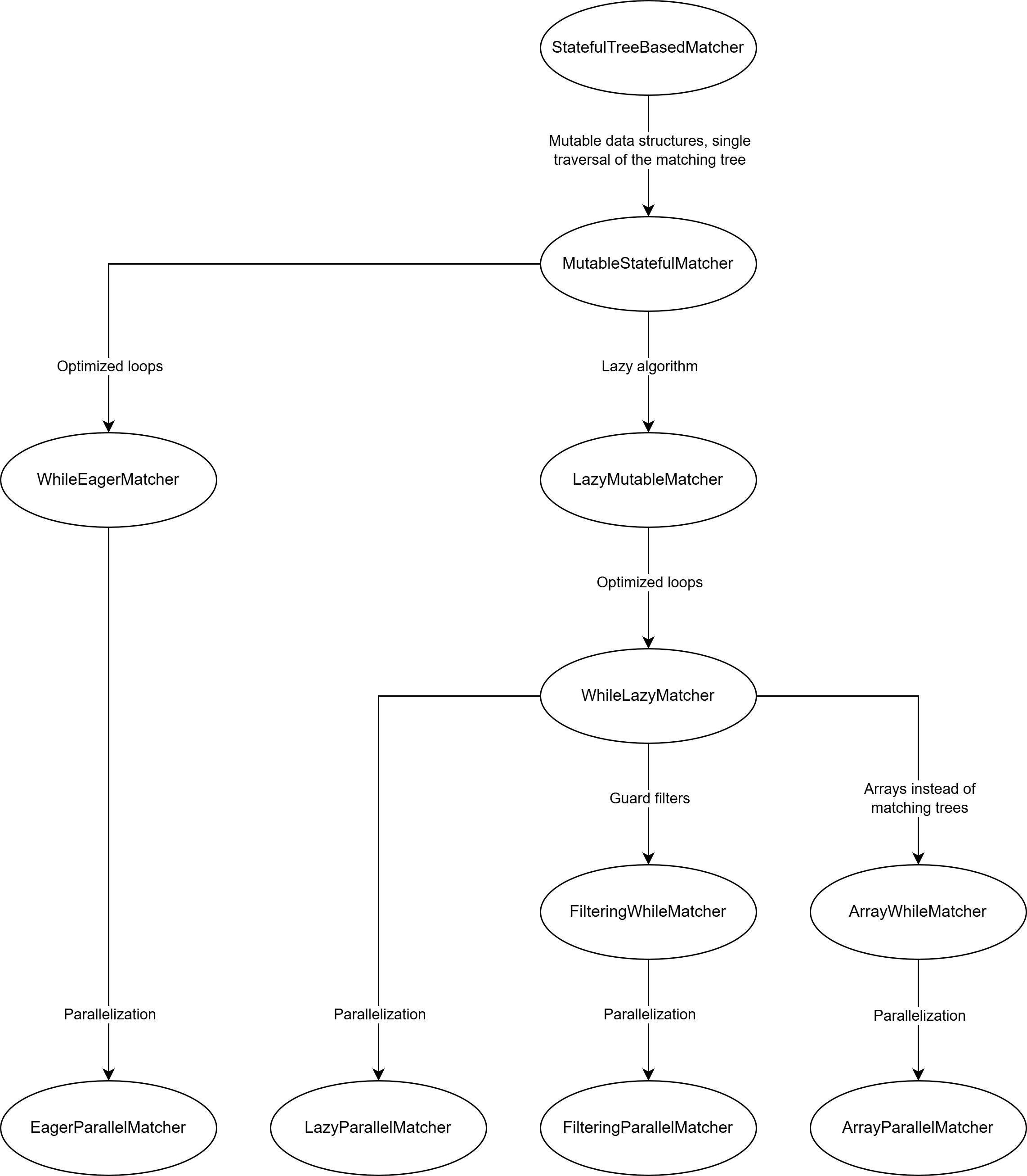}
    \caption{The evolution of our matcher implementations}
    \label{fig:matcher_evolution}
    \vspace{-2mm}
\end{figure}

We use the extensible structure of the existing library to implement our optimizations. For every optimization other than the use of cache-friendly data structures (Section \ref{sec:cache_friendly_data_structures}), we create a new class implementing the \texttt{Matcher} trait and a new case of the \texttt{MatchingAlgorithm} enumeration linked to this class. Some of our optimizations are cumulative, implemented by extending the previous \texttt{Matcher} with the new optimization. However, as our later optimizations are less generally applicable, we eschew this linear structure on some matchers. Figure \ref{fig:matcher_evolution} shows the evolution tree of our optimizations, where each node stands for one of our new \texttt{Matcher} implementations, and each edge is labeled with the optimization added from one matcher to the next.

\subsection{Design}
\label{sec:design}

\subsubsection{Mutability and single traversal}
\label{sec:mutability_and_single_traversal}

\begin{figure}[h!]
    \centering
    \includegraphics[width=1\linewidth]{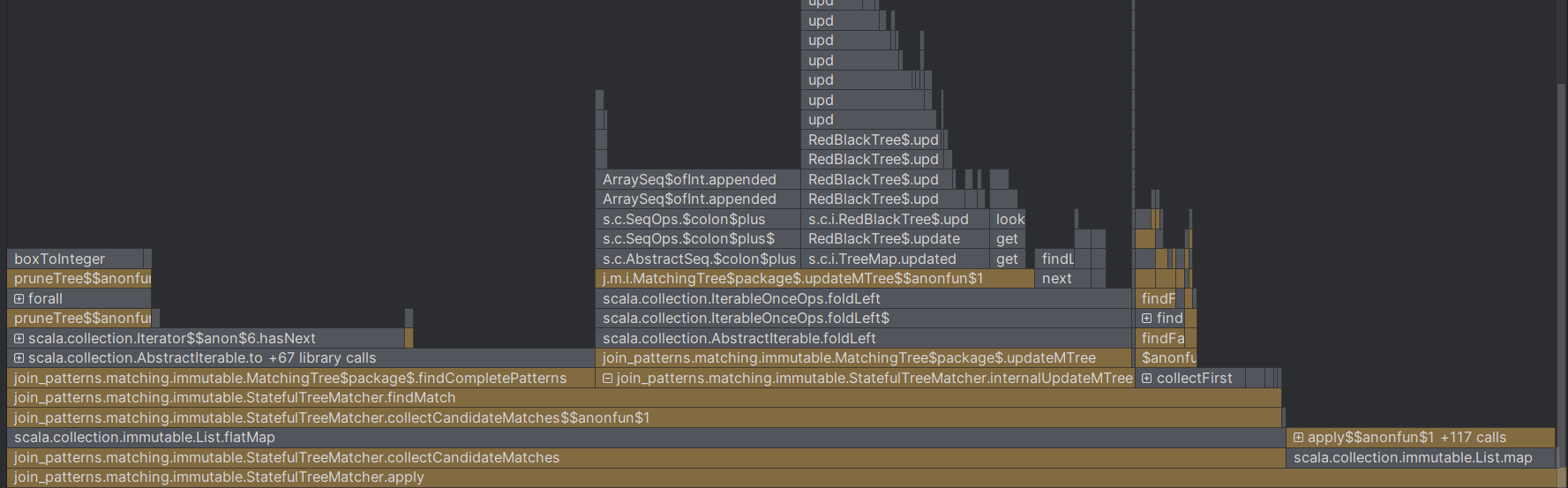}
    \caption{A flame graph of the original stateful tree-based algorithm on the Simple Smart House benchmark}
    \label{fig:immutable_flame_graph}
\end{figure}

The first optimization we considered was mentioned in the ``Future Work" section of the predecessor paper, namely the seemingly low-hanging fruit of converting the matching tree data structure from an immutable functional style to a mutable object-oriented style. This change resulted in a slight, but mostly insubstantial, performance improvement. In the pursuit of a more substantial improvement, we used IntelliJ's profiling tool\footnote{\url{https://www.jetbrains.com/pages/intellij-idea-profiler/}} to examine the time spent in each portion of the algorithm. Figure \ref{fig:immutable_flame_graph} shows a part of a typical flame graph of the existing immutable algorithm on the Simple Smart House benchmark.

We noticed that over half of the runtime is spent in two separate traversals of the matching tree: the \texttt{updateMTree} function and the \texttt{findCompletePatterns} function (seen on the 7th and 6th lines from the bottom of the flame graph). \texttt{updateMTree} ramifies the matching tree when a new message arrives, adding new nodes containing the new message index wherever possible; \texttt{findCompletePatterns}, called soon thereafter, searches the newly ramified tree to find \emph{complete} nodes, i.e., nodes where every pattern index has been assigned a message index and hence a match can take place. We merged these two traversals into one: while ramifying the tree, we check if new nodes being added have complete patterns, and if so, we report them. This change yielded a substantial performance improvement. Thus, our first optimization consists of using a mutable matching tree and traversing it only once.

\subsubsection{Lazy ramification}
\label{sec:lazy_ramification}

The original implementation of the matching tree algorithm for fair join pattern matching exhibits some laziness in its tree traversal, but not as much as it could, leading to performance losses. When a new message arrives, the algorithm ramifies the entire matching tree. Then, it obtains the complete nodes from the tree in sorted order of fairness. It searches these complete nodes one by one for an optimal permutation of message indices that satisfies the guard of the pattern. If it finds such a permutation, it stops the search and reports this permutation as the fairest match for the pattern; if not, it moves on to the next complete node. Finally, it finds the fairest match among all patterns.

The single-traversal optimization we discussed improves this by finding optimal matches while ramifying the tree, but it still ramifies the entire tree before checking any complete node for valid permutations, resulting in redundant computations. If we instead check for permutations immediately after finding complete nodes, we can stop the ramification procedure whenever we find a complete node that has a valid permutation. This approach is correct for two reasons. Firstly, if we find a complete node with a valid permutation, we have in effect found the fairest valid match for this pattern using the messages up to this point, and hence there is no need to continue searching for a match. Secondly, the newly added message will not be used in any future matches, as it is now guaranteed to be consumed, using either the valid match found in the current matching tree or a match from another pattern if a fairer match is found. Thus, all nodes containing the index of the message will be pruned once the overall fairest match is found, so there is no point in continuing ramification and adding further nodes containing the index.

\begin{figure}[h!]
    \centering
    \includegraphics[width=0.75\linewidth]{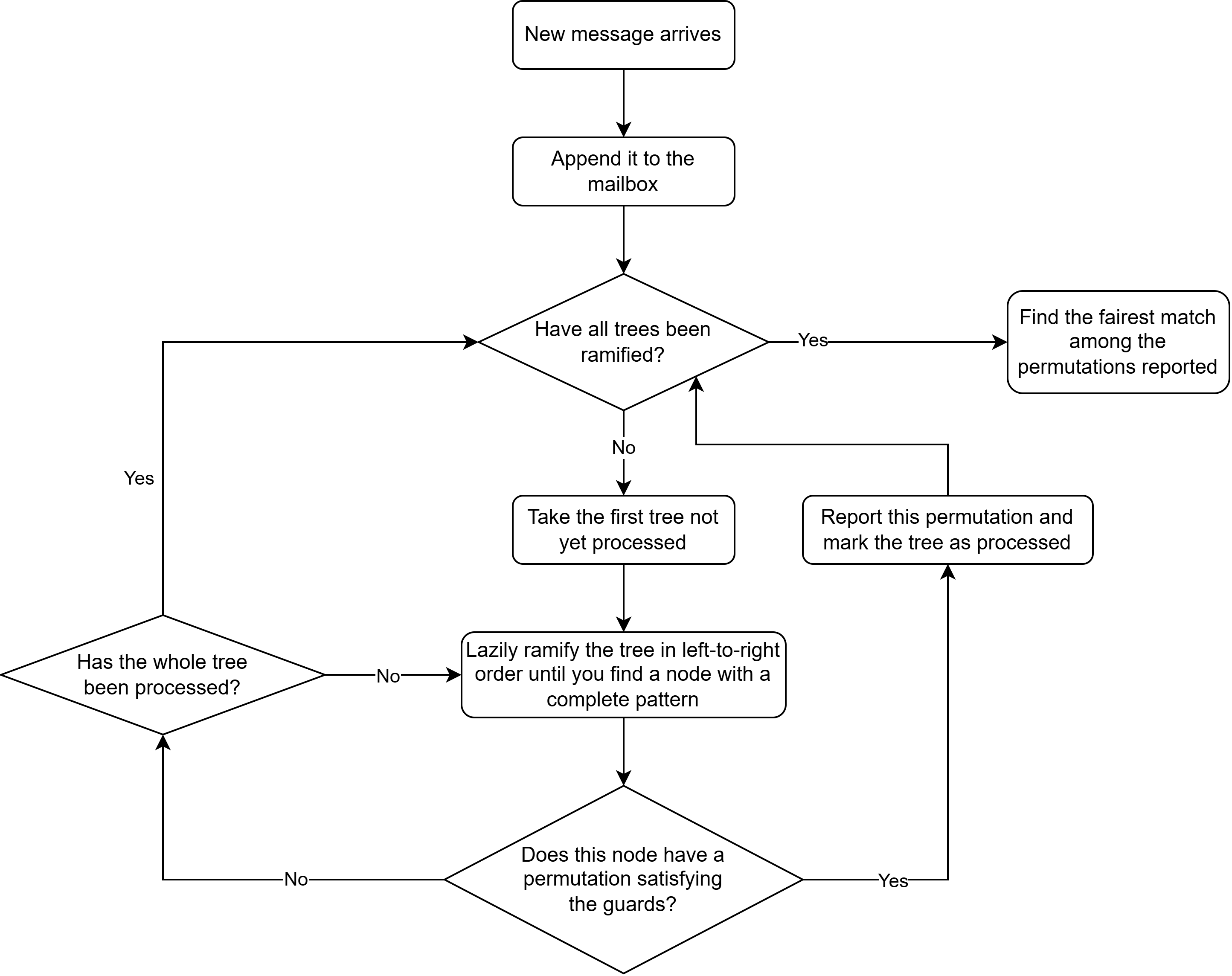}
    \caption{A flow diagram showcasing the operation of the lazy tree-based algorithm}
    \label{fig:lazy_algorithm}
\end{figure}

Our algorithm thus works as follows. We ramify the matching trees in order. While ramifying a tree, if we construct a complete node, we check it for valid permutations satisfying the guard. If we find such a permutation, we stop the ramification, report the node, and move on to the next tree; if not, we continue the traversal. In either case, we do not add the complete node to the tree, as it serves no purpose in any case. Once we have found the fairest match from each tree, we find the fairest match overall and then prune the trees, as before. Figure \ref{fig:lazy_algorithm} shows a flow diagram of this algorithm.

\subsubsection{Optimized loops}
\label{sec:optimized_loops}

All the tree-based algorithms up to this point have a performance-critical \texttt{for} loop that iterates over the nodes of the matching tree and performs ramification. However, Scala's implementation of \texttt{for} loops is quite inefficient, as it compiles them into a call to the \texttt{foreach} method parameterized with a closure \cite{scala_for_inefficiency}. This results in a needless performance loss in our code.

Thankfully, Scala also has \texttt{while} loops, which are compiled similarly to loops in Java. Therefore, we can use these to improve performance. The most interesting part of this optimization is in the implementation details: in the corresponding implementation section, Section \ref{sec:while_lazy_matcher}, we describe how we achieve this improved performance without sacrificing code readability with explicit \texttt{while} loops.

\begin{figure}[h!]
    \centering
    \includegraphics[width=1\linewidth]{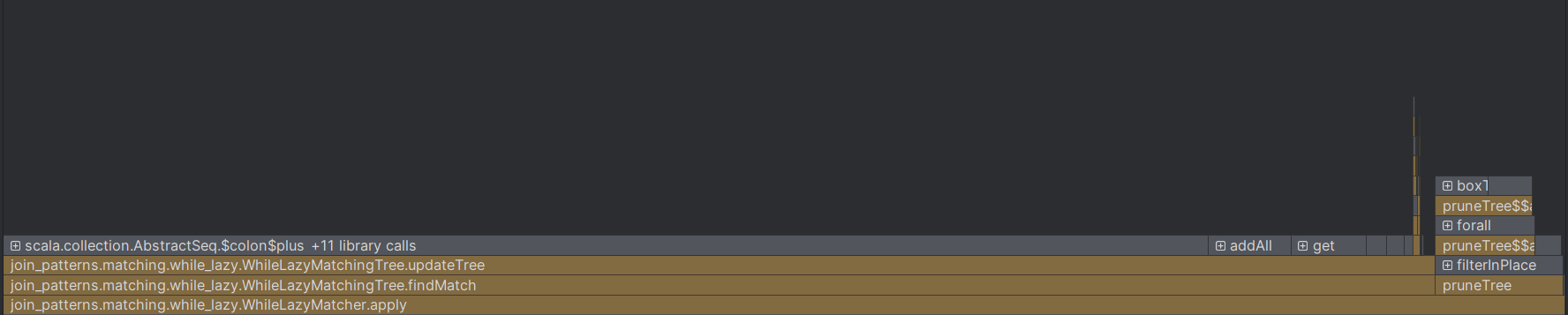}
    \caption{A flame graph for \texttt{WhileLazyMatcher} on the Simple Smart House benchmark}
    \label{fig:while_lazy_flame_graph}
\end{figure}

The hitherto discussed optimizations are generally applicable, so our implementations of these optimizations build on top of each other. However, from this point onward, the optimizations are less generally applicable, so the evolution of our matchers begins to branch, as Figure \ref{fig:matcher_evolution} shows. Furthermore, some of the following optimizations are partially inspired by the performance of the matcher implementing the optimizations up to now, called \texttt{WhileLazyMatcher}. All this considered, we can think of \texttt{WhileLazyMatcher} as a new baseline. Figure \ref{fig:while_lazy_flame_graph} shows a flame graph for the main \texttt{apply} method of this matcher, again on the Simple Smart House benchmark. We use this flame graph to argue for some of our next optimizations.

\subsubsection{Multithreading}
\label{sec:multithreading}

All the algorithms up to this point have been single-threaded, whereas Evrete makes heavy use of multithreading in its implementation of the Rete algorithm. Motivated by this, we investigated how we could use multithreading to improve our performance. It is not trivial to reconcile the previously discussed lazy ramification procedure with multithreading. Therefore, we first made a parallel matching algorithm that does not make use of laziness. The algorithm still iterates through the matching trees in sequence. However, to process each tree, it splits the tree into a number of even partitions, and assigns one thread to each partition to fully ramify it. Then, it analyzes the complete nodes reported from each partition.

\begin{figure}[h!]
    \centering
    \includegraphics[width=0.75\linewidth]{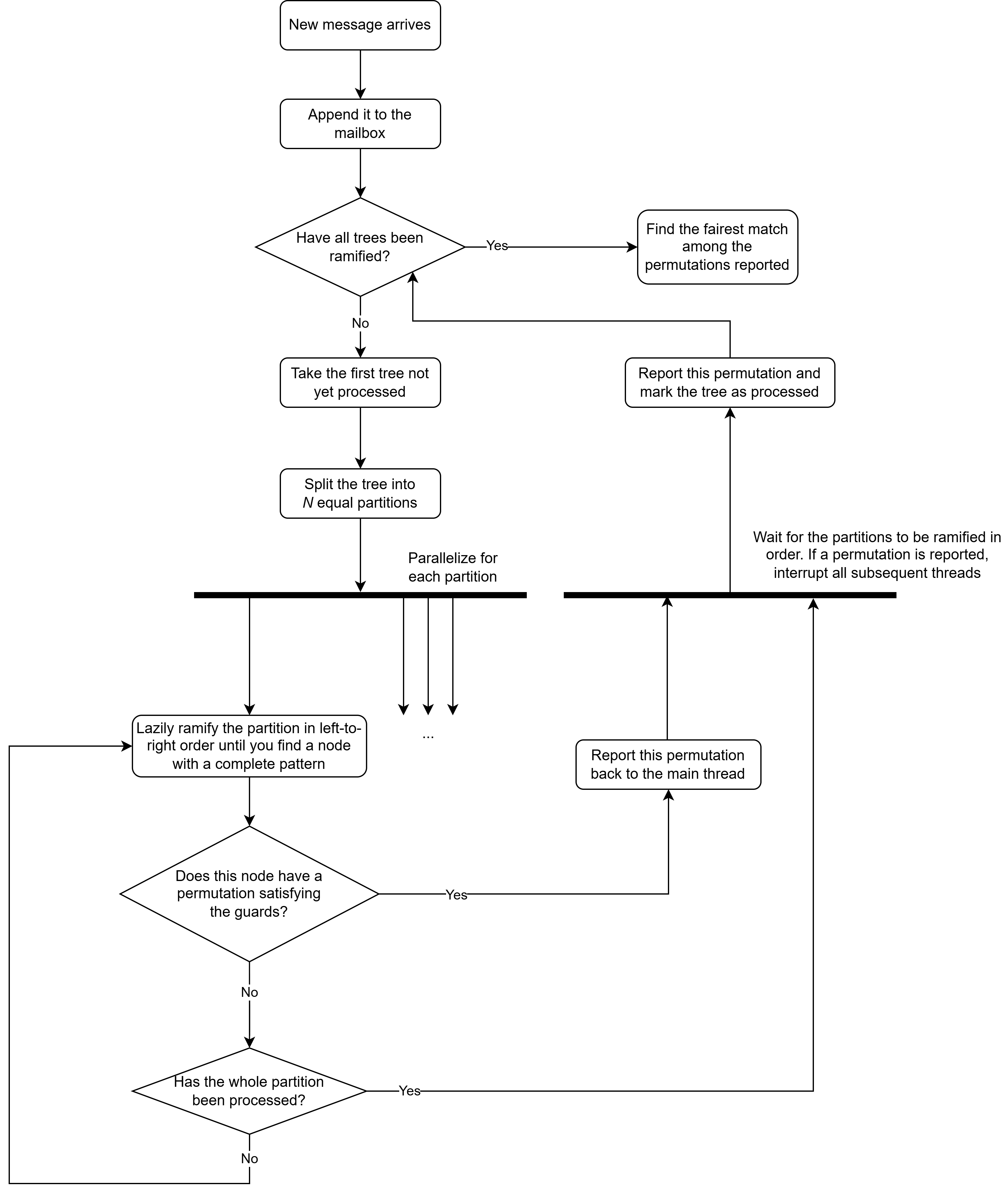}
    \caption{A flow diagram showcasing the operation of the lazy parallel algorithm}
    \label{fig:lazy_parallel_algorithm}
\end{figure}

After this, we focused on reconciling multithreading with laziness. Our lazy parallel algorithm again splits the workload among multiple threads. However, this time, whenever one worker thread finds a complete pattern, it finds valid permutations and checks the guard, as in the singlethreaded lazy algorithm. If the guard can be satisfied, it reports this to the main thread. The main thread waits for all the worker threads to finish in order of the portion of the tree they are operating on. Namely, it first waits for the thread operating on the portion of the tree that can yield the fairest matches, then the thread working on the portion with the next fairest matches, and so on. If one worker thread reports a complete node, the main thread does not wait for any subsequent worker threads, as these threads can only yield less fair matches. Instead, it interrupts those subsequent worker threads and immediately uses the match that was reported to it. Figure \ref{fig:lazy_parallel_algorithm} shows a flow diagram of this lazy parallel algorithm. \

\subsubsection{Message filtering}
\label{sec:message_filtering}

The flame graph for \texttt{WhileLazyMatcher} in Figure \ref{fig:while_lazy_flame_graph} shows that the vast majority of the runtime is spent on the \texttt{updateTree} method, which is used to ramify matching trees. From a theoretical standpoint, given a large enough join pattern with only one message type, every new ramification doubles the size of the matching tree. This results in exponential performance loss. Therefore, the ability to avoid certain ramifications would greatly help the performance of the algorithm. This motivated us to search for a method to avoid ramification whenever possible.

As mentioned previously, the Rete algorithm \cite{rete_original}, implemented in the Evrete library for Java, solves a very similar problem to the problem of join pattern matching. One intriguing characteristic of the Rete algorithm is the possibility of attaching conditions directly to the facts/objects (in our case, messages) before they are joined together. This allows for facts not satisfying a condition to be filtered out before any attempt is made to combine them with any other facts. In our case, a feature like this would allow us to skip tree ramification for certain messages, with a very positive impact on performance.

We design an algorithm that analyzes the Boolean structure of guard expressions and extracts conditions that can be applied directly to incoming messages. The algorithm checks for guards that consist of a conjunction of Boolean subexpressions (for simplicity, we call these subexpressions clauses no matter their form). If a guard is a conjunction of clauses, the algorithm checks which variables are bound in each clause, and which messages those variables come from. Whenever a clause only uses payloads from one message type, and that message type only appears once in the pattern, the clause is reported as a \textit{filtering clause} for that message type. Thus, filtering clauses are clauses that can be made false by a single message, thereby making the whole guard evaluate to false.

For example, consider the join pattern in Listing \ref{lst:filtering_clauses_example}. The guard on line 3 consists of four clauses: $x > 1$, $y = z$, $x \leq y$, and $p = 1$. The first clause, $x > 1$, only uses a variable bound in a message pattern with type \texttt{A}, and the type \texttt{A} only appears once in the pattern. Therefore, $x > 1$ is a filtering clause for messages of type \texttt{A}. The second clause, $y = z$, only uses variables from a message pattern with type \texttt{B}, and \texttt{B} only appears once in the pattern. Therefore, $y = z$ is also a filtering clause, but this time for messages of type \texttt{B}. The third clause, $x < y$, uses variables from two different message types, and is therefore not a filtering clause. Finally, the fourth clause, $p = 1$, only uses one variable, and this variable is bound in a message pattern with type \texttt{C}. However, the type \texttt{C} appears twice in the pattern, and so $p = 1$ is not a filtering clause.

\begin{listing}[h!]
\begin{minted}[bgcolor=bg, linenos]{scala}
receive[Msg, Unit] { _ => {
  case A(x) &:& B(y, z) &:& C(p) &:& C(q)
  if x > 1 && y == z && x <= y && p == 1 => ...
\end{minted}
\caption{An example of a join pattern with filtering clauses}
\label{lst:filtering_clauses_example}
\end{listing}

If we require that every variable in a clause comes from a single message in order for the clause to be a filtering clause, this unfairly excludes externally defined variables and functions. Therefore, we instead look at every combination of clause and single-occurence message type, and check that \textit{no} variable in the clause comes from \textit{any other} message type. If this is true, then we report the clause as a filtering clause for that message type.

We then make use of these filtering clauses in the matching algorithm. Whenever we receive a message, we test it against the corresponding filtering clauses for that message type, if there are any. If there is a clause which the message does not satisfy, we discard the message and avoid a tree ramification.

\subsubsection{Replacing matching trees with arrays}
\label{sec:replacing_matching_trees}

The flame graph for \texttt{WhileLazyMatcher} in Figure \ref{fig:while_lazy_flame_graph} shows a considerable amount time spent on the \texttt{addAll} and \texttt{pruneTree} methods of the \texttt{TreeMap} type. This \texttt{TreeMap} is a structure implemented as a red-black tree that we use to represent matching trees, as previously discussed. We reconsider this matching tree data structure with the aim of reducing the time spent on these methods. The tree-based matching algorithm essentially consists of traversing through all the nodes, and for each node potentially creating a new node. If the new node is complete, it can be investigated for a match, and if not, it can be added to the tree while maintaining a sorted order. No part of this algorithm specifically depends on the underlying data structure being a tree.

In this optimization, we replace the matching tree with a sorted array. When we receive a new message, we traverse this array, and build up a second array of nodes that should be added. Then, we perform a sorted merge algorithm on these two arrays (the \textit{merge} part of the mergesort algorithm), using the same special node ordering that we normally use in the \texttt{TreeMap}. Adding $n$ nodes to a red-black tree of size $n$ takes $O(n\log(n))$ time, while performing a sorted merge of two arrays of size $n$ takes $O(n)$ time. Therefore, this optimization decreases the theoretical time complexity of the ramification procedure.

\subsubsection{Cache-friendly data structures}
\label{sec:cache_friendly_data_structures}

Here we describe one last optimization which we implemented early on in the common code of all matchers, and therefore cannot be said to build on top of any matcher.

Upon inspecting the flame graphs for some of our matching algorithms, we noticed that a lot of time was spent modifying the \texttt{MessageIdxs} and \texttt{PatternIdxs} data structures. \texttt{MessageIdxs} is the structure we use for storing the sets of message indices which make up a matching tree, as explained in Section \ref{sec:stateful_tree_based_algorithm}. \texttt{PatternIdxs} is similar, but it is used by the satellite data of matching trees to store sets of constructor pattern indices.

Both of these types were originally type aliases for linked list data structures: more specifically, \texttt{MessageIdxs} was defined as a Scala \texttt{Queue}\footnote{\url{https://scala-lang.org/api/3.x/scala/collection/immutable/Queue.html}} and \texttt{PatternIdxs} was defined as a Scala \texttt{List}\footnote{\url{https://scala-lang.org/api/3.x/scala/collection/immutable/List.html}}. We changed both of these definitions so that they are instead represented using the Scala \texttt{ArraySeq} class,\footnote{\url{https://scala-lang.org/api/3.x/scala/collection/immutable/ArraySeq.html}} a simple immutable array structure. This yields two performance benefits:

\begin{enumerate}
    \item Arrays are more cache-friendly than linked lists, as they abide by the principle of locality. Therefore, less time is spent on memory fetch instructions when using them.

    \item Scala is a JVM language, and the \texttt{List} and \texttt{Queue} classes in Scala can only hold reference types, similarly to collection classes in Java. When we create a \texttt{List[Int]} or a \texttt{Queue[Int]}, all integers that we store in them are first boxed to object types. This is analogous to e.g. \texttt{List<Integer>} in Java. However, the \texttt{ArraySeq} class uses runtime reflection to automatically specialize to whichever primitive type it is given. Therefore, an \texttt{ArraySeq[Int]} is backed by an \texttt{Array[Int]}, which corresponds to a Java \texttt{int[]}, as opposed to an \texttt{Integer[]}. This way, we both avoid allocating objects and we avoid following object pointers. Avoiding allocations improves performance by itself, but avoiding pointer dereferences also improves performance as it follows the principle of locality.
\end{enumerate}

This change of representation was done in the common code used by all matchers. Therefore, all algorithms benefit from this optimization, and the main branch of the repository unfortunately does not allow for comparison between the performance of the two data structure configurations. However, we provide a repository branch called \texttt{dev-baseline}\footnote{\url{https://github.com/yaniskas/join-actors/tree/dev-baseline}} where the old implementation can be tested on all algorithms.

\subsection{Implementation}
\label{sec:implementation}

\begin{figure}[h!]
    \centering
    \includegraphics[width=0.9\linewidth]{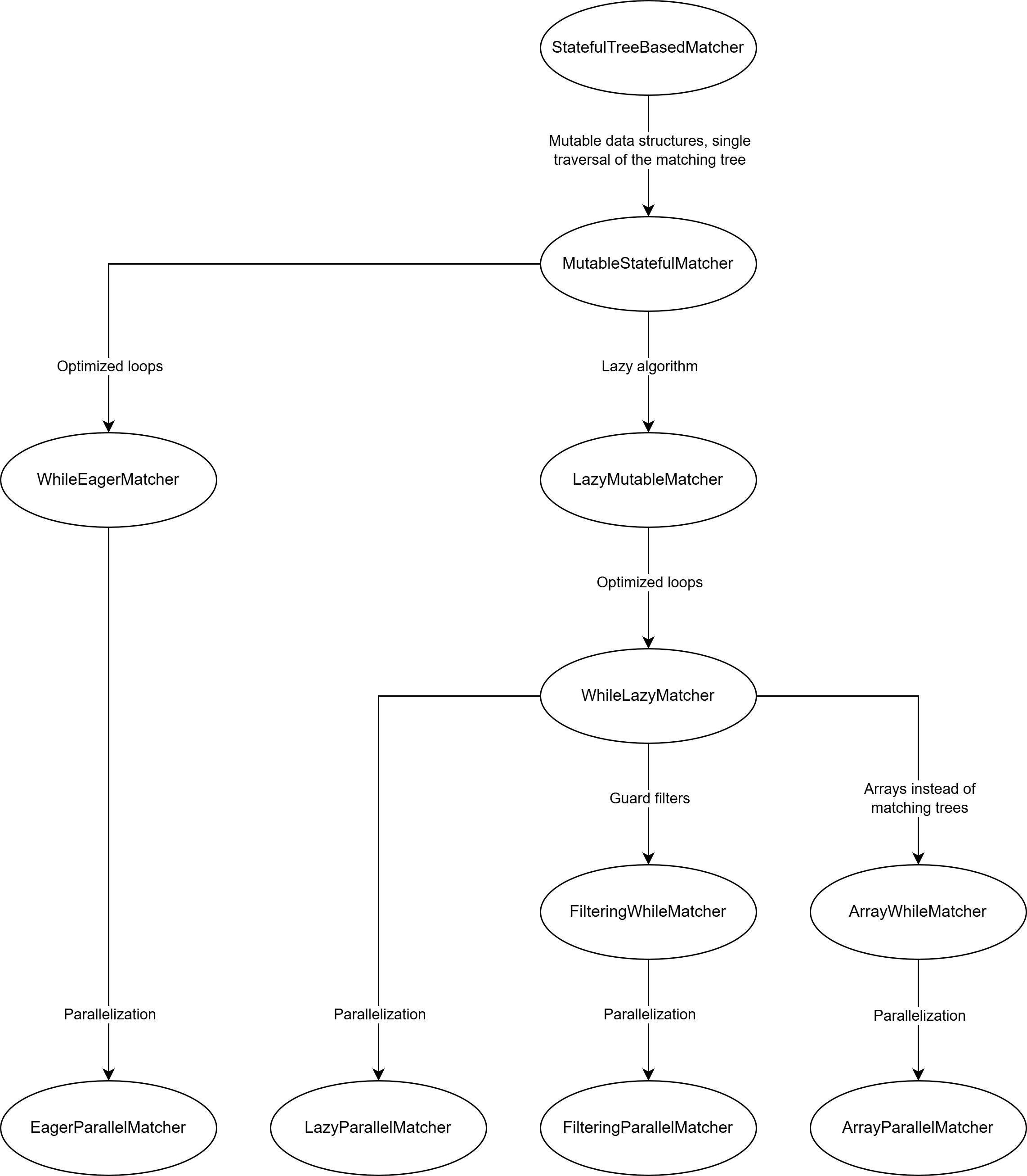}
    \caption{The evolution of our matcher implementations (replicated from Figure \ref{fig:matcher_evolution})}
    \label{fig:matcher_evolution_2}
\end{figure}

Here we discuss the details that we needed to consider to implement the optimizations, as well as the performance improvement shown by each \texttt{Matcher}. For convenience, we replicate our diagram showing the evolution of our matchers in Figure \ref{fig:matcher_evolution_2}. We name the following subsections after the matchers they describe. Namely:

\begin{itemize}
    \item \texttt{MutableStatefulMatcher} is described in Section \ref{sec:mutable_stateful_matcher}.

    \item \texttt{LazyMutableMatcher} is described in Section \ref{sec:lazy_mutable_matcher}.

    \item \texttt{WhileLazyMatcher} is described in Section \ref{sec:while_lazy_matcher}.

    \item \texttt{WhileEagerMatcher}, \texttt{EagerParallelMatcher}, and \texttt{LazyParallelMatcher} are described in Section \ref{sec:parallel_matchers}.

    \item \texttt{FilteringWhileMatcher} and \texttt{FilteringParallelMatcher} are described in Section \ref{sec:filtering_matchers}.

    \item \texttt{ArrayWhileMatcher} and \texttt{ArrayParallelMatcher} are described in Section \ref{sec:array_matchers}.
\end{itemize}

We use our new refactored benchmark suite for all benchmarks (described in detail in Section \ref{sec:benchmark_suite_improvements}). For most of our comparisons between matchers, we use the Simple Smart House benchmark described in Section \ref{sec:simple_smart_house}, with the number of matches set to 25 and the maximum number of prefix messages set to 20. We chose Simple Smart House because it is the most deterministic benchmark that has a fully configurable length, and we found that its plots show the performance improvements in our algorithms more clearly than the other benchmarks. We show the commands used to run all of our benchmarks in Appendix \ref{sec:benchmark_commands}, and we provide these commands in a file called \texttt{report\_benchmarks.txt} in the \texttt{benchmarks} project of the repository.

We run all our benchmarks on a computer with an Intel Core i7-10710U CPU and 16 GB of RAM running Windows 10. We use Scala 3.6.3 and JDK 23 with the maximum heap size set to 8 GB.

\begin{figure}[h!]
    \centering
    \includegraphics[width=0.75\linewidth]{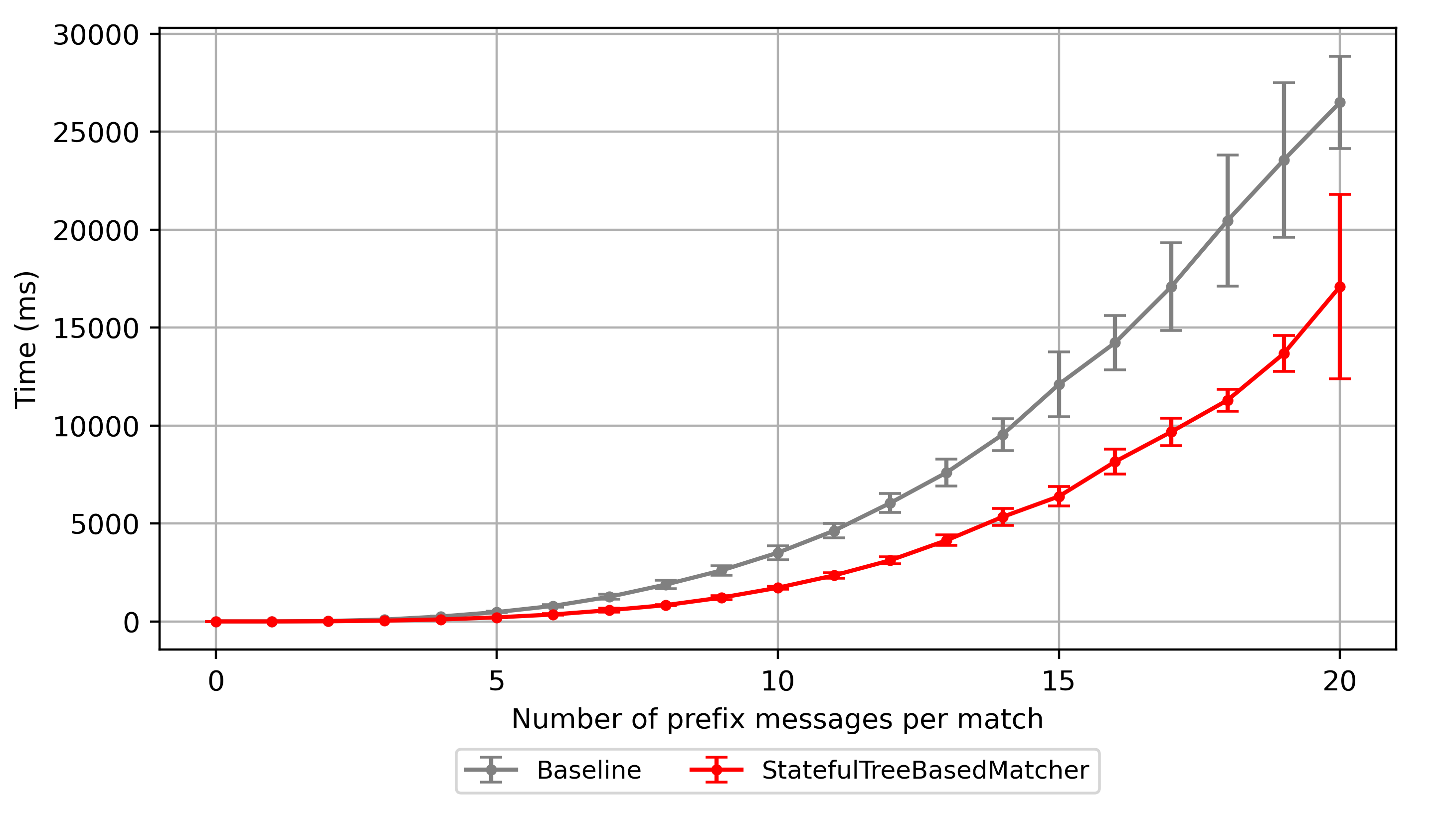}
    \caption{A comparison of the current \texttt{StatefulTreeBasedMatcher} against the baseline measurement (\texttt{StatefulTreeBasedMatcher} without the data structure optimization)}
    \label{fig:baseline_vs_stateful}
\end{figure}

We begin by establishing a baseline against which we can compare our matchers. As we discussed in Section \ref{sec:cache_friendly_data_structures}, our optimization related to cache-friendly data structures was performed in the common code shared by all matchers, so all matchers benefit from it in the main branch of the repository. Therefore, in order to create a baseline measurement, we use the \texttt{dev-baseline} branch of the repository where this change is reverted, and run a benchmark on the existing \texttt{StatefulTreeBasedAlgorithm}. Figure \ref{fig:baseline_vs_stateful} shows a comparison between this baseline measurement and the \texttt{StatefulTreeBasedAlgorithm} in the \texttt{main} branch of the repository, which includes the \texttt{ArraySeq} optimization. This way, we show the effect of the \texttt{ArraySeq} optimization by itself. As can be seen, this optimization alone reduces the time spent on the last parameter of the benchmark by approximately a third. Apart from this baseline measurement, all of our benchmarks from here on out include the \texttt{ArraySeq} optimization.

\subsubsection{\texttt{MutableStatefulMatcher}}
\label{sec:mutable_stateful_matcher}

The \texttt{MutableStatefulMatcher} class is a version of the old \texttt{StatefulTreeBasedMatcher} with the improvements discussed in section \ref{sec:mutability_and_single_traversal}, namely the use of mutable data structures and only traversing the matching tree once. This matcher ramifies the tree, and whenever it creates a new node, it checks whether this node is complete. If so, it adds the node to a list. Once the ramification is done, it returns this list of complete patterns for further processing.

\begin{figure}[h!]
    \centering
    \includegraphics[width=0.75\linewidth]{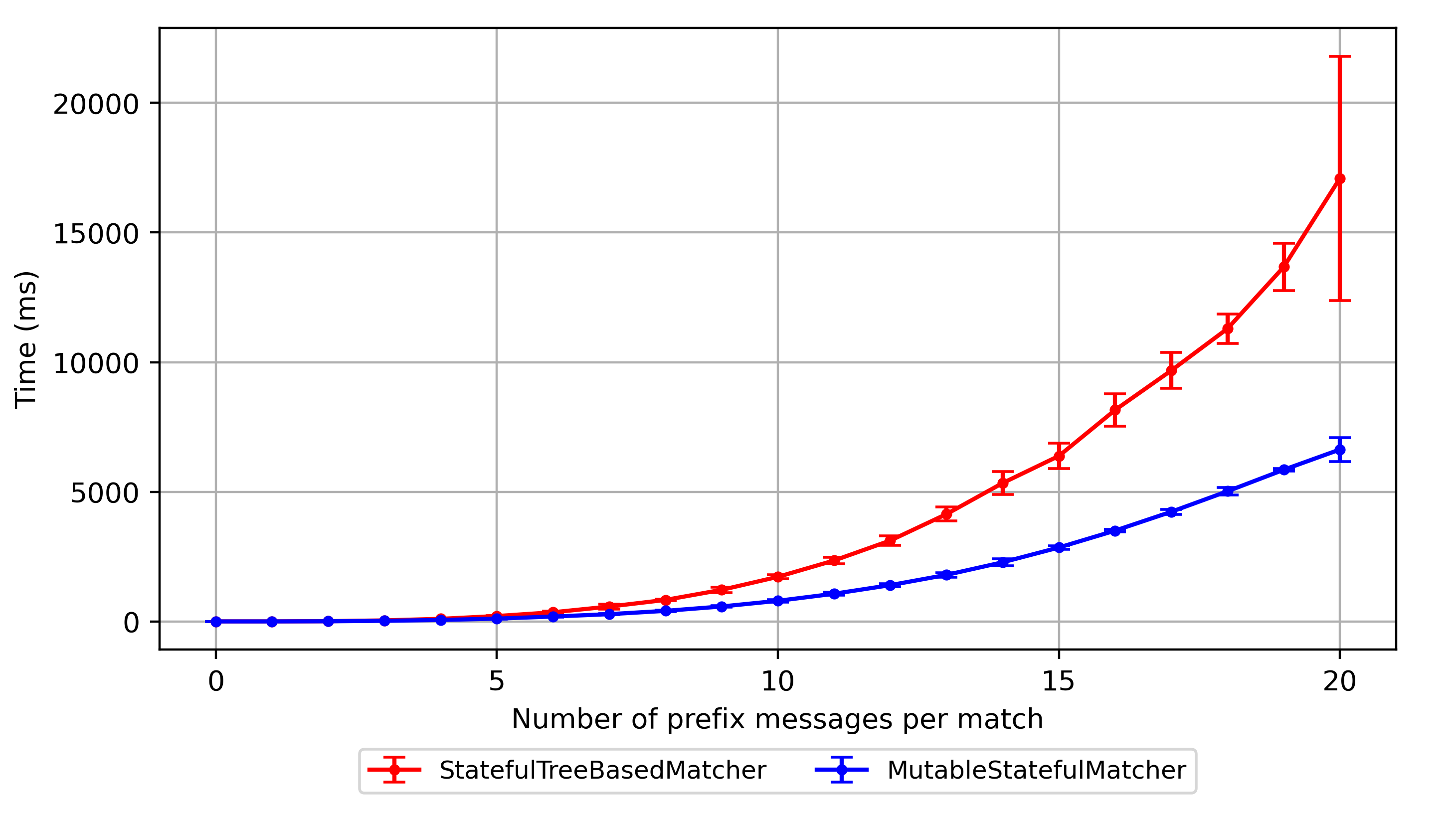}
    \caption{A comparison of \texttt{MutableStatefulMatcher} against \texttt{StatefulTreeBasedMatcher} on the Simple Smart House benchmark}
    \label{fig:immutable_vs_mutable}
\end{figure}

Figure \ref{fig:immutable_vs_mutable} shows the result of this optimization on the Simple Smart House benchmark. This optimization results in approximately double performance on the last parameter of the benchmark.

\subsubsection{\texttt{LazyMutableMatcher}}
\label{sec:lazy_mutable_matcher}

The \texttt{LazyMutableMatcher} class implements the optimization discussed in section \ref{sec:lazy_ramification}, namely lazy ramification of the matching tree. When a node with complete patterns is found, we immediately check for valid permutations that satisfy the guard. If the check is successful, we stop the algorithm and report the fairest valid permutation. If not, we continue the traversal. Either way, we do not add the complete node to the tree.

A nontrivial implementation detail of this optimization is how to achieve the stopping of the algorithm once a valid permutation is found. One option would be the functional approach: implementing the algorithm as a recursive function, and encoding the possibility of early stopping in the return type of the function. However, this would introduce performance overhead, and we had already committed to a performant mutable coding style with the previous optimization, so we did not want to undo that work.

Instead, we decided to opt for the boundary/break API introduced in Section \ref{sec:boundary_break_API}. The advanced expression-oriented \texttt{break} statement defined by that API precisely matches our use case: we want to provide a value exactly when we break out of the algorithm.

We wrap the main loop of our algorithm in a \texttt{boundary} expression, and we assign a value with the \texttt{Option} type to this expression. If we find a valid match, we use \texttt{break} to provide information about the match wrapped in the \texttt{Some} case of \texttt{Option}. We also include the \texttt{None} case at the end of the boundary, and this is what the boundary evaluates to if no \texttt{break} is encountered.

\begin{figure}[h!]
    \centering
    \includegraphics[width=0.75\linewidth]{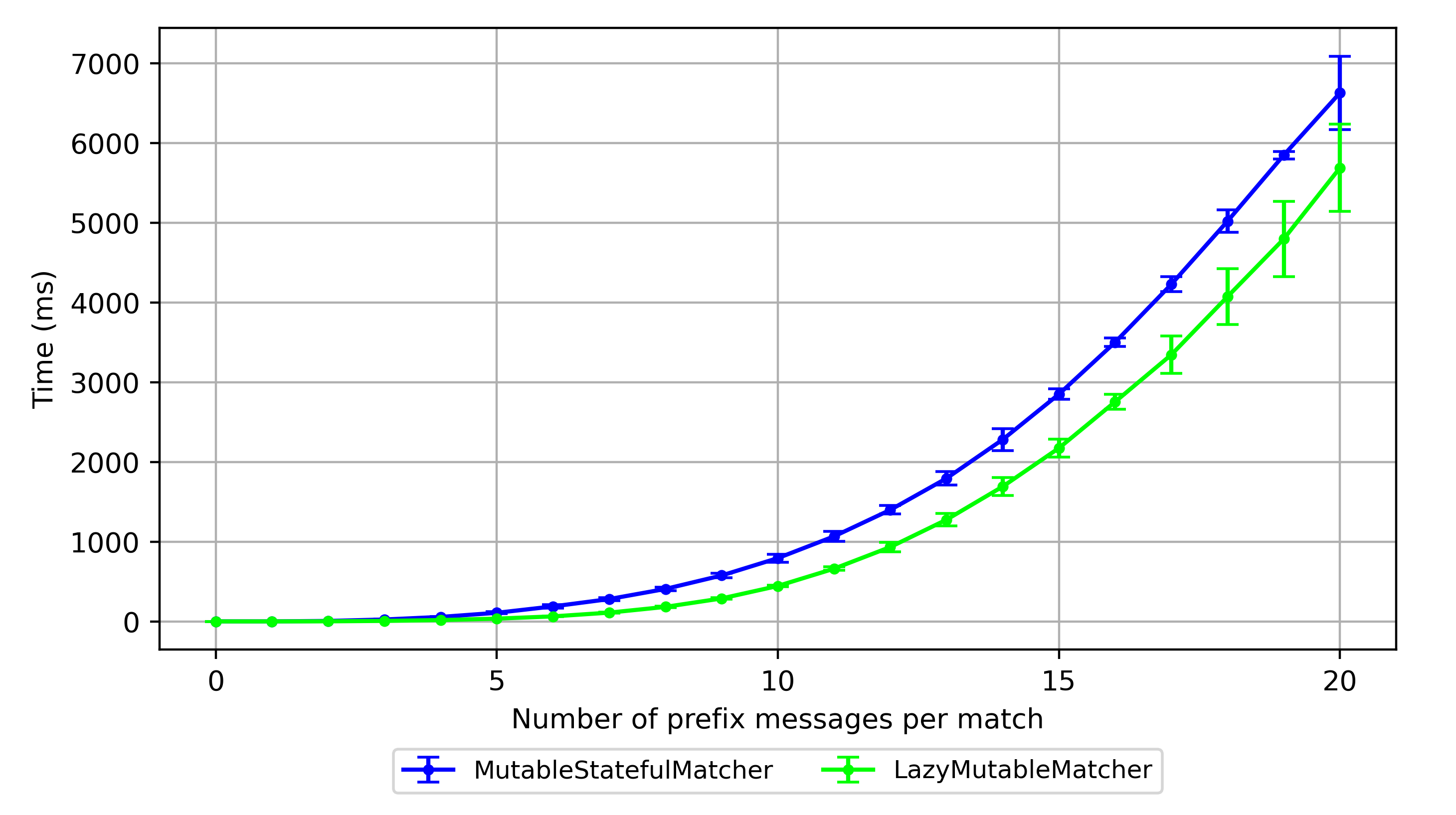}
    \caption{A comparison of \texttt{LazyMutableMatcher} against \texttt{MutableStatefulMatcher} on the Simple Smart House benchmark}
    \label{fig:mutable_vs_lazy}
\end{figure}

Figure \ref{fig:mutable_vs_lazy} shows the effect of this optimization. While not as drastic as the effects of the cache-friendly \texttt{ArraySeq} data structure or the single-traversal optimization, it is still substantial.

\subsubsection{\texttt{WhileLazyMatcher}}
\label{sec:while_lazy_matcher}

The \texttt{WhileLazyMatcher} class implements the optimization discussed in Section \ref{sec:optimized_loops}, namely optimized loops. By default, Scala for loops are compiled into a call to a \texttt{foreach} method with a lambda function, which has a substantial performance impact. For example, a code snippet like

\begin{center}
    \texttt{for v <- coll do f(v)}
\end{center}

is converted by the compiler into

\begin{center}
    \texttt{coll.foreach(v => f(v))}
\end{center}

Since Scala also offers \texttt{while} loops that are compiled more efficiently, we would like to use these instead. However, we are able to avoid explicitly writing cumbersome \texttt{while} loops. Instead, we make use of macros and the extensions to the type system in Scala 3 to write a conversion method for collections that can turn \texttt{for} loops on that collection into \texttt{while} loops. This way, there is very little impact on code readability. Our method is inspired by the discussion in \cite{scala_for_inefficiency}, as well as the suggestions in \cite{c_style_loop_in_scala}.

We define the conversion as a macro extension method called \texttt{fast} on the type \texttt{IterableOnce}. \texttt{IterableOnce} is a standard library type which is the supertype of all collection types as well as all iterators. Using this macro, a code snippet like

\begin{center}
    \texttt{for v <- coll.fast do f(v)}
\end{center}

is converted into

\begin{center}
    \texttt{val it = coll.iterator; while it.hasNext do f(it.next())}
\end{center}

The code we use to achieve this conversion is shown in Listing \ref{lst:fast_iterable}. We first define an \textit{opaque type}\footnote{\url{https://docs.scala-lang.org/scala3/book/types-opaque-types.html}} called \texttt{FastIterable} on line 1. This is an alias of \texttt{IterableOnce}, but since it is opaque, it appears as a separate type outside of this file. On lines 3-7 we define an extension method on \texttt{FastIterable} called \texttt{foreach}, which takes a function parameter and returns \texttt{Unit}. This method signature allows it to be used as the implementation for a Scala \texttt{for} comprehension on the \texttt{FastIterable} type. In this \texttt{foreach} method, we take the iterator of the \texttt{IterableOnce} and use it in a \texttt{while} loop. We define this \texttt{foreach} method as an \texttt{inline} method\footnote{\url{https://docs.scala-lang.org/scala3/guides/macros/inline.html}} with its parameter also \texttt{inline}, thereby making it a simple macro. Finally, on lines 9-10, we define the extension method \texttt{fast} on \texttt{IterableOnce}, which simply changes the static type of the object calling the method into \texttt{FastIterable}.

\begin{listing}[h!]
\begin{minted}[bgcolor=bg, linenos]{scala}
opaque type FastIterable[T] = IterableOnce[T]

extension[T] (self: FastIterable[T])
  inline def foreach[U](inline f: T => U): Unit =
    val it = self.iterator
    while it.hasNext do
      f(it.next())

extension[T] (self: IterableOnce[T])
  def fast: FastIterable[T] = self
\end{minted}
\caption{The \texttt{FastIterable} type and related methods we use to optimize \texttt{for} loops}
\label{lst:fast_iterable}
\end{listing}

When we use a \texttt{for} comprehension on a collection converted with the \texttt{.fast} method, the effect is the following:

\begin{enumerate}
    \item Scala sees that the \texttt{for} comprehension is used on a value of type \texttt{FastIterable}. Since this type has a \texttt{foreach} method with a valid signature, the \texttt{for} comprehension is replaced with a call to this \texttt{foreach} method.

    \item Since the \texttt{foreach} method on \texttt{FastIterable} is a macro, its body is substituted into the call site. Since the function parameter is also \texttt{inline}, its body is substituted into the \texttt{while} loop coming from the \texttt{foreach} method.

    \item Since \texttt{FastIterable} is an opaque type and thereby a type alias, it is erased and has no effect on the generated bytecode.
\end{enumerate}

Thus, the fact that \texttt{FastIterable} is an opaque type instead of a regular class, combined with the fact that its \texttt{foreach} method is a macro, means that there is zero abstraction cost for using this system instead of manually writing a \texttt{while} loop.

\begin{figure}[h!]
    \centering
    \includegraphics[width=0.75\linewidth]{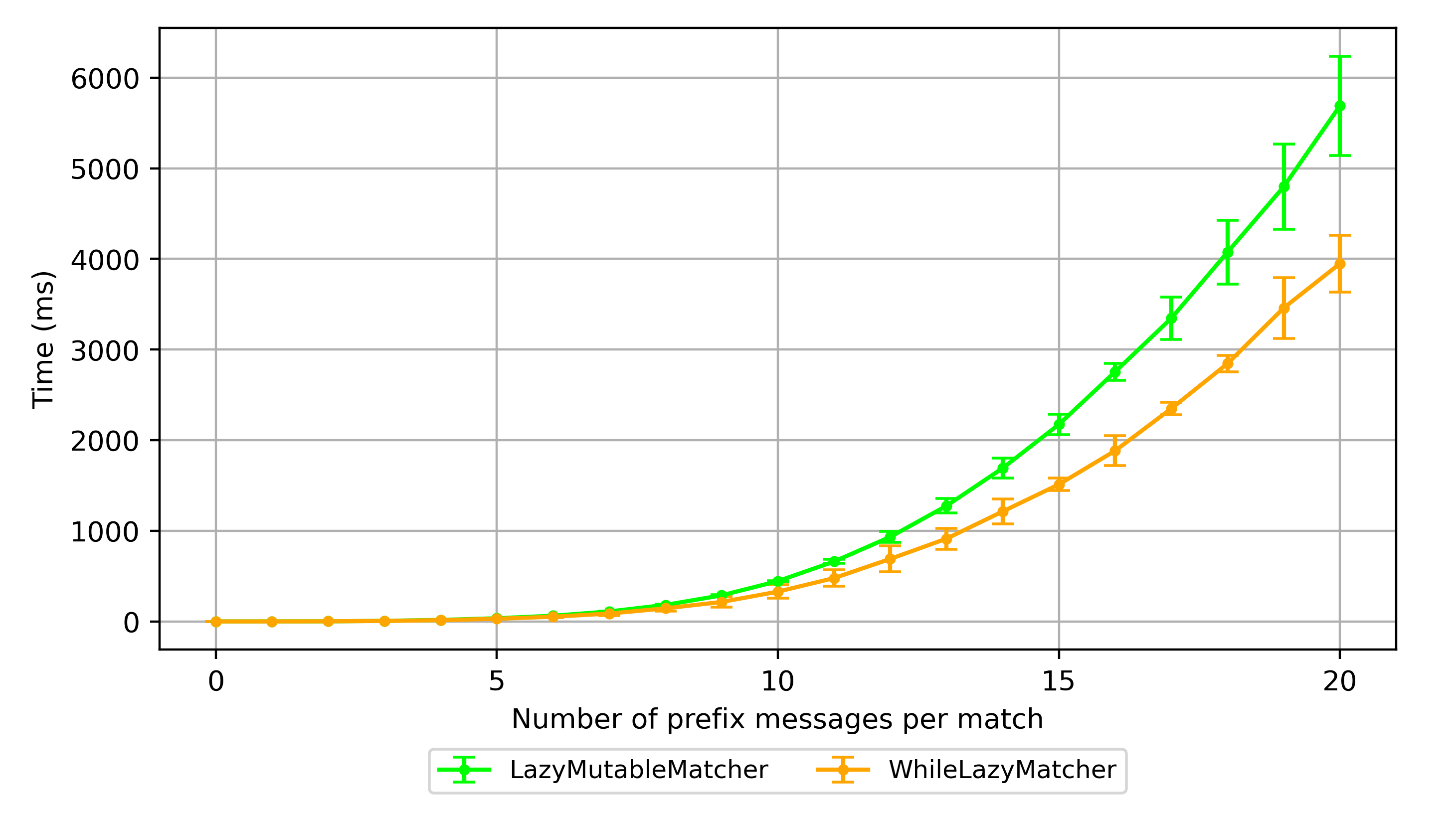}
    \caption{A comparison of \texttt{WhileLazyMatcher} against \texttt{LazyMutableMatcher} on the Simple Smart House benchmark}
    \label{fig:lazy_vs_while}
\end{figure}

The effect of this optimization is shown in Figure \ref{fig:lazy_vs_while}. As can be seen, the runtime is decreased by approximately a third on the last parameter.

\subsubsection{\texttt{EagerParallelMatcher}, \texttt{WhileEagerMatcher}, and \texttt{LazyParallelMatcher}}
\label{sec:parallel_matchers}

\begin{figure}[h!]
    \centering
    \includegraphics[width=0.75\linewidth]{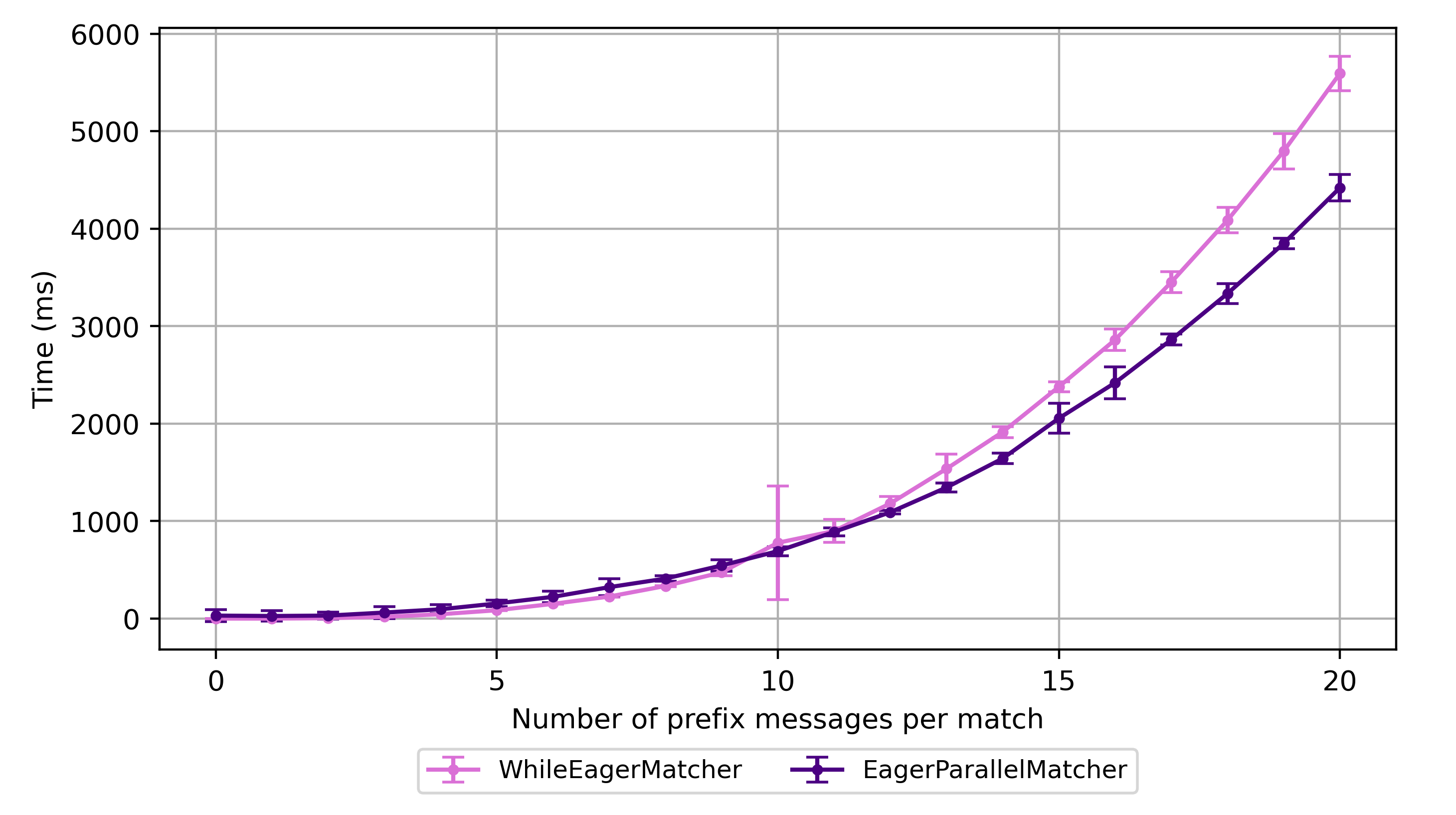}
    \caption{A comparison of \texttt{EagerParallelMatcher} against \texttt{WhileEagerMatcher} on the Simple Smart House benchmark}
    \label{fig:while_eager_vs_eager_parallel}
\end{figure}

As discussed in section \ref{sec:multithreading}, we implement an eager multithreaded matching algorithm as well as a lazy one. The eager version is found in the class \texttt{EagerParallelMatcher}. Here, we use Java's \texttt{Spliterator} API\footnote{\url{https://docs.oracle.com/en/java/javase/23/docs/api/java.base/java/util/Spliterator.html}} to split the matching tree into $N$ partitions, and then spawn one thread to fully ramify on each partition, where $N$ is a configurable parameter. In order to have a fair comparison for this eager multithreaded algorithm, we also provide a singlethreaded algorithm with optimized loops but without laziness in the class \texttt{WhileEagerMatcher}. Figure \ref{fig:while_eager_vs_eager_parallel} shows a comparison between these two algorithms. In all our benchmarks concerning multithreaded algorithms, here and in all subsequent plots, we configure the multithreaded algorithms to use 8 threads. This is because the \texttt{Spliterator} class can only evenly split the tree into multiples of 2, and the computer we use for the benchmarks has 6 cores. Therefore, 8 is the smallest choice that evenly splits the tree while still allowing all CPU cores to be occupied.

As can be seen, parallelization does result in a performance improvement for the eager algorithm, but it is a very slight improvement. We speculate that synchronization overhead plays a role in this.

We also implement the lazy multithreaded algorithm specified in Section \ref{sec:multithreading} in the class \texttt{LazyParallelMatcher}. This algorithm again uses the \texttt{Spliterator} class to parallelize the workload. However, this time, whenever one worker thread finds a complete pattern, it checks if a match can be produced. If so, it reports this to the main thread. The main threads waits for all the worker threads to finish in order of the portion of the tree they are operating on. If one worker thread reports a complete node, the main thread interrupts all subsequent worker threads and uses the match that was reported to it.

All synchronization in \texttt{ParallelEagerMatcher} is implemented using Scala's \texttt{Promise} type. In \texttt{ParallelLazyMatcher}, we make use of Java's \texttt{ExecutorService} API to allow for thread interruptions. \texttt{ExecutorService} works with the Java \texttt{Future} class, so we use these for synchronization instead of Scala promises.

\begin{figure}[h!]
    \centering
    \includegraphics[width=0.75\linewidth]{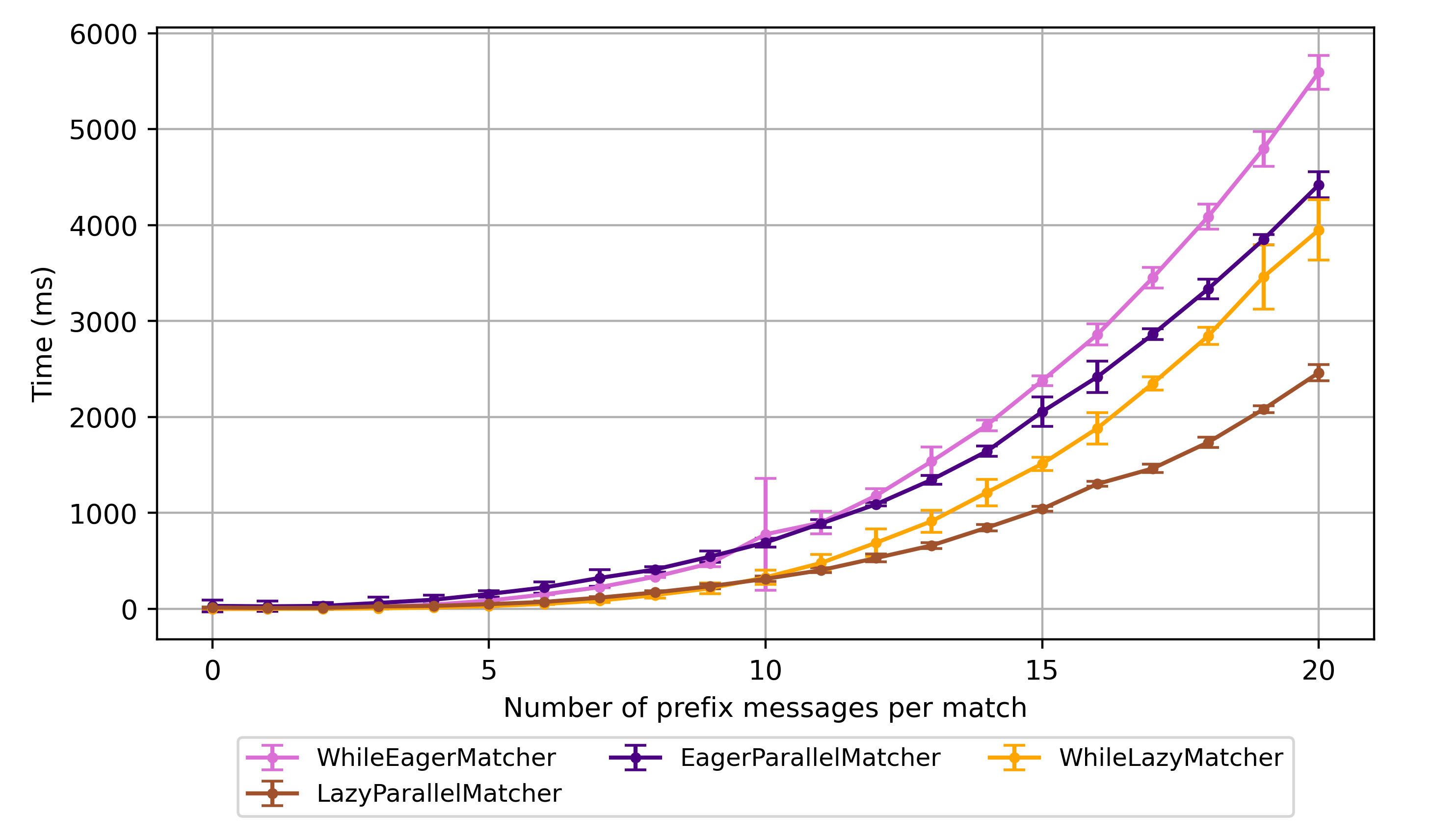}
    \caption{A comparison of \texttt{WhileEagerMatcher}, \texttt{EagerParallelMatcher}, \texttt{WhileLazyMatcher}, and \texttt{LazyParallelMatcher} on the Simple Smart House benchmark}
    \label{fig:while_lazy_vs_lazy_parallel}
\end{figure}

Figure \ref{fig:while_lazy_vs_lazy_parallel} shows a comparison between this lazy parallel matcher, the corresponding singlethreaded matcher (\texttt{WhileLazyMatcher}), and the two matchers from the previous plot. The lazy algorithm benefits more from parallelization than the eager algorithm, as the performance almost doubles on the last parameter. However, as the computer running the benchmarks has 6 cores, the theoretical optimal speedup is much higher than this. This shows that synchronization overhead still plays a major role.

\subsubsection{\texttt{FilteringWhileMatcher} and \texttt{FilteringParallelMatcher}}
\label{sec:filtering_matchers}

The classes \texttt{FilteringWhileMatcher} and \texttt{FilteringParallelMatcher} implement the guard filtering functionality discussed in Section \ref{sec:message_filtering} in a singlethreaded and multithreaded fashion respectively. In order to implement this functionality, we expand on our use of the powerful macro system in Scala 3. We enhance the \texttt{receive} macro with code that analyzes the abstract syntax trees of pattern guards. We check for a Boolean structure consisting of a conjunction of clauses, and then we analyze each clause to see where its variables come from. As previously discussed, we look at every combination of clause and single-occurrence message type, and report the clause as filtering if no variables in the clause come from any other message type.

Once we have identified the filtering clauses, we provide closures called \textit{filtering lambdas} for all the message types in the pattern. If filtering clauses exist, the lambda evaluates the conjunction of the filtering clauses, whereas if there are no filtering clauses, the lambda simply returns \texttt{true}. These lambdas act as tests that can be used by a matcher on all incoming messages. If a message makes its corresponding filtering lambda evaluate to \texttt{false}, it can be discarded.

As an added note, it is worth mentioning that the method we use for detecting conjunctions works for expressions returned from a macro. Therefore, one can avoid writing long expressions in pattern guards by hiding them behind a simple macro implemented as an inline function with inline arguments.

We provide two matcher implementations that use these filtering lambdas, one singlethreaded and one multithreaded. \texttt{FilteringWhileMatcher} is a version of \texttt{WhileLazyMatcher} with filtering functionality, and \texttt{FilteringParallelMatcher} is a version of \texttt{LazyParallelMatcher} with filtering functionality. In both cases, whenever a new message arrives, the matcher tests the message against the corresponding filtering lambda. If the lambda returns \texttt{false}, the message is immediately discarded. The message is only considered for addition to the matching tree if the lambda returns \texttt{true}. In all other respects, the matchers are identical to their non-filtering counterparts.

\begin{figure}[h!]
    \centering
    \includegraphics[width=0.75\linewidth]{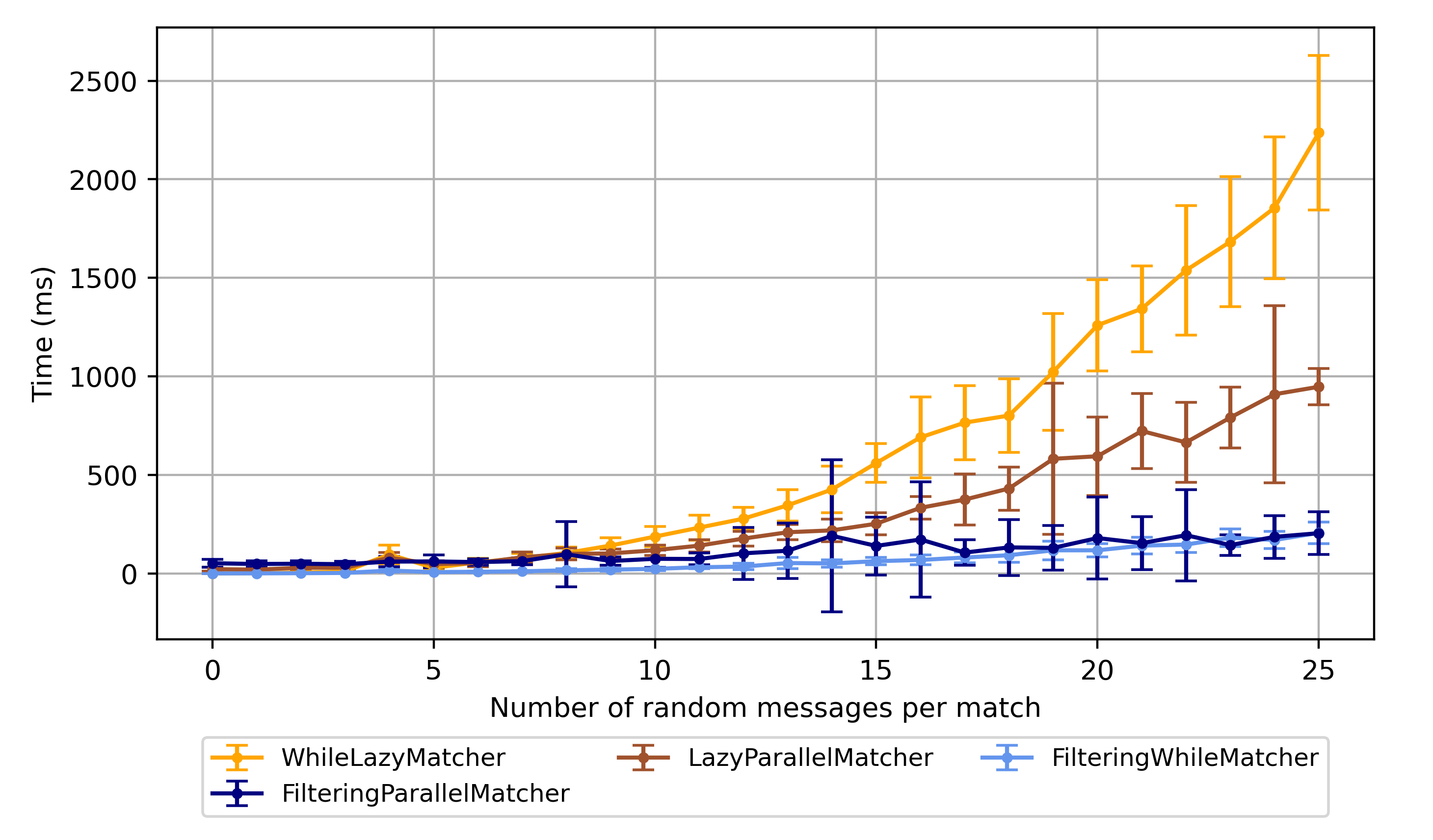}
    \caption{A comparison of the filtering matching algorithms against their non-filtering counterparts on the Complex Smart House benchmark}
    \label{fig:filtering_matchers}
\end{figure}

\begin{figure}[h!]
    \centering
    \includegraphics[width=0.75\linewidth]{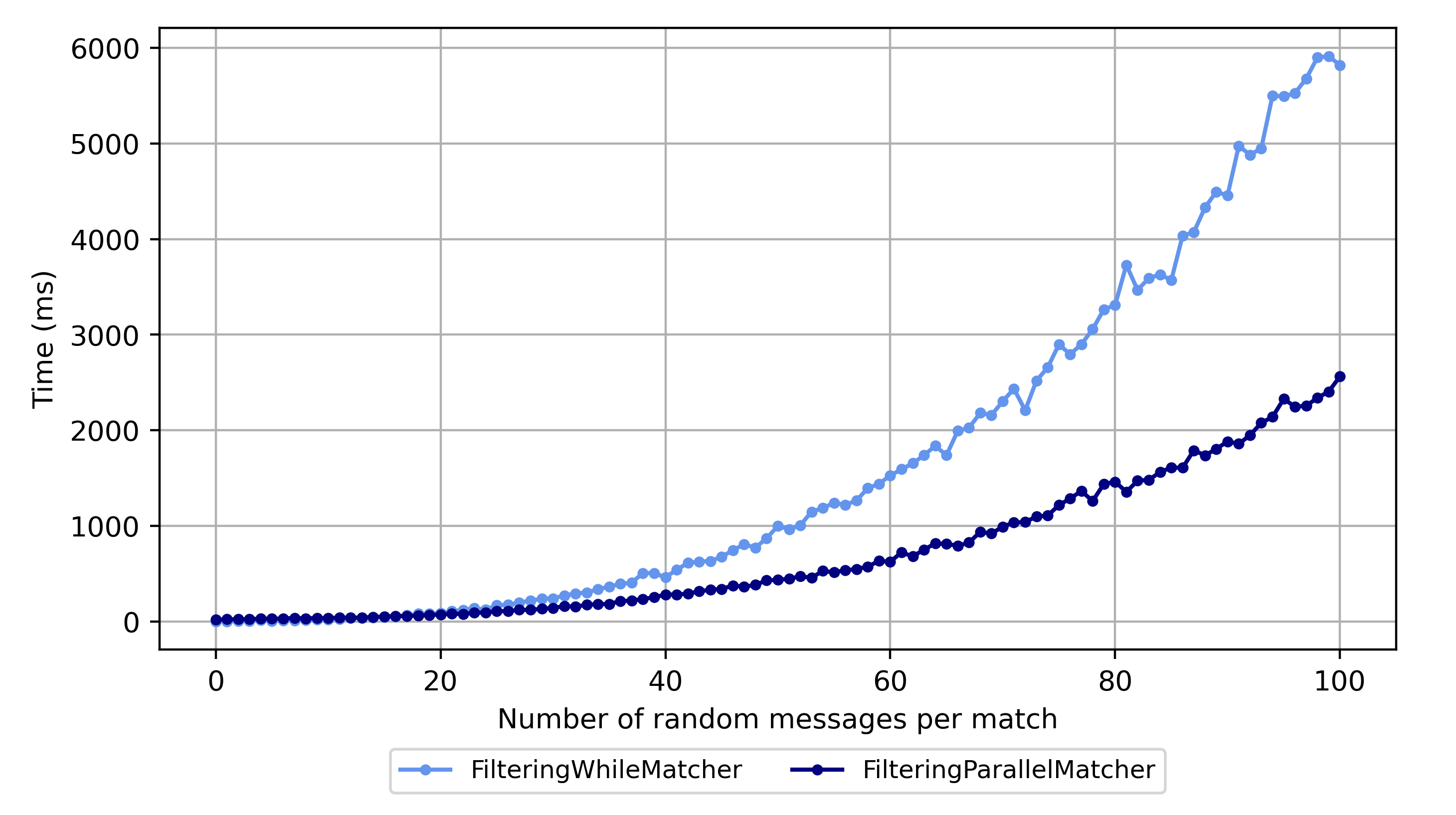}
    \caption{A comparison of the filtering matching algorithms against each other on a longer version of the Complex Smart House benchmark}
    \label{fig:filtering_matchers_long}
\end{figure}

So far we have been using the Simple Smart House benchmark for our comparisons; however, this benchmark never uses messages that are rejected by a guard. In fact, the only benchmark that uses such messages and has filtering clauses in its guards (as we have defined them) is the Complex Smart House benchmark, described in Section \ref{sec:complex_smart_house}. Therefore, this is the only benchmark that benefits from these filtering algorithms. Figure \ref{fig:filtering_matchers} shows the performance of these filtering algorithms on this benchmark compared to their non-filtering counterparts. Both filtering algorithms vastly outperform the previous algorithms, with even the singlethreaded one achieving performance approximately five times faster on the last parameter than the fastest multithreaded algorithm, \texttt{LazyParallelMatcher}. This graph does not show the performance difference between \texttt{FilteringWhileMatcher} and \texttt{FilteringParallelMatcher}, so we also run a longer benchmark comparing these two matchers against each other; this is shown in Figure \ref{fig:filtering_matchers_long}. We omit error bars in this plot due to the number of data points, to avoid cluttering it. Similarly to our comparison between \texttt{WhileLazyMatcher} and \texttt{LazyParallelMatcher}, the parallel algorithm once again achieves approximately double the performance of the singlethreaded one. Again, the commands used to run these benchmarks are provided in Appendix \ref{sec:benchmark_commands}.

\subsubsection{\texttt{ArrayWhileMatcher} and \texttt{ArrayParallelMatcher}}
\label{sec:array_matchers}

The classes \texttt{ArrayWhileMatcher} and \texttt{ArrayParallelMatcher} implement the optimization discussed in Section \ref{sec:replacing_matching_trees}, namely replacing matching trees with arrays, in a singlethreaded and multithreaded fashion respectively.

Due to the limited applicability of the filtering optimization, we do not include it in these matchers. We build these array optimizations on top of the \texttt{WhileLazyMatcher} from Section \ref{sec:while_lazy_matcher} and the \texttt{LazyParallelMatcher} from Section \ref{sec:parallel_matchers}. In \texttt{ArrayWhileMatcher}, we replace the matching tree with an array that we maintain sorted. Every time we receive a new message, we construct an array of additions by initializing an empty array with the same size as the node array, and then populating it by ramifying the node array. Then, we create an array that contains the original nodes merged with the nodes to be added. Finally, we set the array of nodes to this merged array. In \texttt{ArrayParallelMatcher}, we again have a sorted array instead of a tree. Like in \texttt{LazyParallelMatcher}, when we receive a new message, we use the \texttt{Spliterator} API to split the array into partitions and spawn a worker thread to work on each partition. If a worker reports a complete node, the main thread uses this and interrupts the other threads. Otherwise, the worker threads build up new arrays of additions, and then the main thread combines these into one array, and performs a sorted merge with the original node array. In both matchers, we ``prune" the array simply by creating a new array with the correct elements removed.

\begin{figure}[h!]
    \centering
    \includegraphics[width=0.75\linewidth]{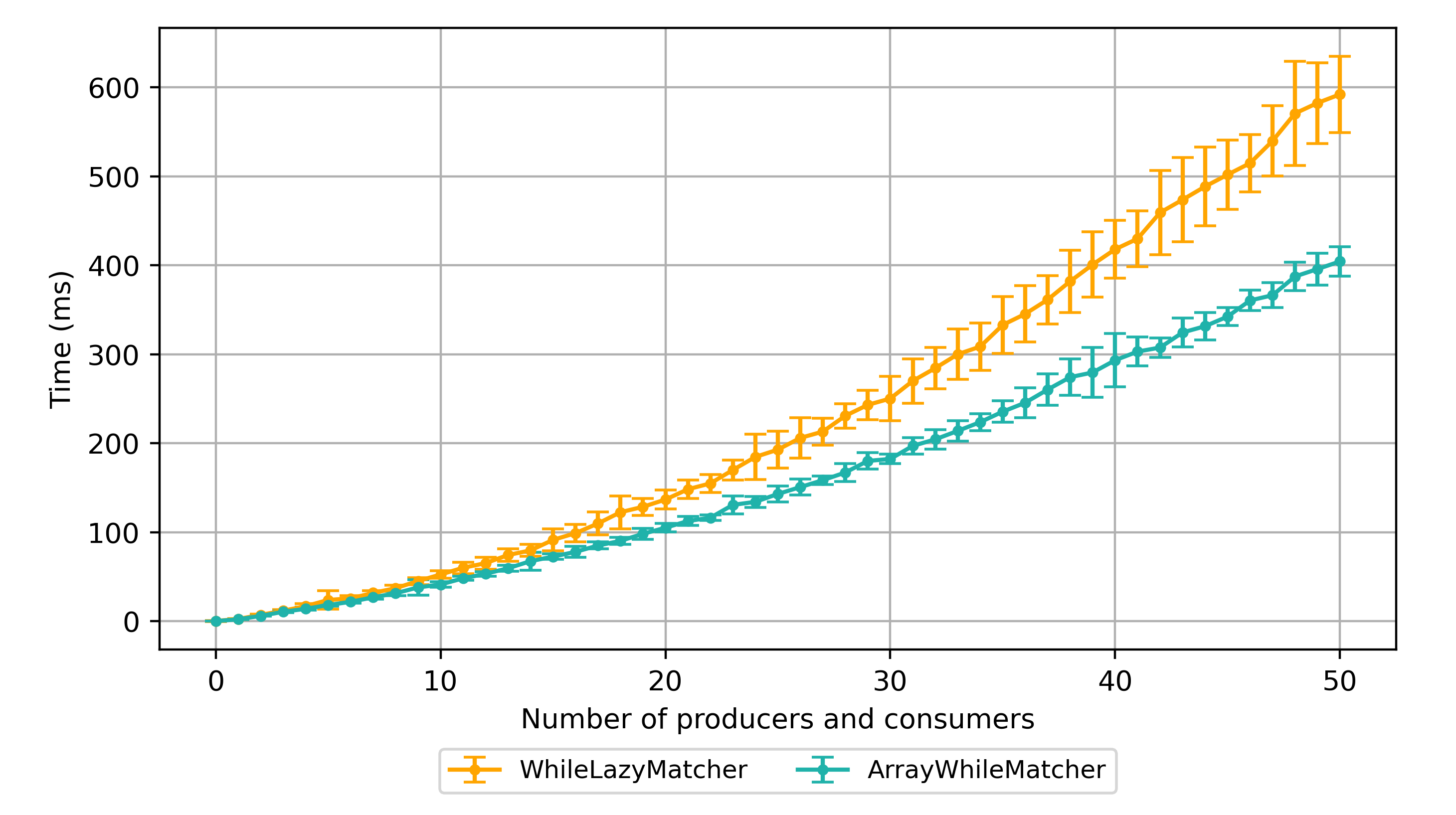}
    \caption{A comparison of \texttt{ArrayWhileMatcher} against \texttt{WhileLazyMatcher} on the Bounded Buffer benchmark}
    \label{fig:while_vs_array}
\end{figure}

\texttt{ArrayWhileMatcher} achieves better performance than \texttt{WhileLazyMatcher} on the Bounded Buffer benchmark described in Section \ref{sec:bounded_buffer}; this is shown in Figure \ref{fig:while_vs_array}. However, it achieves similar performance on Complex Smart House, and worse performance on Simple Smart House. Thus, this is another optimization that appears to have limited applicability.

The main distinguishing feature between Simple Smart House and Bounded Buffer is that Simple Smart House allows the matching tree to grow indefinitely, while Bounded Buffer does not. Therefore, we can speculate that the worse performance on Simple Smart House is due to the large size of the ramifications that need to be performed. For every ramification, \texttt{ArrayWhileMatcher} needs to allocate a large amount of space twice: once to store the nodes that need to be added, and once to store the new merged array. Meanwhile, \texttt{WhileLazyMatcher} allocates a dynamic array to store the nodes to be added, and then it can add nodes to the tree with small allocations. On Bounded Buffer, messages are constantly both added and consumed, so the array in \texttt{ArrayWhileMatcher} can stay small and benefit from memory locality. Meanwhile, \texttt{WhileLazyMatcher} needs to constantly allocate many small memory locations with no locality.

\begin{figure}[h!]
    \centering
    \includegraphics[width=0.75\linewidth]{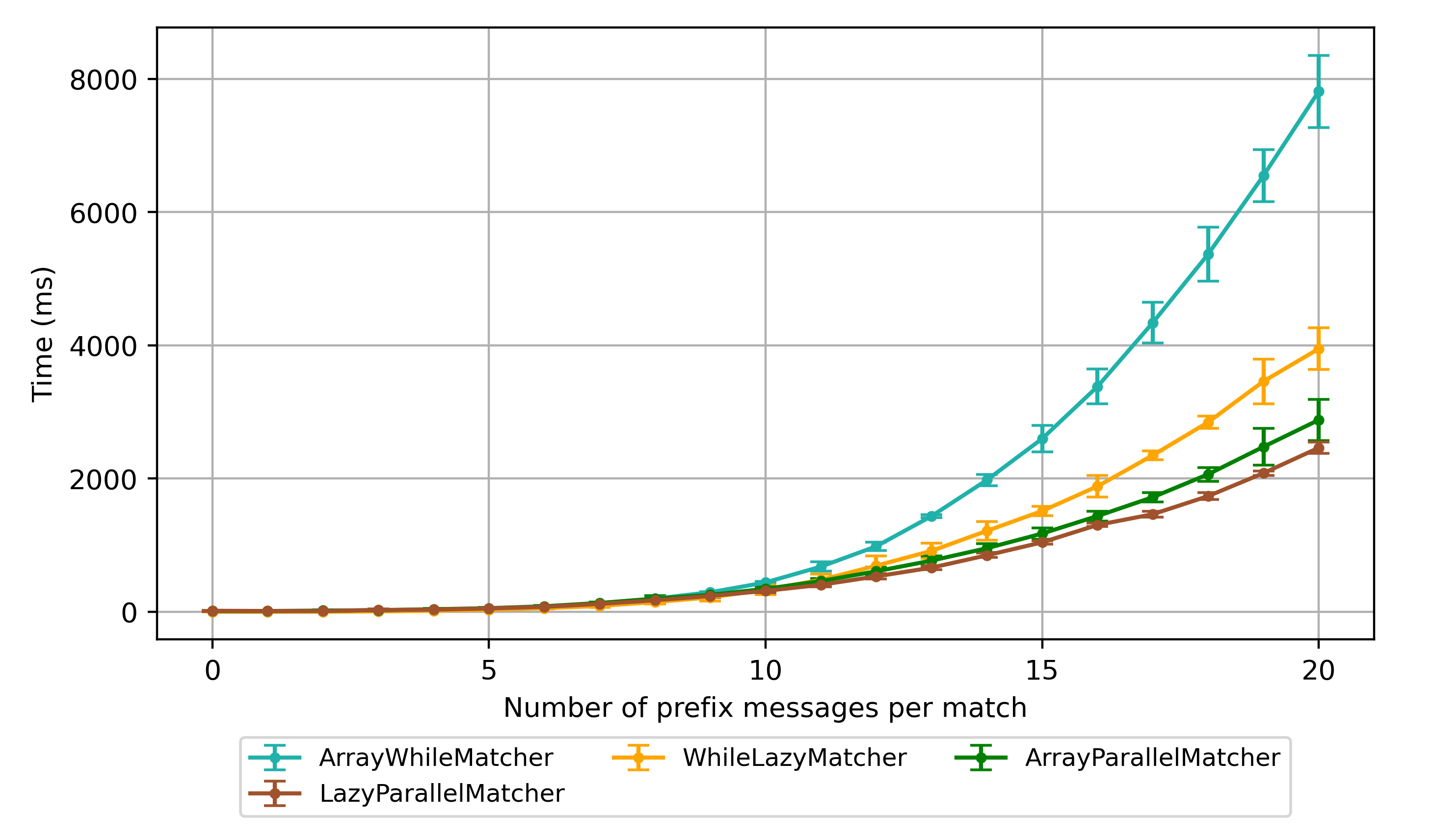}
    \caption{A comparison of the two array-based matchers against their tree-based counterparts on the Simple Smart House benchmark}
    \label{fig:array_parallel_vs_array_while}
\end{figure}

We also benchmark \texttt{ArrayParallelMatcher}. However, the Bounded Buffer benchmark, which appears to be the only one that benefits from this array optimization, is inherently parallel, so parallel algorithms perform very inconsistently on it. Therefore, \texttt{ArrayParallelMatcher} does not perform visibly better than \texttt{LazyParallelMatcher} on any benchmark. Nevertheless, it still performs better than \texttt{ArrayWhileMatcher} on other benchmarks; Figure \ref{fig:array_parallel_vs_array_while} shows the results on Simple Smart House, comparing both of the array-based matchers to both of the optimized tree-based ones. \texttt{ArrayParallelMatcher} achieves performance quite close to that of the fastest algorithm, \texttt{LazyParallelMatcher}. This is quite different from \texttt{ArrayWhileMatcher}, which is considerably slower than its tree-based counterpart. We can expand on our hypothesis about the performance of arrays to attempt to explain this difference. \texttt{ArrayParallelMatcher} splits the workload of ramifying the matching array into 8 even chunks, where each chunk is operated on by a thread. Each thread builds a new array of nodes to be added to the main array. It is possible that the matching array is split into small enough pieces for the arrays of node additions to benefit more from memory locality, thereby making the algorithm achieve better performance.

\subsection{Overall evaluation}

\subsubsection{Comparison between our algorithms}
\label{sec:overall_comparison}

We now show plots comparing many of our algorithms at the same time on all of the benchmarks in the new benchmark suite. We run all of our benchmarks with twenty repetitions for each algorithm and each $x$-axis value. Some of the benchmarks have a lot of inherent randomness, resulting in a lot of variation in results between repetitions. Therefore, to avoid clutter, we omit error bars on some of our plots. All the benchmarks are described in Section \ref{sec:benchmarks}.

As mentioned previously, we run all our benchmarks with the new benchmark suite, on a computer with an Intel Core i7-10710U CPU and 16 GB of RAM running Windows 10. We use Scala 3.6.3 and JDK 23 with the maximum heap size set to 8GB. We again provide all the commands use to run these benchmarks in Appendix \ref{sec:benchmark_commands}.

\begin{figure}[h!]
    \centering
    \includegraphics[width=0.75\linewidth]{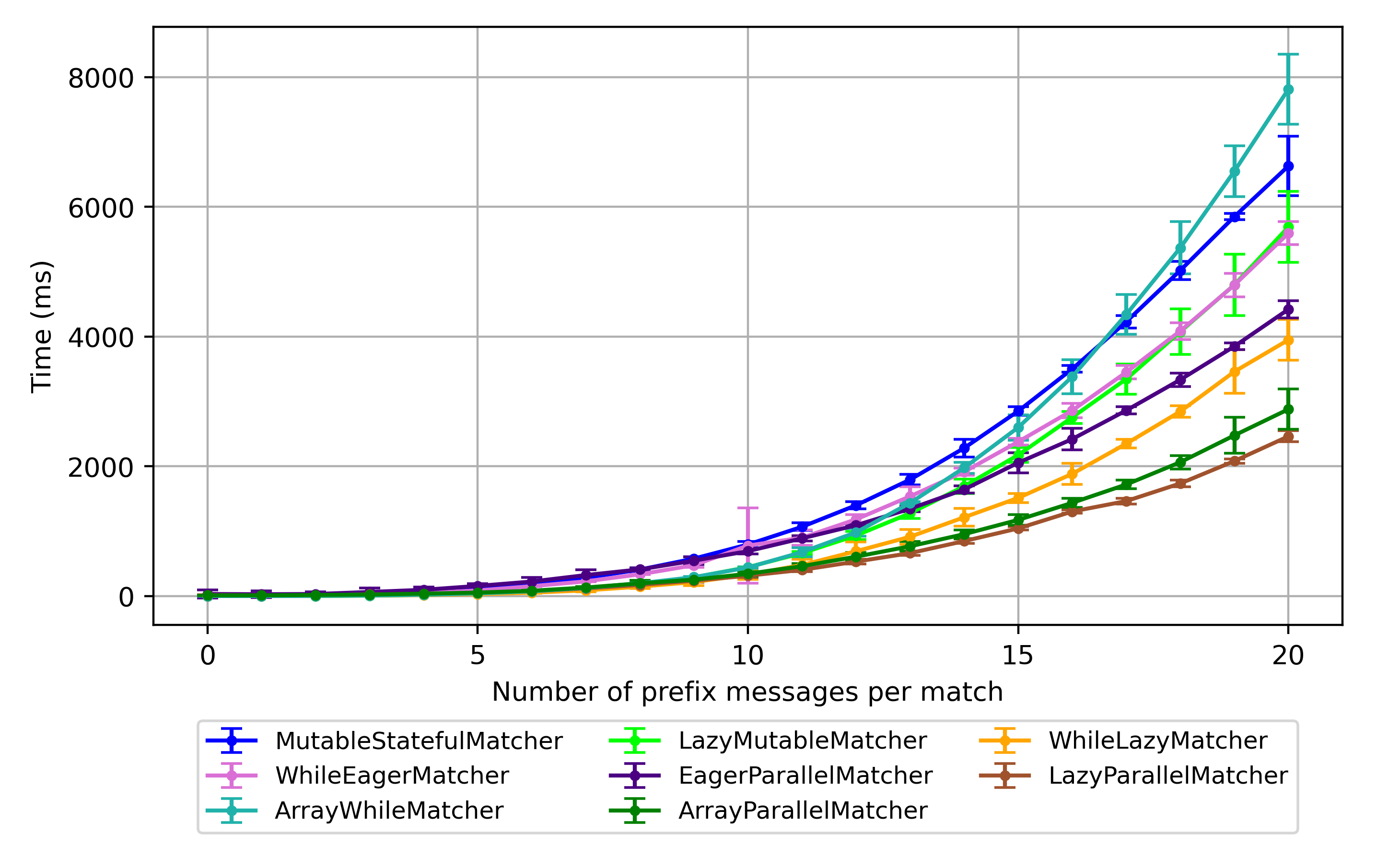}
    \caption{A comparison of our optimized algorithms against each other on the Simple Smart House benchmark}

    \label{fig:simple_smart_house_long}
\end{figure}

Figure \ref{fig:simple_smart_house_all} shows the performance of most of the algorithms in the repository on the Simple Smart House benchmark. We exclude the guard filtering algorithms, as this benchmark does not include any filtering clauses. \texttt{LazyParallelMatcher}, the fastest matcher, achieves performance around ten times faster than the baseline. Figure \ref{fig:simple_smart_house_long} shows a version of this plot that excludes the baseline and the current version of \texttt{StatefulTreeBasedMatcher} for a more precise comparison. Both of these plots gather the data that was shown previously in the Simple Smart House plots in the Implementation section.

\begin{figure}[h!]
    \centering
    \includegraphics[width=0.75\linewidth]{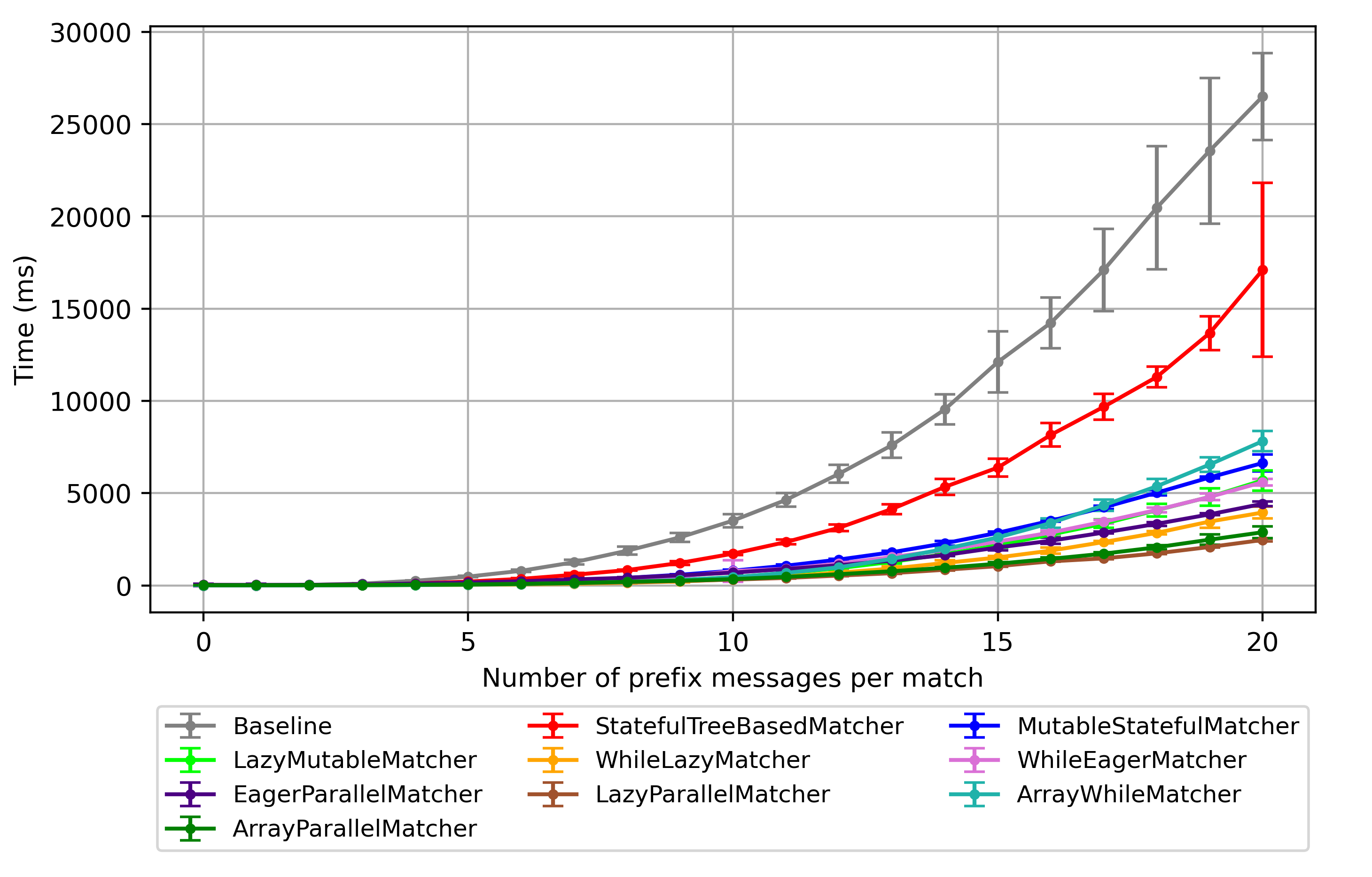}
    \caption{A comparison of the baseline against most of our optimizations on the Simple Smart House benchmark}

    \label{fig:simple_smart_house_all}
\end{figure}

\begin{figure}[h!]
    \centering
    \includegraphics[width=0.75\linewidth]{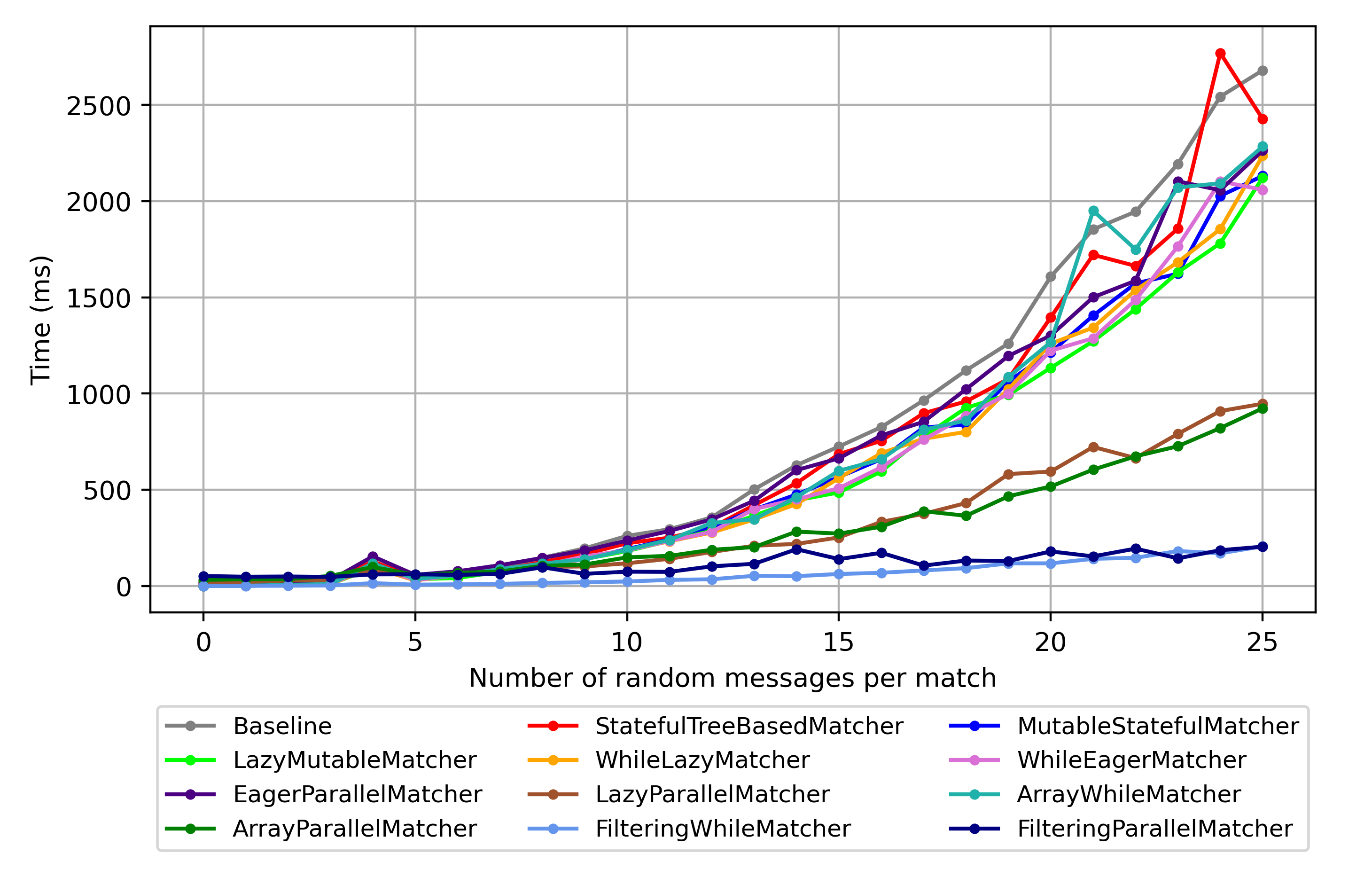}
    \caption{A comparison of the baseline against all of our optimizations on the Complex Smart House benchmark}
    
    \label{fig:complex_smart_house_all}
\end{figure}

Figure \ref{fig:complex_smart_house_all} shows the performance of all the algorithms on the Complex Smart House benchmark. This is the only benchmark that has filtering clauses, so it is the only one where we include the filtering algorithms. The results here are quite different: it seems the algorithms can be categorized into three groups based on their performance. All the singlethreaded algorithms as well as \texttt{EagerParallelMatcher} have performance very similar to the baseline. Then, the two parallel algorithms that do not make use of guard filters, \texttt{LazyParallelAlgorithm} and \texttt{ArrayParallelAlgorithm}, perform over twice as well as the baseline on the last parameter. Finally, the guard filtering algorithms show the best performance, around ten times faster than the baseline on the last parameter. We have previously shown a longer comparison between the filtering algorithms on this benchmark, in Figure \ref{fig:filtering_matchers_long}.

\begin{figure}[h!]
    \centering
    \includegraphics[width=0.75\linewidth]{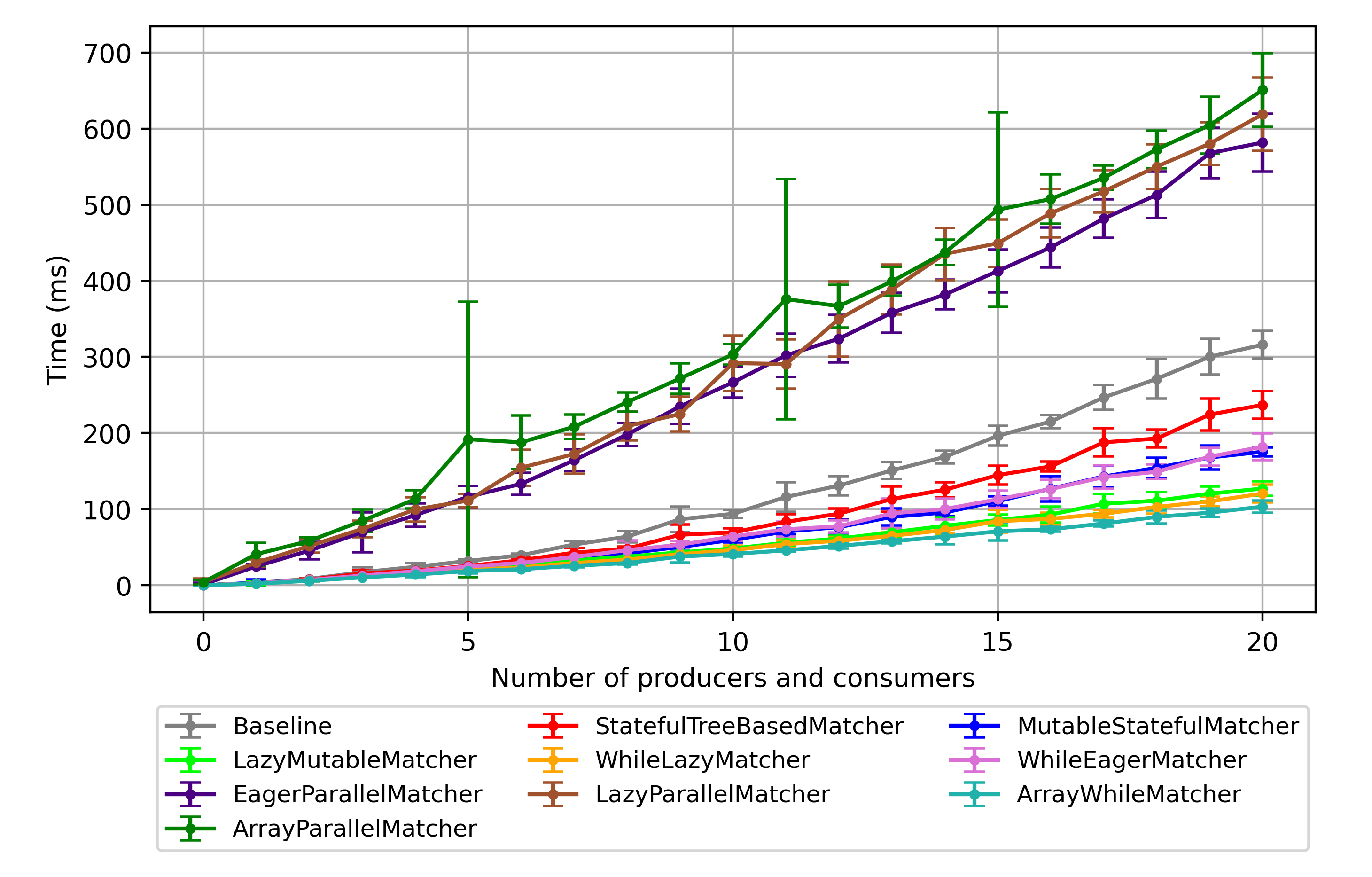}
    \caption{A comparison of the baseline against most of our optimizations on the Bounded Buffer benchmark}
    \label{fig:bounded_buffer_all}
\end{figure}

\begin{figure}[h!]
    \centering
    \includegraphics[width=0.75\linewidth]{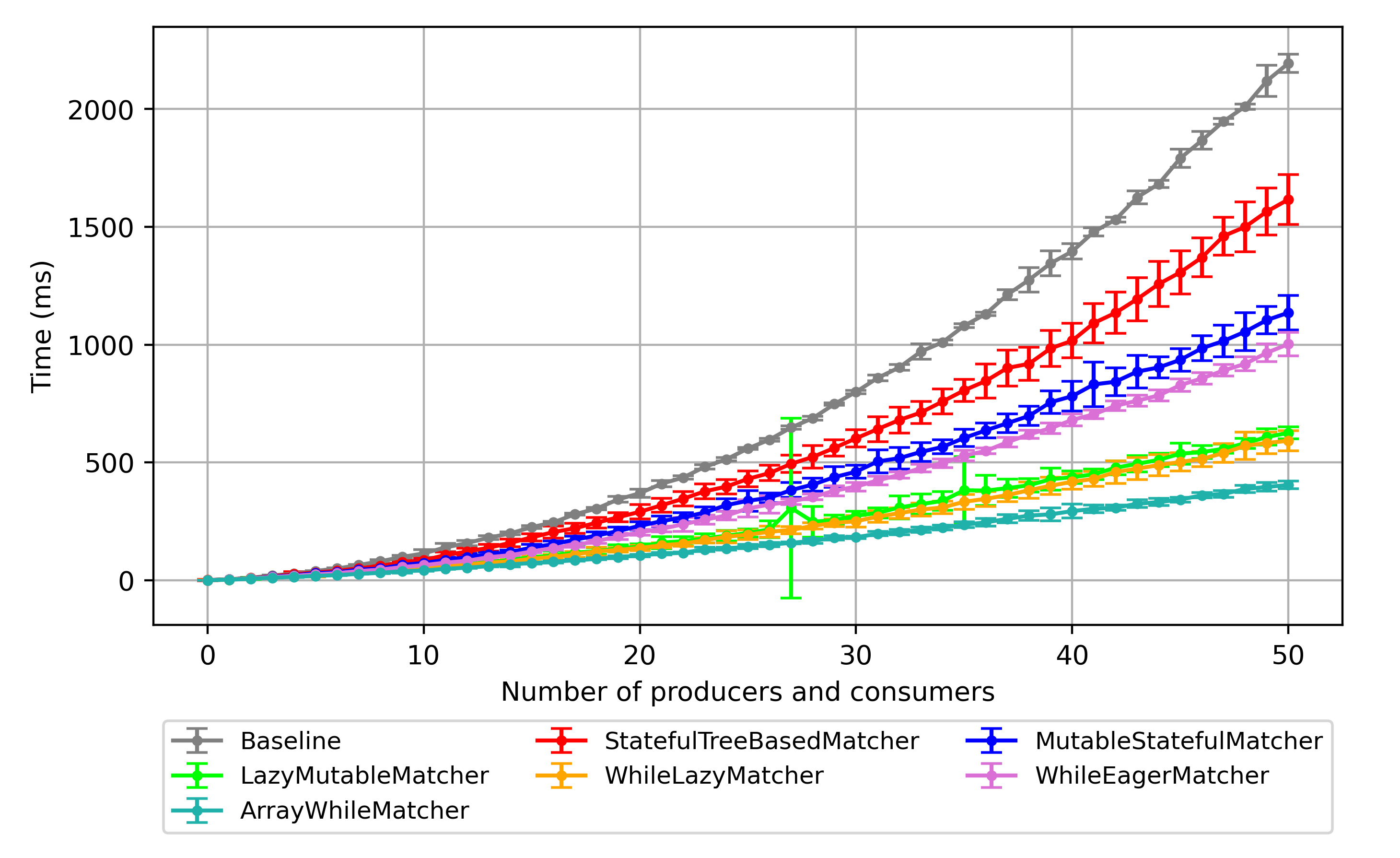}
    \caption{A comparison of the baseline against most of our singlethreaded algorithms on a longer version of the Bounded Buffer benchmark}
    \label{fig:bounded_buffer_long}
\end{figure}

Figure \ref{fig:bounded_buffer_all} shows the performance of most of our algorithms on the Bounded Buffer benchmark. Here the parallel algorithms have the worst performance by far. This is because this benchmark spawns many threads to act as producers and consumers, so using a multithreaded matcher negatively impacts performance as it leads to much more competition between threads. In an ideal scenario, the bounded buffer as well as every producer and consumer would be running on separate machines so that this is not an issue. Figure \ref{fig:bounded_buffer_long} shows the results from a longer version of the benchmark excluding the multithreaded algorithms. Here the results are quite similar to the Simple Smart House results. The notable differences are the following:

\begin{itemize}
    \item \texttt{LazyMutableMatcher} has significantly better performance than \texttt{WhileEagerMatcher}

    \item \texttt{WhileLazyMatcher} does not appear to improve upon \texttt{LazyMutableMatcher}

    \item \texttt{ArrayWhileMatcher} has the best performance out of all the matchers, as previously discussed and analyzed. Its performance is almost five times faster than the baseline on the last parameter.
\end{itemize}

\begin{figure}[h!]
    \centering

    \begin{subfigure}[b]{0.49\textwidth}
        \centering
         \includegraphics[width=\textwidth]{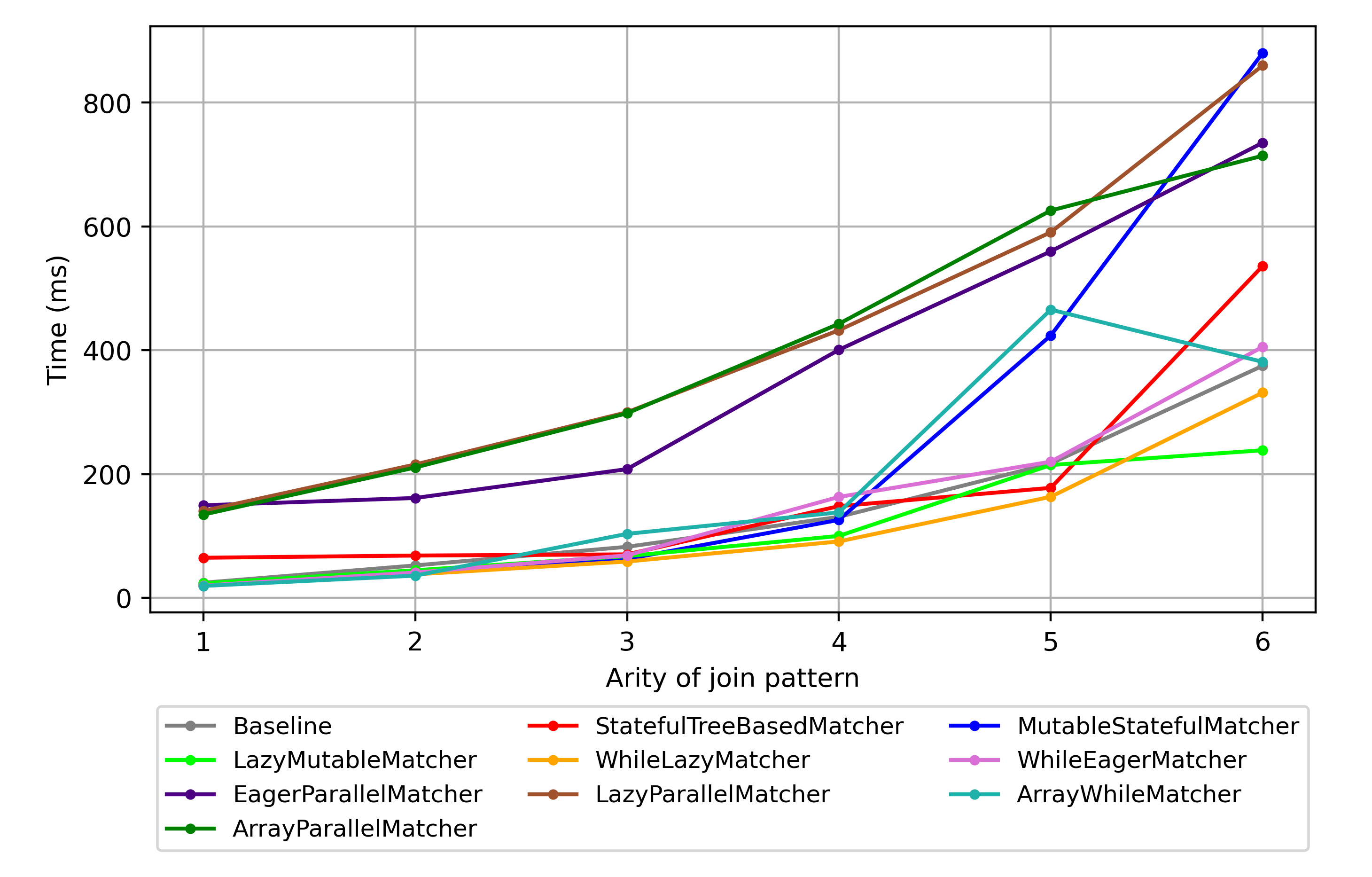}
         \caption{Size}
    \end{subfigure}
    \hfill
    \begin{subfigure}[b]{0.49\textwidth}
        \centering
         \includegraphics[width=\textwidth]{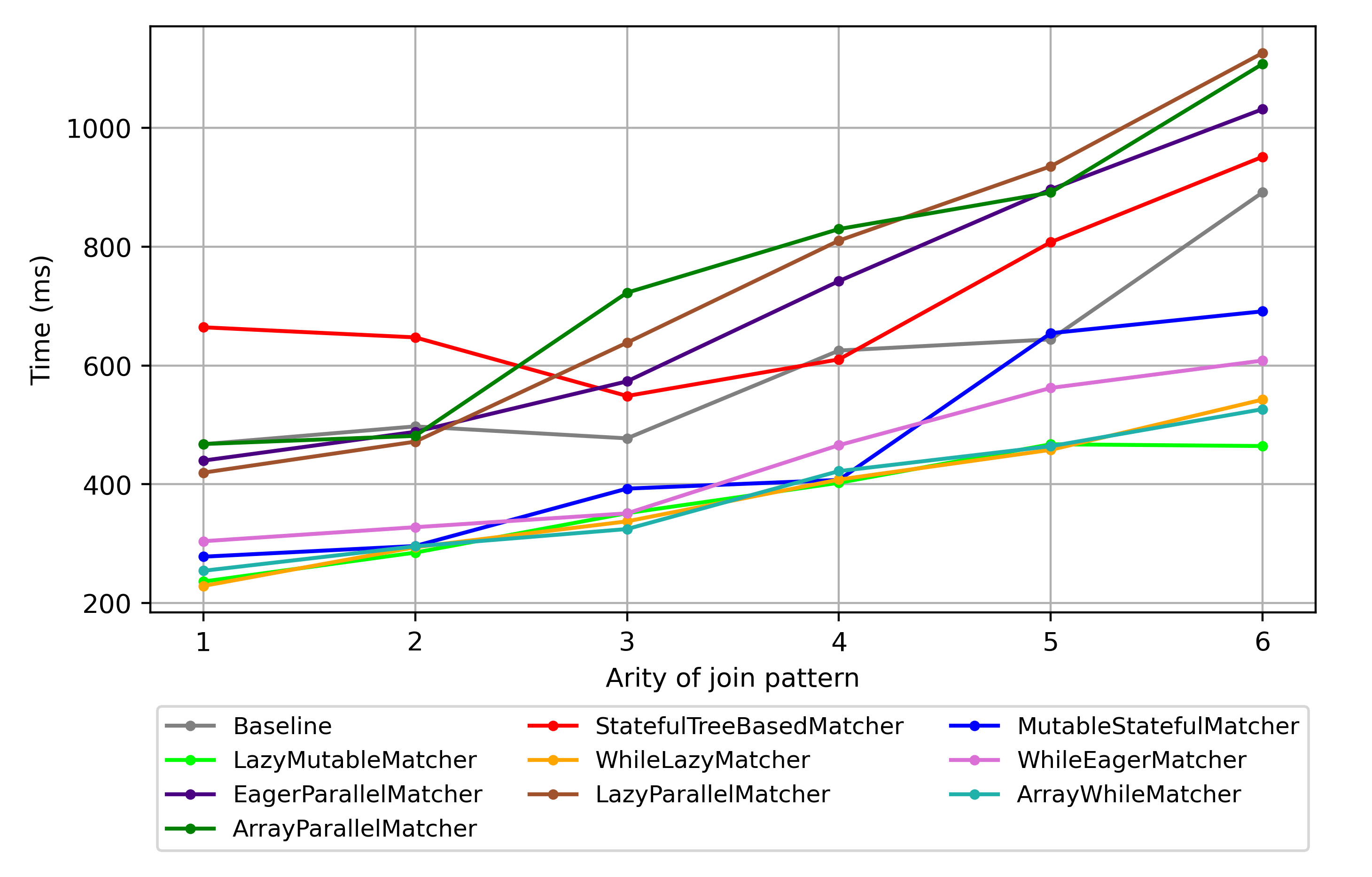}
         \caption{Size with noise}
    \end{subfigure}
    \hfill
    \begin{subfigure}[b]{0.49\textwidth}
        \centering
         \includegraphics[width=\textwidth]{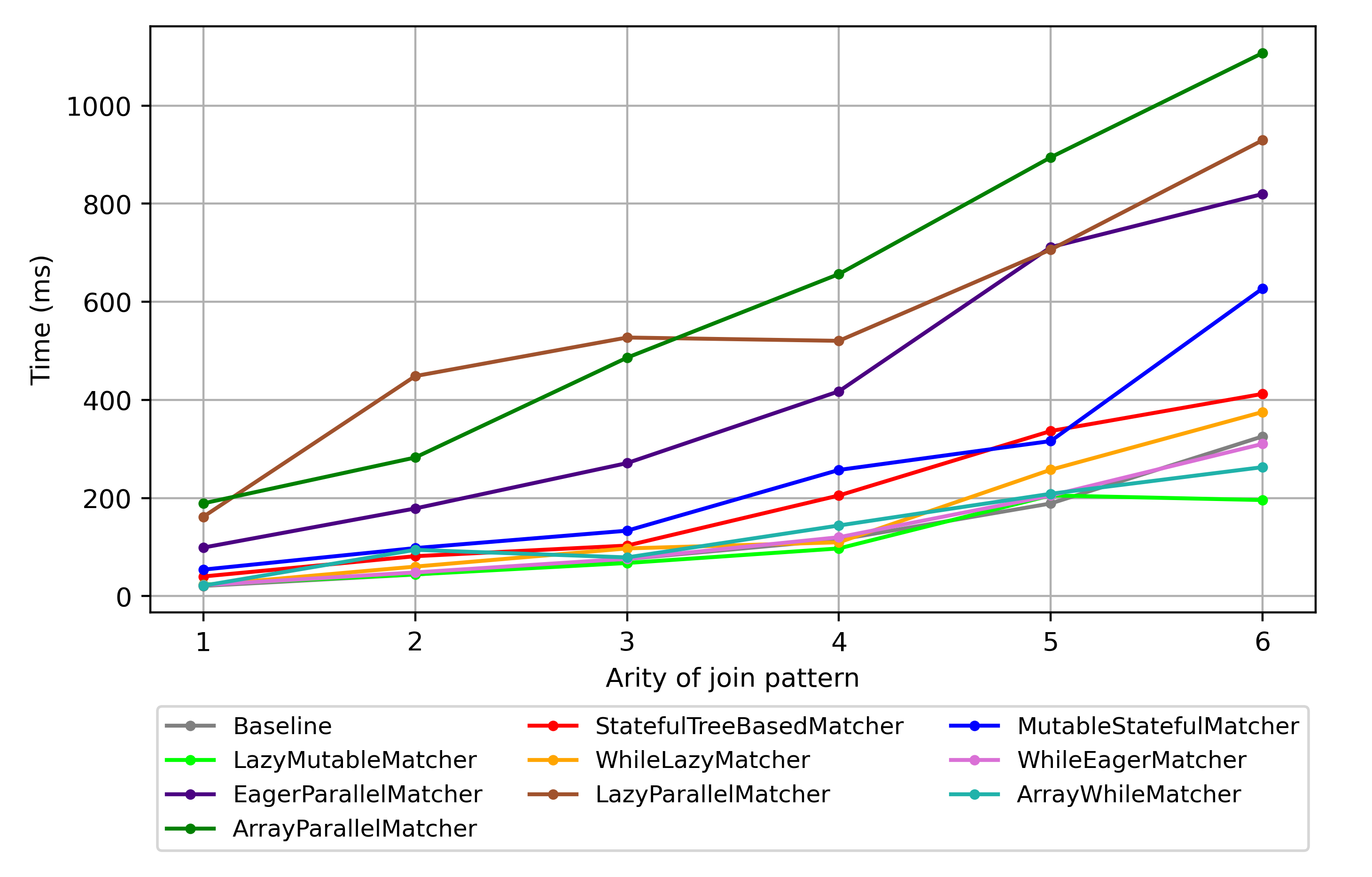}
         \caption{Size with guards}
    \end{subfigure}
    \hfill
    \begin{subfigure}[b]{0.49\textwidth}
        \centering
         \includegraphics[width=\textwidth]{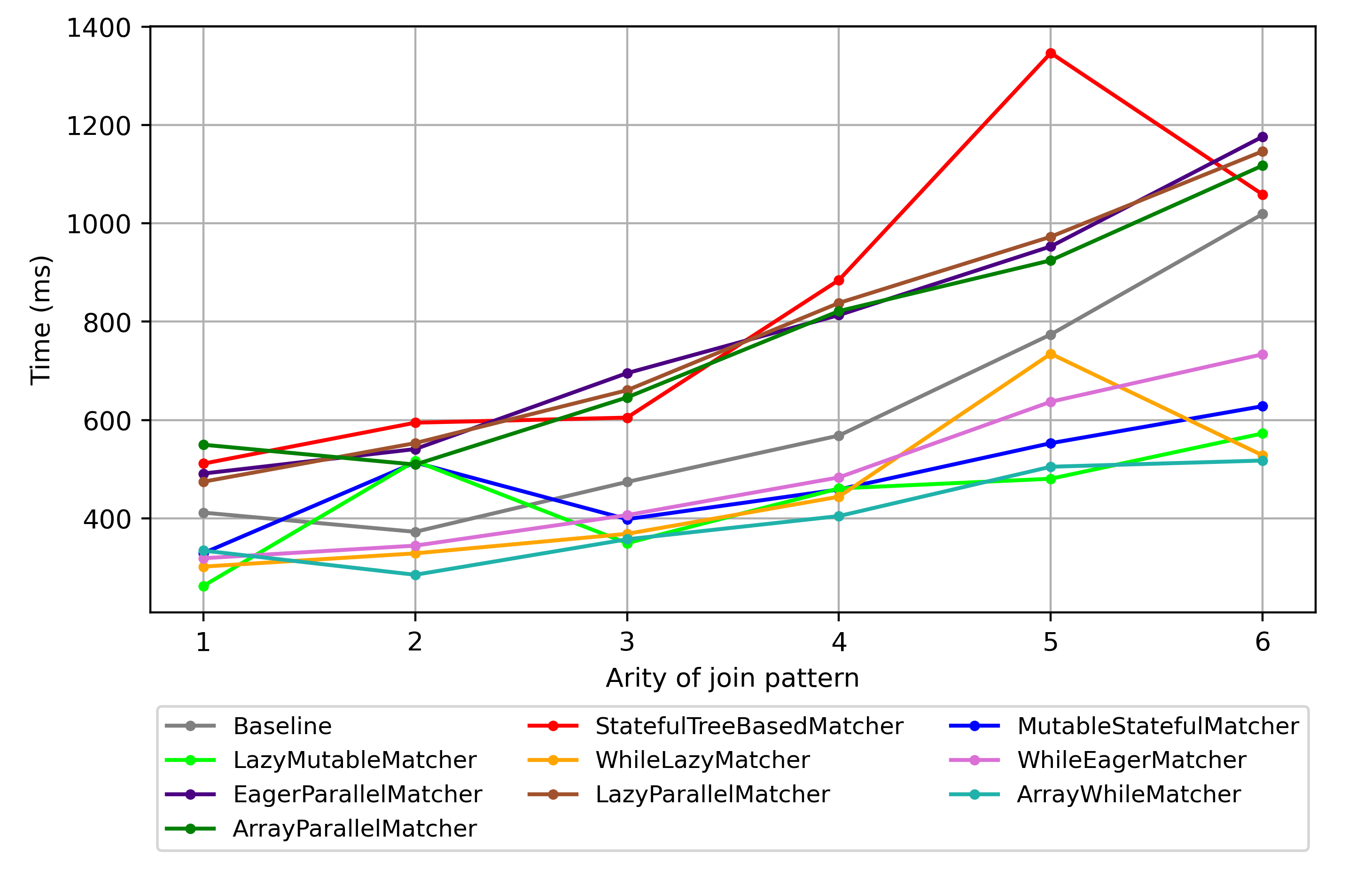}
         \caption{Size with guards and noise}
    \end{subfigure}
    \hfill
    \begin{subfigure}[b]{0.75\textwidth}
        \centering
         \includegraphics[width=\textwidth]{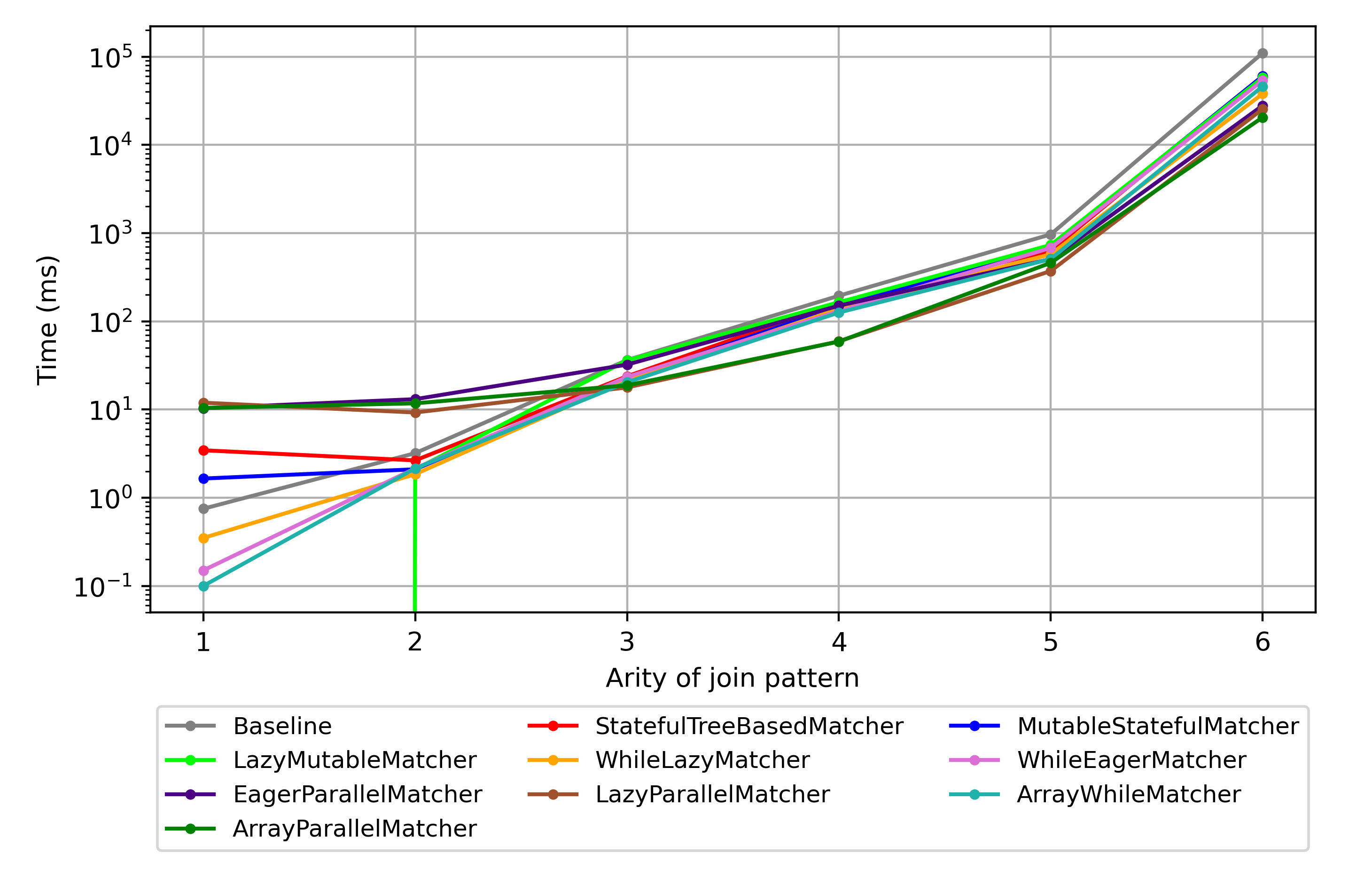}
         \caption{Size with guards and non-satisfying payloads}
    \end{subfigure}
    
    \caption{A comparison of the baseline against most of our algorithms on the ``Size" and ``Size with guards" benchmarks}
    \label{fig:size_benchmarks}
\end{figure}

Our results on the ``Size" and ``Size with guards" benchmarks are much more noisy and do not leave us with much useful information; however, we still show them for completeness in Figure \ref{fig:size_benchmarks}. We use a logarithmic scale for the last benchmark, which includes guards as well as payloads which do not satisfy the guards. This is because the last measurement is significantly higher than the previous ones. The last plot shows a strange vertical line at the beginning of the \texttt{LazyMutableMatcher} data; this is because the first data point is 0 due to measurement inaccuracies, and 0 cannot be represented on a logarithmic scale.

\subsubsection{Comparison to Evrete}
\label{sec:evrete_comparison}

The predecessor paper compares its stateful tree-based algorithm to a Rete-based implementation using the Evrete library on the Simple Smart House benchmark. Rete is an algorithm used in the domain of expert systems, and the problem solved by it has similarities to the problem of join pattern matching. However, the problem solved is not the exact same, and a lot of adaptation is needed to capture the semantics of fair join pattern matching. The predecessor paper uses an implementation that is hard-coded to the Simple Smart House benchmark. The most important difference between Evrete and the stateful tree-based matching algorithm is that Evrete has no concept of fairness. Therefore, much like the brute-force algorithm, the Evrete implementation exhaustively computes all possible matches, evaluates the guard for all matches, and then finds the fairest among these.

We keep this Evrete-based Simple Smart House implementation mostly the same. However, we equip it with the ability to generate the messages needed by the benchmark by itself, instead of having to obtain them from the main benchmark suite. In addition, we give it a richer interface similar to that of the refactored benchmark suite. Our repository with these changes can be found at \url{https://github.com/yaniskas/evrete-smarthouse}

The predecessor paper runs two comparisons: one with the standard benchmark, and one with the so-called ``heavy guard" flag enabled. This flag artificially slows down the time taken to evaluate the guard of the main pattern in the join definition, simulating a computationally intensive guard. Since the Evrete implementation has to evaluate guards for all potential matches, the heavy guards flag significantly diminishes the performance of Evrete. In the original results, Evrete is over ten times faster on the variant without heavy guards, while the matching tree algorithm is faster on the variant with heavy guards.

\begin{figure}[h!]
    \centering
    \begin{subfigure}[b]{0.49\textwidth}
        \centering
         \includegraphics[width=\textwidth]{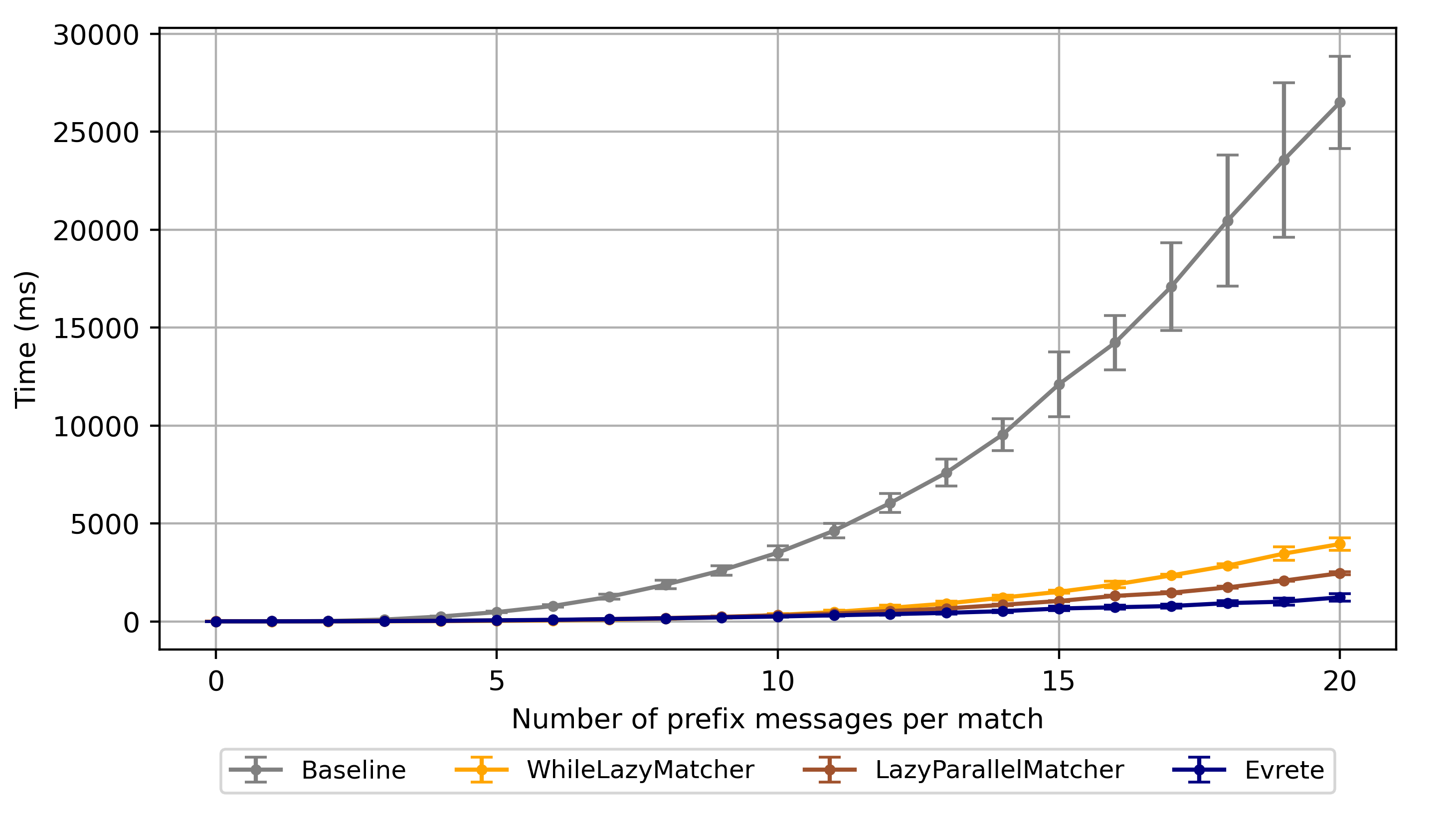}
         \caption{Baseline included}
         \label{fig:evrete_vs_baseline}
    \end{subfigure}
    \hfill
    \begin{subfigure}[b]{0.49\textwidth}
        \centering
         \includegraphics[width=\textwidth]{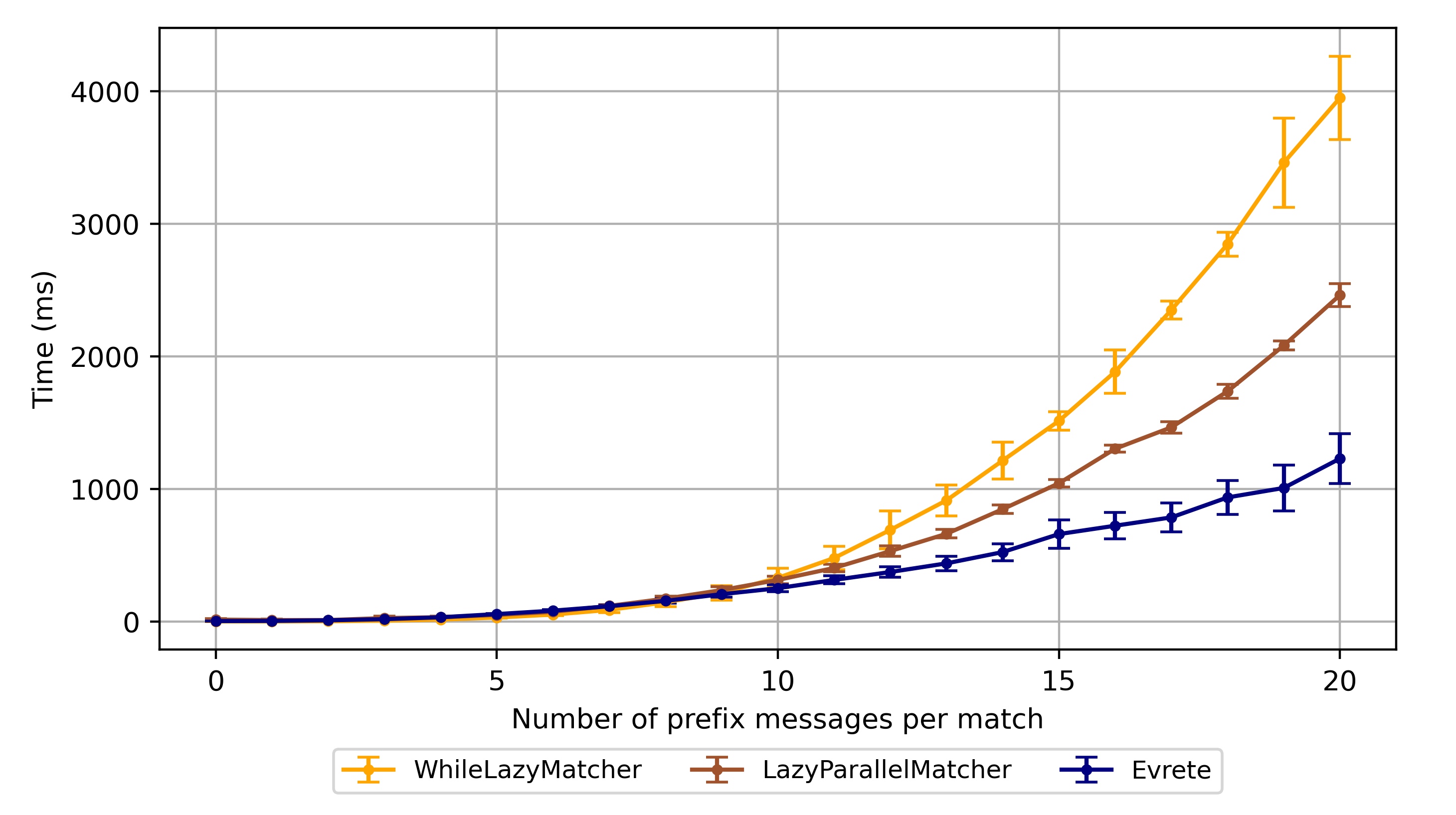}
         \caption{Baseline not included}
         \label{fig:evrete_vs_optimized}
    \end{subfigure}
    
    \caption{A comparison between the Evrete implementation, \texttt{WhileLazyMatcher}, \texttt{LazyParallelMatcher}, and the baseline on the Simple Smart House benchmark}

    \label{fig:evrete_comparison}
\end{figure}

We first repeat the same Simple Smart House benchmark we have already conducted with the Evrete-based implementation. We compare the performance of Evrete to three algorithms: the baseline, our \texttt{WhileLazyMatcher}, which we consider a new baseline, and our \texttt{LazyParallelMatcher}, which is our fastest algorithm on this benchmark. Figure \ref{fig:evrete_vs_baseline} shows the results. As can be seen, our optimized implementations are still slower than Evrete, but they are much closer to the performance of Evrete than the baseline is. Figure \ref{fig:evrete_vs_optimized} shows a version of the same graph with the baseline excluded. Evrete performs approximately twice as fast as \texttt{LazyParallelMatcher} on the last parameter, and around four times as fast as \texttt{WhileLazyMatcher}.

\begin{figure}[h!]
    \centering
    \begin{subfigure}[b]{0.49\textwidth}
        \centering
         \includegraphics[width=\textwidth]{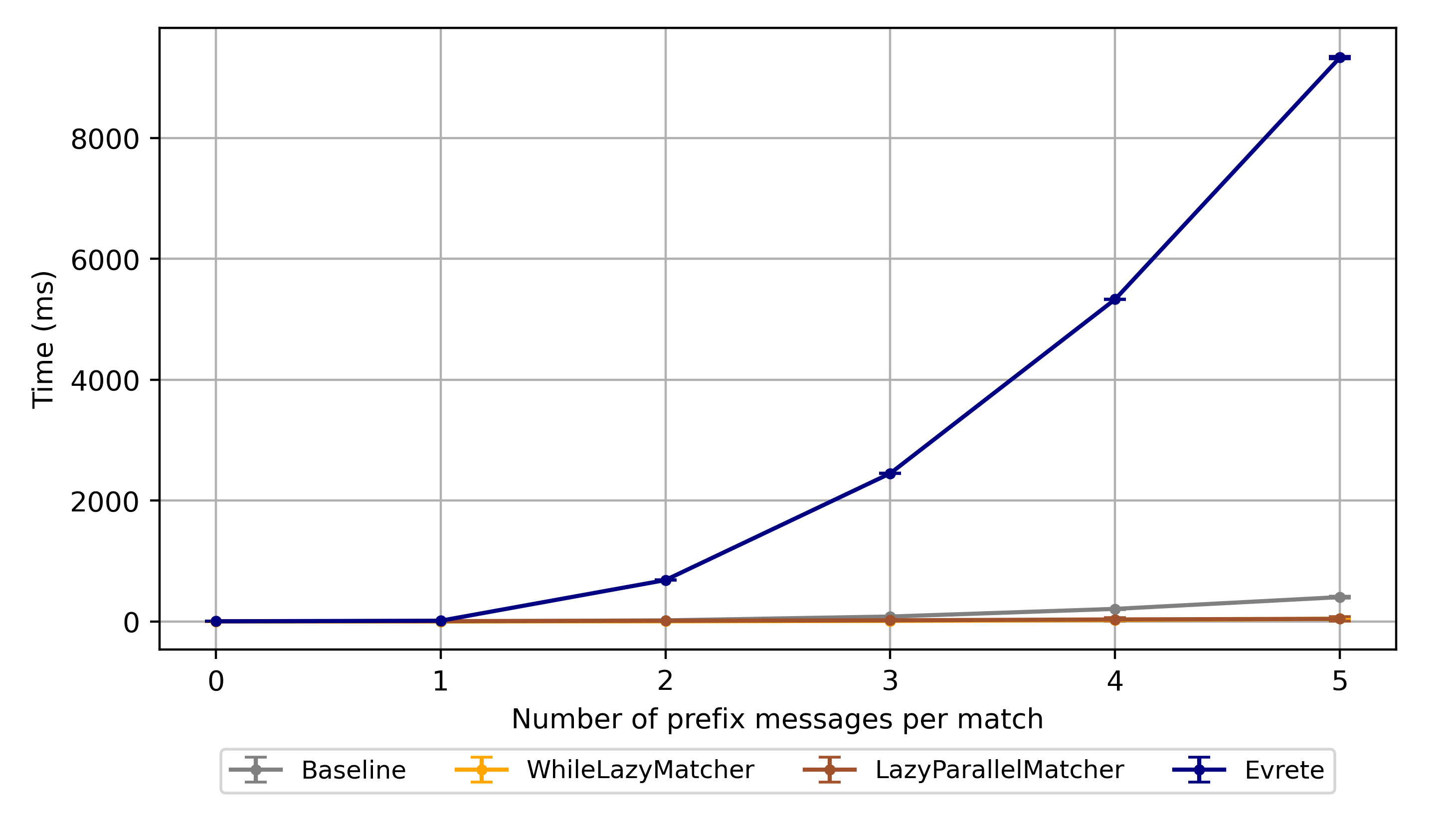}
         \caption{Evrete included}
         \label{fig:evrete_heavy_guards}
    \end{subfigure}
    \hfill
    \begin{subfigure}[b]{0.49\textwidth}
        \centering
         \includegraphics[width=\textwidth]{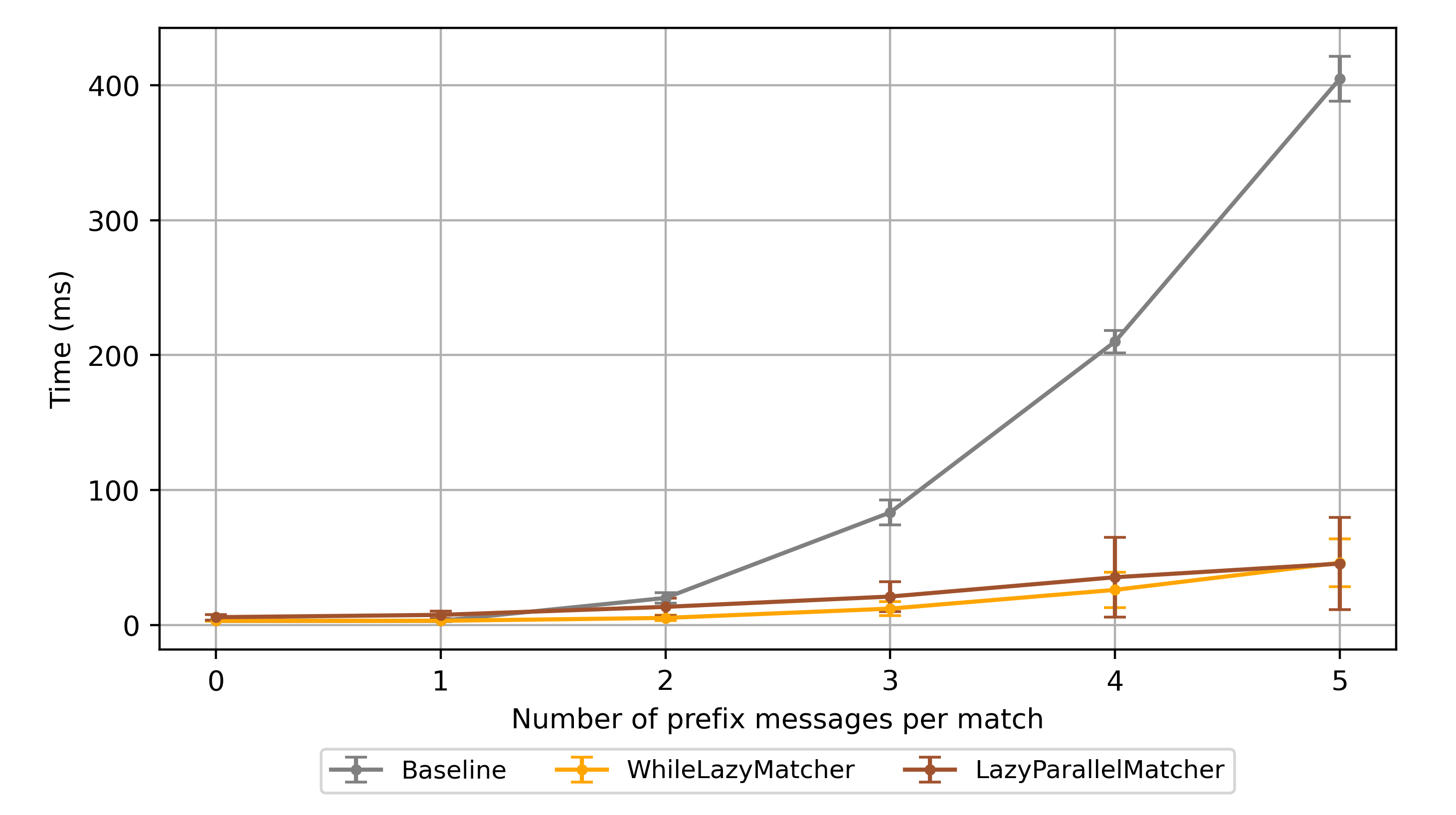}
         \caption{Evrete not included}
         \label{fig:evrete_heavy_guards_no_evrete}
    \end{subfigure}
    
    \caption{A comparison between the Evrete implementation, \texttt{WhileLazyMatcher}, \texttt{LazyParallelMatcher}, and the baseline on the Simple Smart House benchmark with heavy guards}

    \label{fig:evrete_comparison_heavy_guards}
\end{figure}

We also run a version of the Simple Smart House benchmark with the \texttt{heavy-guard} flag given. We make the benchmark much shorter, as Evrete takes a very long time to complete with this flag set. Figure \ref{fig:evrete_heavy_guards} shows the results, and \ref{fig:evrete_heavy_guards_no_evrete} shows the results without the Evrete data. As in the predecessor paper, the baseline vastly outperforms Evrete in this scenario. Moreover, \texttt{WhileLazyAlgorithm} and \texttt{LazyParallelAlgorithm} again perform much better than the baseline. \texttt{WhileLazyMatcher} seems to slightly outperform \texttt{LazyParallelMatcher}; this is probably due to the short length of the benchmark increasing the relative effect of synchronization overhead.

\subsection{Other attempted optimizations}
\label{sec:failed_optimizations}

Here we describe optimizations that we implemented and tested, but which did not lead to a consistent measurable performance benefit.

\subsubsection{Further memory optimizations for \texttt{MessageIdxs}}

\paragraph{Design}

As previously mentioned, the flame graph of \texttt{WhileLazyMatcher} in Figure \ref{fig:while_lazy_flame_graph} shows that the majority of its runtime is spent on the \texttt{:+} method of \texttt{ArraySeq}. We introduced this data structure in Section \ref{sec:cache_friendly_data_structures}, where we used it to replace the linked list structures we were using previously. The \texttt{:+} method is used to create a copy of an \texttt{ArraySeq} with a new element appended; we use this method every time we create a child node while ramifying a matching tree. The implementation of \texttt{:+} in \texttt{ArraySeq} simply creates a new \texttt{ArraySeq} that has the new element added. This results in a lot of unnecessary duplication of data, since we never modify nodes after they are created, and whenever we remove a node we also remove all of its children. Therefore, with the aim of reducing the time spent on this copy-and-append operation, we try to define a data structure that implements the operation efficiently.

We define a data structure which we call a \emph{growing array reference} (GAR) that avoids this unnecessary copying of data. A GAR always stores a reference to a dynamic array along with a \emph{view length}. We define the following main operations on this structure:

\begin{itemize}
    \item \textbf{Comparison:} When comparing two GARs, we compare elements of the dynamic arrays, but we only consider elements up to the view length.

    \item \textbf{Construction:} We can create a GAR with a collection of elements. The dynamic array is set to contain these elements, and the view length is set to the length of the array.

    \item \textbf{Copy-and-append (\texttt{:+}):} When we call \texttt{:+} on a newly constructed GAR, we do not copy the array. Instead, we expand the dynamic array with this new element and return a new GAR that refers to the same underlying array, but which has an incremented view length. If we then call \texttt{:+} on this new GAR, we repeat this process, and end up with three references backed by a single array. In general, whenever we perform this operation on a GAR with a view length that equals the length of the underlying array, we can perform the operation this way. However, if we call \texttt{:+} on a GAR with a view length that does not equal the length of the underlying array, we necessarily create a copy of the array and then append the new element to this copy, and return a GAR that is backed by this new array.
\end{itemize}

The actual implementation requires methods implementing a few more operations, such as obtaining iterators; we omit these for conciseness. In general, such operations consider the underlying array only up to the view length.

\paragraph{Implementation}

The \texttt{dev-growing-array-ref} branch of the repository\footnote{\url{https://github.com/yaniskas/join-actors/tree/dev-growing-array-ref}} contains an implementation of this optimization. We define a class called \texttt{GrowingIntArrayRef} that implements the GAR data structure, specialized for integers. We then redefine the \texttt{MessageIdxs} type alias to this type. This makes it so all matching tree nodes are represented as GARs, so every time we traverse a tree and call \texttt{:+} to create a new child node, we use the special \texttt{:+} implementation of GARs that avoids memory allocations as much as possible.

\paragraph{Evaluation}

\begin{figure}[h!]
    \centering

    \begin{subfigure}[b]{0.49\textwidth}
        \centering
         \includegraphics[width=\textwidth]{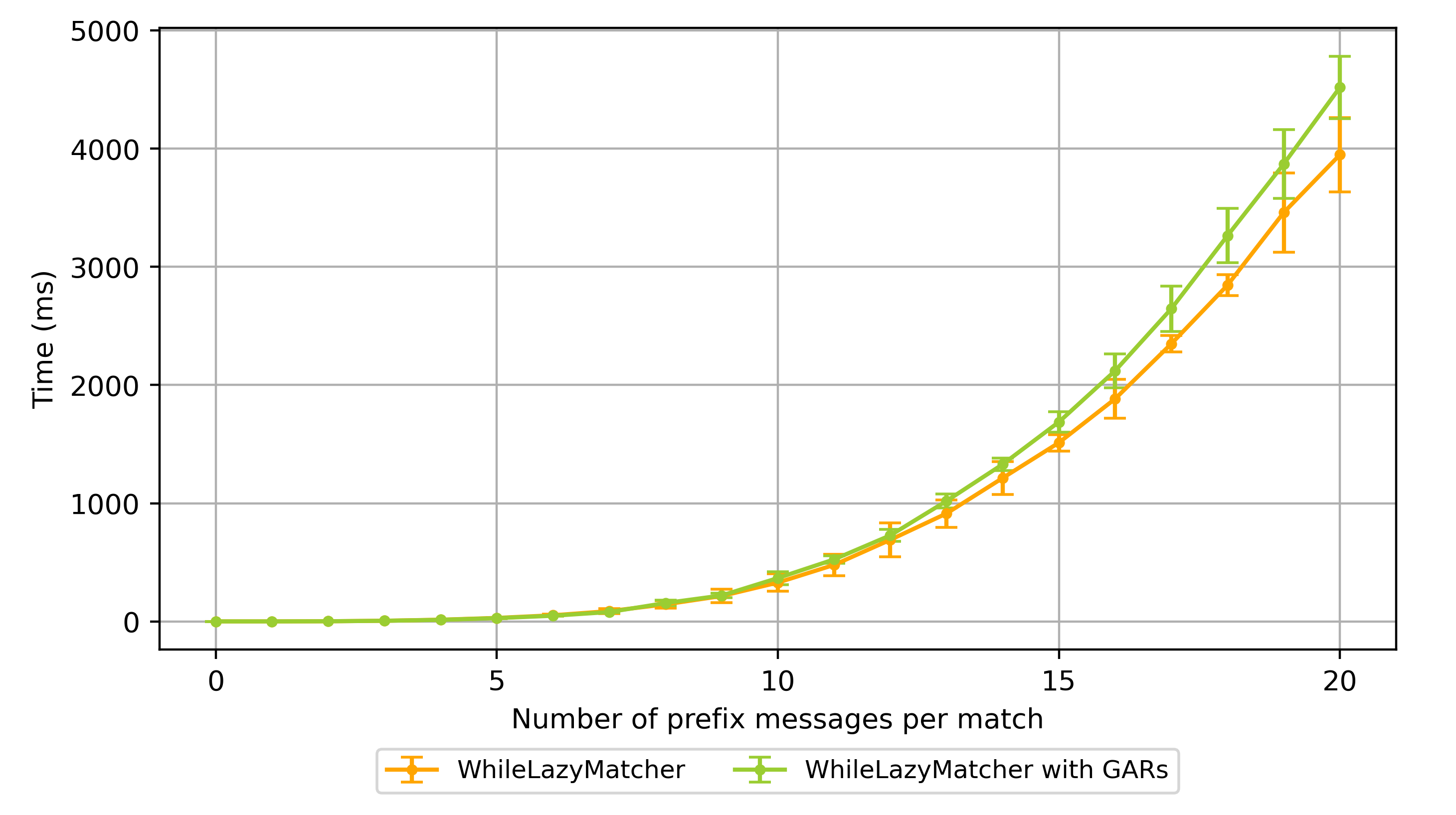}
         \caption{Simple Smart House}
    \end{subfigure}
    \hfill
    \begin{subfigure}[b]{0.49\textwidth}
        \centering
         \includegraphics[width=\textwidth]{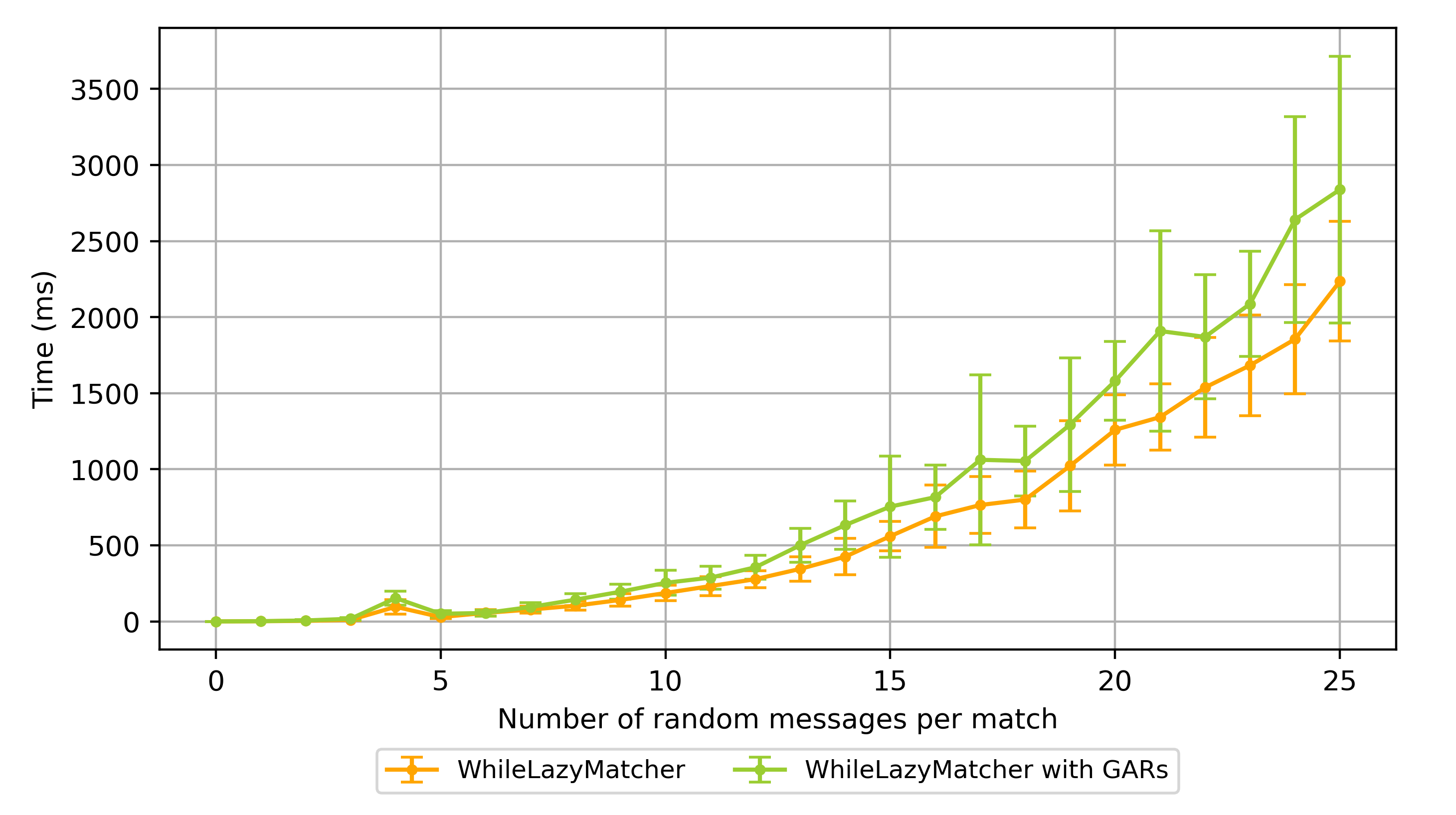}
         \caption{Complex Smart House}
    \end{subfigure}
    \hfill
    \begin{subfigure}[b]{0.49\textwidth}
        \centering
         \includegraphics[width=\textwidth]{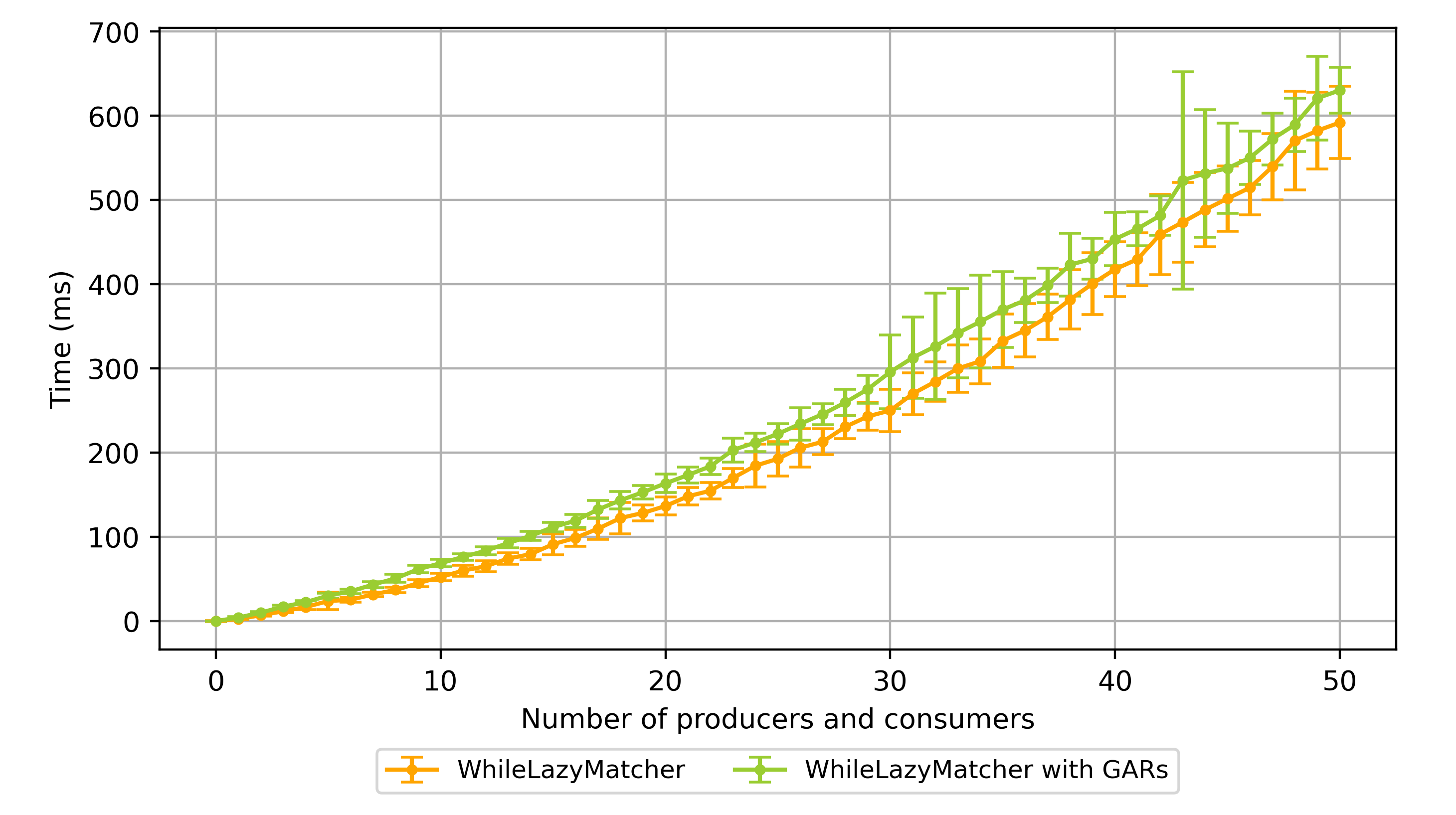}
         \caption{Bounded Buffer}
    \end{subfigure}
    
    \caption{A comparison of \texttt{WhileLazyMatcher} with the growing array reference change against \texttt{WhileLazyMatcher} without it}
    \label{fig:gar_benchmarks}
\end{figure}

Since we consider \texttt{WhileLazyMatcher} to be a new baseline, we test the regular version of this matcher to the one on the \texttt{dev-growing-array-ref} branch that is affected by this optimization. Figure \ref{fig:gar_benchmarks} shows the results on the three main benchmarks. This change results in slightly worse performance on all benchmarks. A reason for this may be that the GAR data structure introduces a layer of indirection, which is not cache-friendly.

\subsubsection{In-place sorted merge}

\paragraph{Design}

In Section \ref{sec:replacing_matching_trees}, we described the concept that became the \texttt{ArrayWhileMatcher} in Section \ref{sec:array_matchers}. However, since it achieves worse performance on the Simple Smart House benchmark, we attempt to improve upon it. We try this by implementing a more memory-efficient sorted merge algorithm that does not involve allocating a third array. This is possible to do when the initial array has enough space allocated to hold all the new elements. We treat the empty space at the end of the first array as the third array, and perform the sorted merge backwards.

\paragraph{Implementation}

The \texttt{dev-buffer} branch of the repository\footnote{\url{https://github.com/yaniskas/join-actors/tree/dev-buffer}} contains an implementation of this optimization. The class \texttt{BufferWhileMatcher} implements an attempt to improve our \texttt{ArrayWhileMatcher} by using a partially in-place algorithm. Here, we maintain an \texttt{ArrayBuffer} of nodes. When we receive a new message, we again construct an array of additions. However, instead of creating a third array, we expand the \texttt{ArrayBuffer} until it has enough space to hold all the additions. This may result in a new memory allocation, but if there is already enough space allocated for expansion, we avoid an allocation. Finally, we perform an in-place sorted merge between the buffer and the array of additions.

\paragraph{Evaluation}

\begin{figure}[h!]
    \centering

    \begin{subfigure}[b]{0.49\textwidth}
        \centering
         \includegraphics[width=\textwidth]{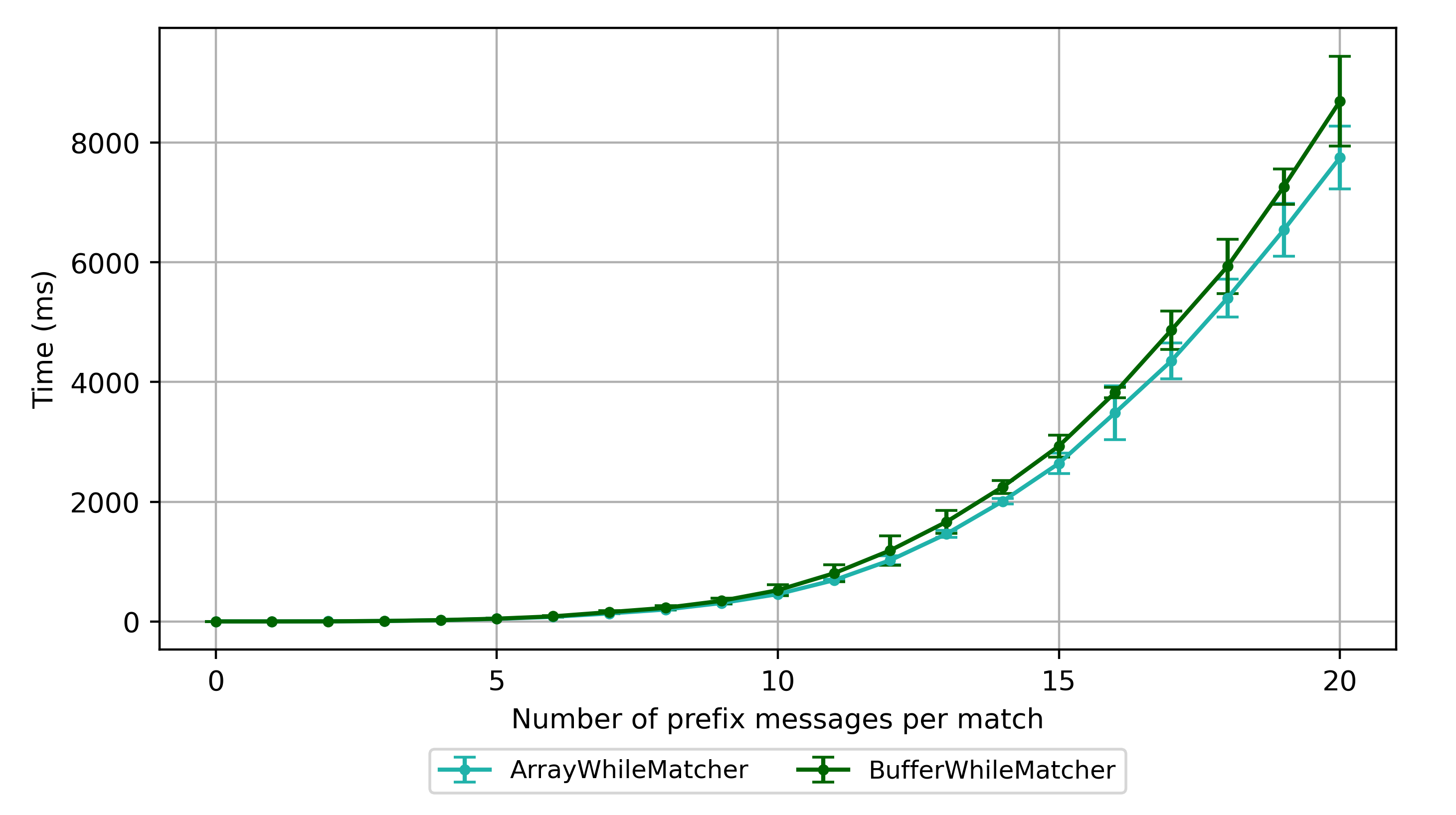}
         \caption{Simple Smart House}
    \end{subfigure}
    \hfill
    \begin{subfigure}[b]{0.49\textwidth}
        \centering
         \includegraphics[width=\textwidth]{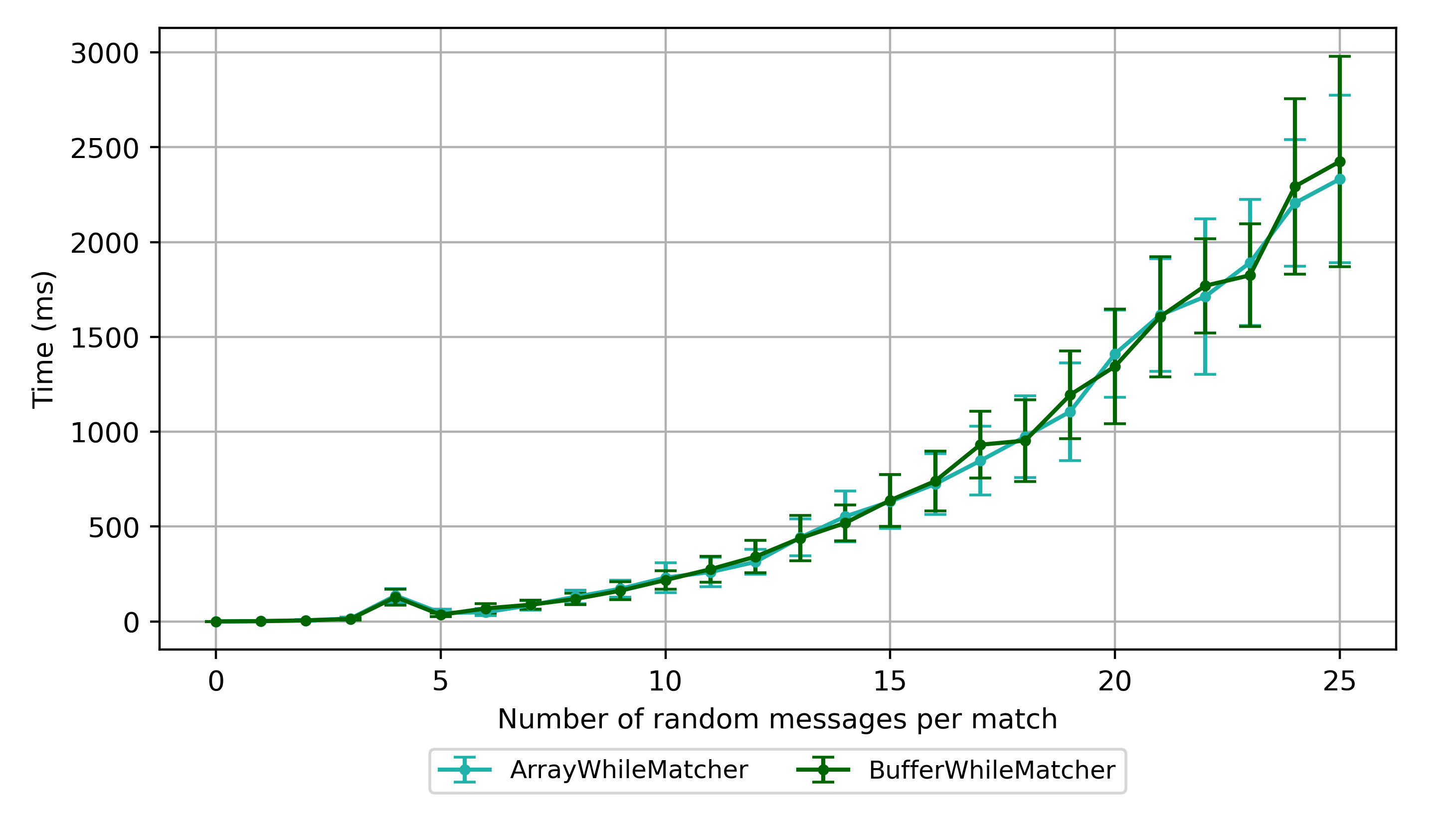}
         \caption{Complex Smart House}
    \end{subfigure}
    \hfill
    \begin{subfigure}[b]{0.49\textwidth}
        \centering
         \includegraphics[width=\textwidth]{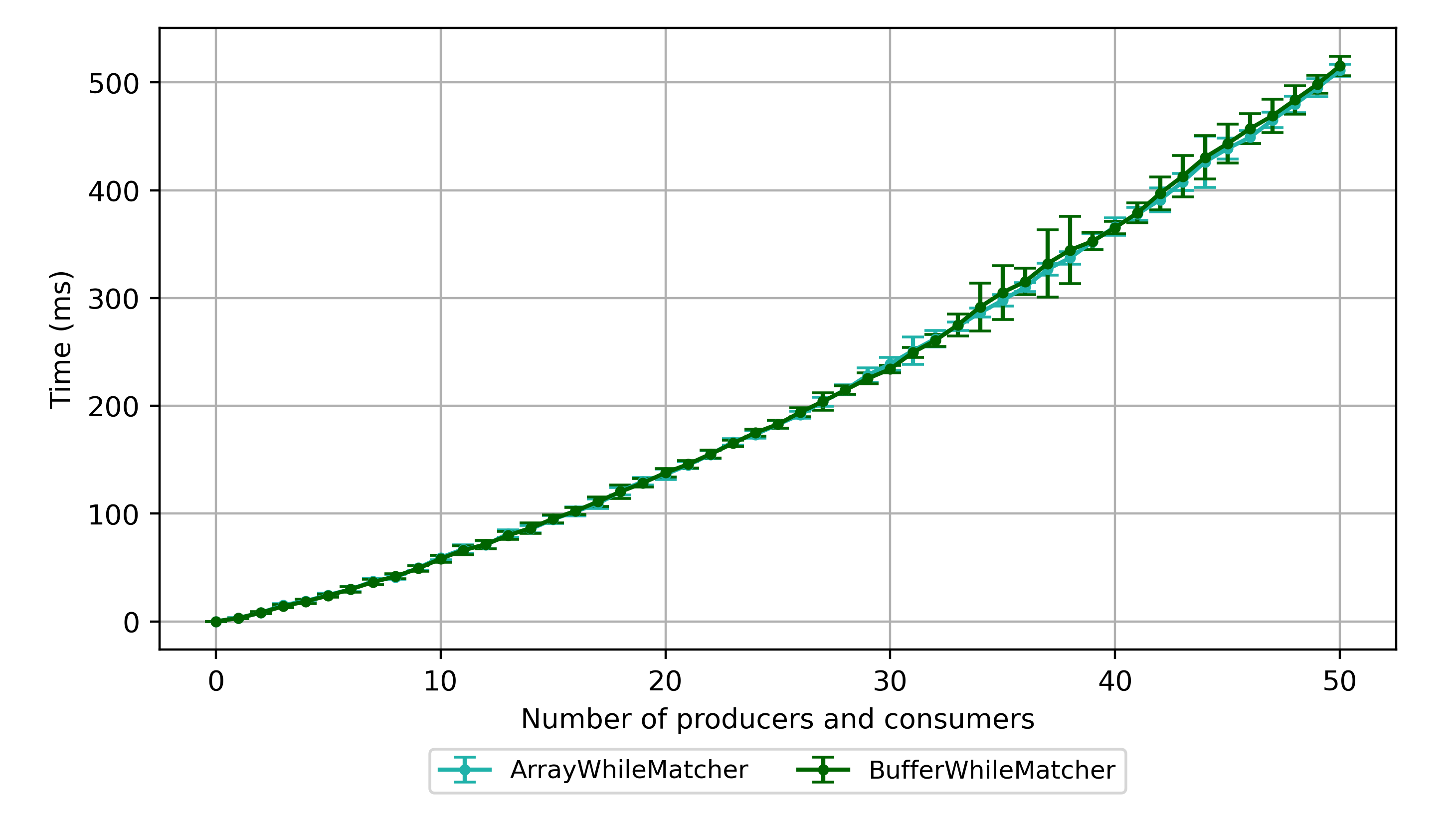}
         \caption{Bounded Buffer}
    \end{subfigure}
    
    \caption{A comparison of \texttt{BufferWhileMatcher} against \texttt{ArrayWhileMatcher}}
    \label{fig:buffer_benchmarks}
\end{figure}

This new matcher showed some promise in the Bounded Buffer benchmark, occasionally outperforming \texttt{ArrayWhileMatcher}. However, we were not able to replicate these results consistently. Figure \ref{fig:buffer_benchmarks} shows typical results from our three main benchmarks. The new matcher achieves slightly worse performance on Simple Smart House, and approximately equal performance on the other benchmarks.

\subsubsection{\texttt{PatternBins} as a hash map}

\paragraph{Design}

The flame graph of \texttt{WhileLazyMatcher} in Figure \ref{fig:while_lazy_flame_graph} shows a considerable amount of time spent on a \texttt{get} method. This is on the \texttt{PatternBins} data type, which is an alias for the Scala \texttt{TreeMap} type. Here we store satellite information in the matching tree about the types of the messages that have been partially matched. The \texttt{get} method is simply the accessor method for this map, taking a key and returning a value. In this considered optimization, we tried to reduce the time spent on this \texttt{get} method. Until this point, we had used a \texttt{TreeMap} to implement this type. This is a red-black tree map, with $O(\log(n))$ access time. Hash maps, on the other hand, have $O(1)$ access time. Therefore, we simply redefined \texttt{PatternIdxs} to use a Scala \texttt{HashMap}.

\paragraph{Implementation}

The \texttt{dev-hashmap} branch of the repository\footnote{\url{https://github.com/yaniskas/join-actors/tree/dev-hash-map}} contains an implementation of this optimization. Thanks to \texttt{PatternIdxs} being a type alias, not much work was involved in implementing this change. We just changed the alias definition, and refactored a few places where the data structure is constructed.

\paragraph{Evaluation}

\begin{figure}[h!]
    \centering

    \begin{subfigure}[b]{0.49\textwidth}
        \centering
         \includegraphics[width=\textwidth]{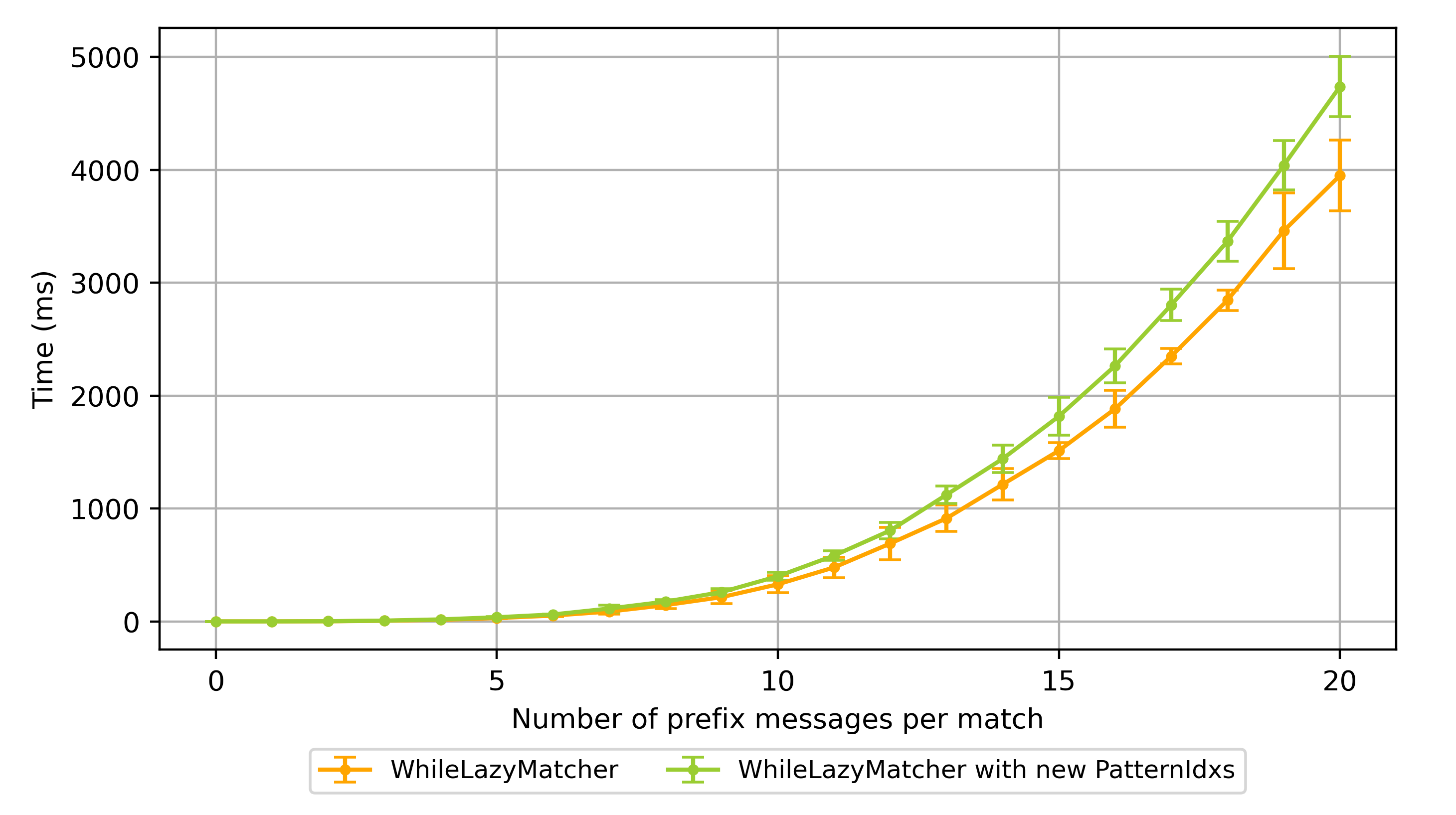}
         \caption{Simple Smart House}
    \end{subfigure}
    \hfill
    \begin{subfigure}[b]{0.49\textwidth}
        \centering
         \includegraphics[width=\textwidth]{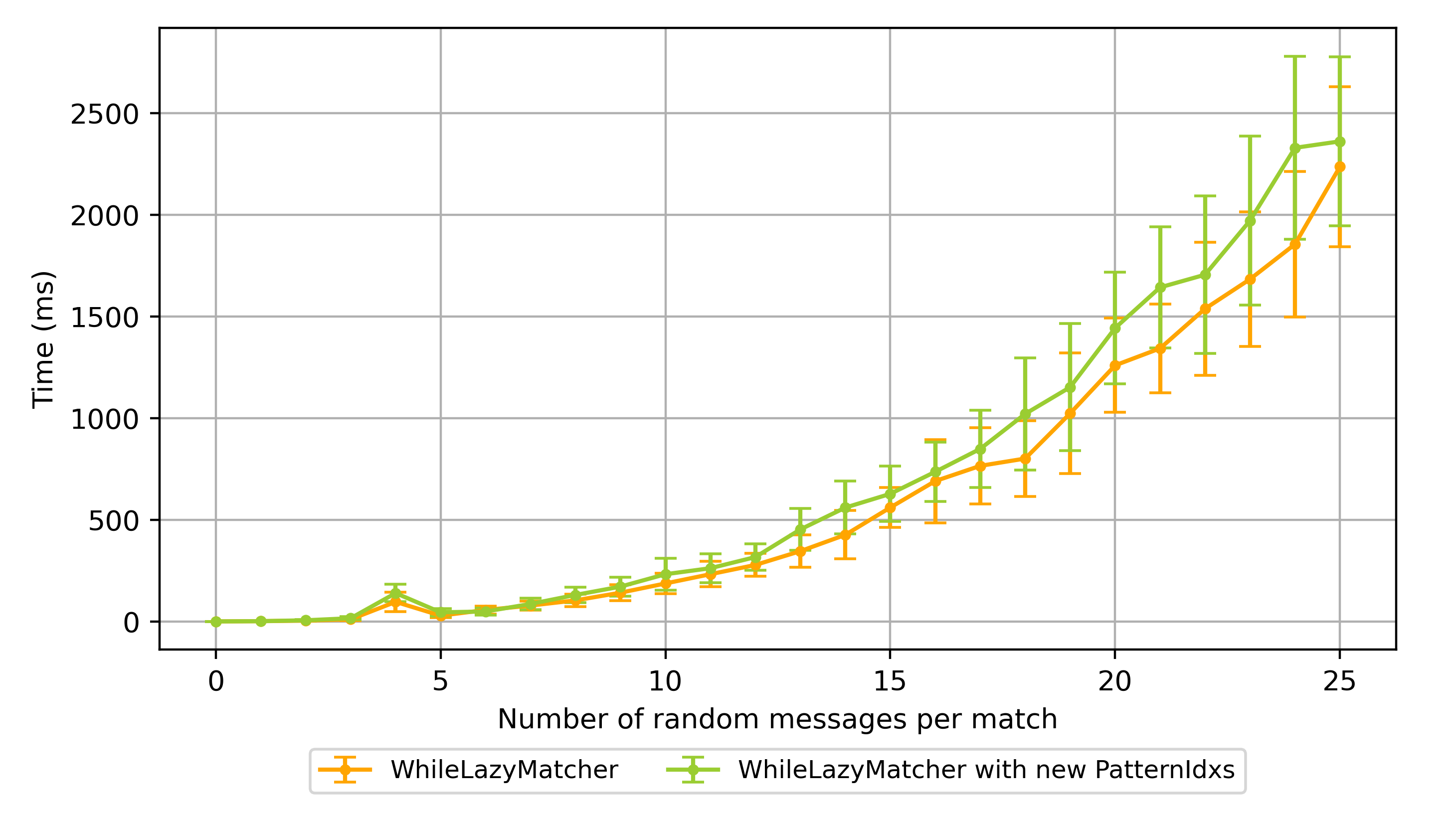}
         \caption{Complex Smart House}
    \end{subfigure}
    \hfill
    \begin{subfigure}[b]{0.49\textwidth}
        \centering
         \includegraphics[width=\textwidth]{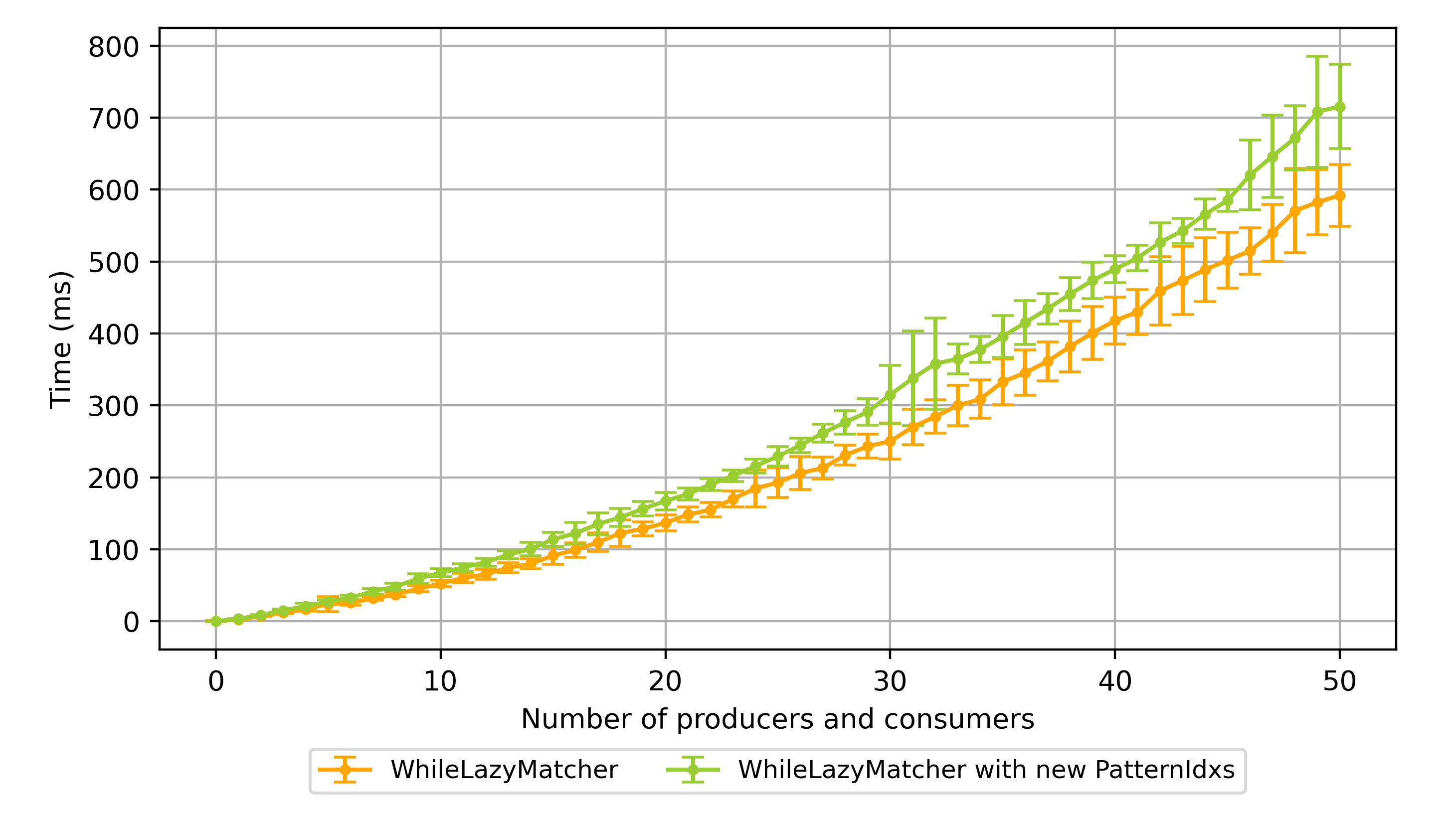}
         \caption{Bounded Buffer}
    \end{subfigure}
    
    \caption{A comparison of \texttt{WhileLazyMatcher} with the PatternIdxs change against \texttt{WhileLazyMatcher} without it}
    \label{fig:hashmap_benchmarks}
\end{figure}

We again test the version of \texttt{WhileLazyMatcher} with this change to the one without it. Figure \ref{fig:hashmap_benchmarks} shows the results. We obtain slightly worse performance on all benchmarks. This is a strange result, but it may be due to the computational cost incurred by computing hash codes.

\section{Benchmark suite improvements}
\label{sec:benchmark_suite_improvements}

The original \texttt{JoinActors} library came with a benchmark suite for evaluating the performance of its join pattern implementations. We create an updated suite that addresses some problems with the old suite and provides new functionalities. The new suite contains the following improvements:

\begin{itemize}
    \item A more consistent interface to the different benchmarks

    \item More readable real-time output while the benchmark is running

    \item More convenient CSV file output of results

    \item Automatic generation of plots from results using the JFreeChart library,\footnote{\url{https://www.jfree.org/jfreechart/}} for quick evaluation of results

    \item Repetitions that are run sequentially instead of in parallel for more reliable results

    \item Warmup repetitions that are all run before the benchmark, instead of before every benchmark repetition

    \item The ability to specify which algorithms to use with a command-line parameter

    \item Default values for all parameters, to greatly simplify usage

    \item A more extensible software structure and unified running logic for all benchmarks, so new benchmarks can be added and immediately benefit from most of the aforementioned features
\end{itemize}

This new suite is found in the \texttt{new\_benchmarks} package within the \texttt{benchmarks} project in the repository. The old benchmarks are still available in the \texttt{old\_benchmarks} package, but their main method is disabled.

\subsection{Interface}

A detailed description of how to use the new benchmark suite can be found in the \texttt{README} file of the \texttt{benchmarks} project in the repository. Here we mention some of the main aspects of the interface, with a focus on the benefits they provide.

Every benchmark now has a \textit{main parameter}, as well as zero or more \textit{configuration parameters}. The main parameter is a parameter that is varied between runs of the benchmark. Configuration parameters, on the other hand, are set by the user and stay the same throughout a benchmark. The range of the main parameter is specified with three command-line arguments: \texttt{min-param}, \texttt{max-param}, and \texttt{param-step}. These are specified in the same way for each benchmark type, providing a consistent interface. Configuration parameters are specific to each benchmark by nature, so these are specified with different command-line arguments depending on the benchmark type. The \texttt{repetitions} argument specifies the number of repetitions to be run for each main parameter value.

The \texttt{algorithms} command-line argument is used to specify a list of matching algorithms that should be tested in the benchmark. The options include all the algorithms presented in Section \ref{sec:implementation}, as well as the \texttt{StatefulTreeBasedAlgorithm} and \texttt{BruteForceAlgorithm} from the predecessor paper.

In the normal mode of operation, we run benchmarks by first iterating through the requested algorithms. For each requested algorithm, we iterate through the requested main parameter values, and for each parameter, we run the requested number of repetitions. The \texttt{smoothen} command-line flag changes this order: we still iterate through algorithms first, but we then iterate through repetitions, and then through main parameter values. The purpose of this is to make random performance dips induced by the operating system less visible. By running different parameter values in sequence, the effect of these performance dips becomes spread over multiple parameter values, instead of being concentrated on one parameter value and turning that data point into an outlier. This flag was used for all the benchmark results presented in this thesis, as can be seen in Appendix \ref{sec:benchmark_commands}.

\begin{figure}[h!]
    \centering
    \includegraphics[width=0.75\linewidth]{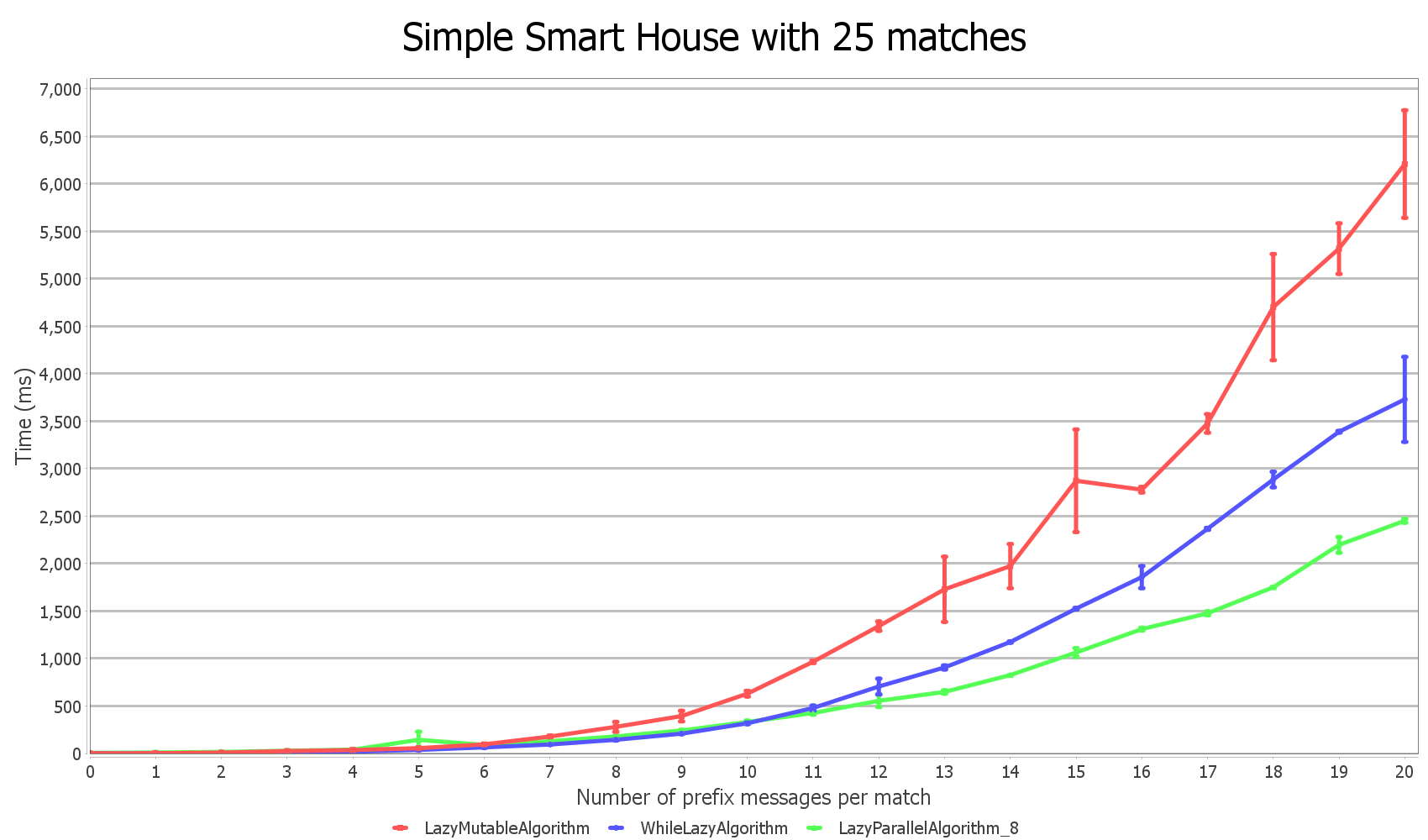}
    \caption{An example of a plot automatically generated by the new benchmark suite}
    
    \label{fig:jfreechart_example}
\end{figure}

The CSV output includes the main parameter in the first column. Subsequent columns include the results from each algorithm and each repetition, as well as the average and standard deviation for each algorithm and main parameter value. The automatically generated performance plots use the main parameter as the $x$-axis and the average performance as the $y$-axis. Figure \ref{fig:jfreechart_example} shows an example of a generated plot.

\subsection{Structure and extension}

The new benchmark suite allows for easy and flexible extension with new benchmarks. Every benchmark is a class implementing the \texttt{Benchmark} trait, with a companion object\footnote{\url{https://docs.scala-lang.org/scala3/book/domain-modeling-tools.html}} implementing the \texttt{BenchmarkFactory} trait. Essentially, \texttt{BenchmarkFactory} only contains information about the ``archetype" of the benchmark, e.g. ``Simple Smart House". By providing a \texttt{BenchmarkFactory} with a matching algorithm and the values of configuration parameters, one can create an instance of the corresponding class implementing \texttt{Benchmark}. \texttt{Benchmark} contains everything needed to run a concrete repetition of a benchmark except the value of the main parameter, which is supplied as a method parameter.

\texttt{Benchmark} defines two methods: \texttt{prepare} and \texttt{run}. \texttt{prepare} takes a value of the main parameter and returns a generic type \texttt{PassPrereqs}. \texttt{run} then takes a \texttt{PassPrereqs} and runs the benchmark, returning \texttt{Unit}. The benchmarks are run by first calling \texttt{prepare}, then starting the timer, then calling \texttt{run}, and then stopping the timer. All preprocessing such as generating messages should be done in \texttt{prepare}, and the data from the preprocessing should be passed to \texttt{run} through the \texttt{PassPrereqs} type. We design it this way so that we can  avoid wasting benchmark time on preprocessing, while at the same time allowing for every benchmark type to perform whatever preparations it needs before the timer starts.

\texttt{BenchmarkFactory} defines an \texttt{apply} method which takes the algorithm to use and an instance of an abstract type\footnote{\url{https://docs.scala-lang.org/tour/abstract-type-members.html}} \texttt{Config}, and returns an instance of another abstract type \texttt{InstanceType}. \texttt{InstanceType} is meant to be set to the corresponding class implementing \texttt{Benchmark}. One more abstract type is defined, \texttt{PassPrereqs}, which should be equal to the \texttt{PassPrereqs} type in \texttt{Benchmark}. This is enforced: we restrict the abstract type \texttt{InstanceType} to subtypes of \texttt{Benchmark[PassPrereqs]}.

All of our implemented benchmarks are located in subpackages of the \texttt{new\_benchmarks} package. Each subpackage contains:

\begin{enumerate}
    \item a class implementing \texttt{Benchmark} with a companion object implementing \texttt{BenchmarkFactory}

    \item a class providing the \texttt{Config} of the benchmark (if necessary)

    \item all other utilities that are used by the benchmark class.
\end{enumerate}

There is also a subpackage called \texttt{mixin} with a trait called \texttt{MessageFeedBenchmark} extending \texttt{Benchmark}. This is a trait that includes an implementation of the \texttt{run} method in order to refactor the most common type of benchmark, where a series of messages is fed to a single actor.

Finally, the package contains a \texttt{Main} object defining the command-line interfaces to the benchmarks, and a series of files containing utility functions for running the benchmarks and outputting results.

Therefore, to add a new benchmark, all one has to do is create a new class-object pair implementing \texttt{Benchmark} and \texttt{BenchmarkFactory}, and register the benchmark in the \texttt{Main} object of the package.

\section{Feature extensions}

\subsection{Operator syntax for joins}
\label{sec:operator_syntax}

The implementation from the predecessor paper uses a tuple-like syntax for join patterns. This may be seen as confusing, as it may be misinterpreted as a pattern consisting of a tuple instead of a join pattern. Therefore, as mentioned previously, we implement a new syntax for joins using a custom operator written as \texttt{\&:\&}. Listing \ref{lst:join_operator_syntax} shows a comparison between the old and new syntax.

\begin{listing}[h!]
\begin{minted}[bgcolor=bg, linenos]{scala}
// Old tuple syntax
receive { (_: ActorRef[Msg]) => {
  case (A(), B(), C()) =>
    println(s"I've received 3 messages: A, B and C :)")
    Continue
...

// New operator syntax
receive { (_: ActorRef[Msg]) => {
  case A() &:& B() &:& C() =>
    println(s"I've received 3 messages: A, B and C :)")
    Continue
...
\end{minted}
\caption{A comparison between the old and new join syntax}
\label{lst:join_operator_syntax}
\end{listing}

We implement the operator as a Scala object that defines an \texttt{unapply} method, also known as an extractor object.\footnote{\url{https://docs.scala-lang.org/tour/extractor-objects.html}} We enhance the \texttt{receive} macro with the ability to recognize when a guard is using this syntax, and extract the constructor patterns in the correct order from the abstract syntax tree.

Nevertheless, we maintain backwards compatibility with the old syntax for convenience. Both syntaxes can be used interchangeably, as long as any single join pattern uses only one syntax.

\subsection{Dynamic pattern switching}
\label{sec:dynamic_patterns}

We enhance the \texttt{JoinActors} library with the ability for an actor to \textit{dynamically switch} the matcher and join definition it is using. Listing \ref{lst:dynamic_pattern_simple} shows a simple example of this feature in use. Here, an actor starts with a matcher called \texttt{matcher1}. This matcher is defined on lines 6-9, while the actor is defined on line 11. This matcher has a single join pattern that takes two messages of types \texttt{M1} and \texttt{M2}. The right-hand side of this pattern indicates that the actor should switch to a different matcher, \texttt{matcher2}. This second matcher is defined before the first matcher on lines 1-4, so that the first matcher can refer to it. The second matcher has a join pattern that takes a message of type \texttt{M3}, and the right-hand side of this stops the actor.

In this example we define the second matcher before the first so that the first can refer to the second. However, this can be avoided using the \texttt{lazy val} declaration in Scala, which allows for forward references. We demonstrate this later in Listing \ref{lst:dynamic_switch_and_switch_back}.

\begin{listing}[h!]
\begin{minted}[bgcolor=bg, tabsize=1, linenos]{scala}
val matcher2 =
  receive[MsgPlain, Unit] { (_) => {
    case M3() => Stop(())
  }}(algorithm)

val matcher1 =
  receive[MsgPlain, Boolean] { (_) => {
    case M1() &:& M2() => Switch(matcher2)
  }}(algorithm)

val actor = Actor(matcher1)
val (futureResult, actorRef) = actor.start()

actorRef ! M1()
actorRef ! M2()
actorRef ! M3()
\end{minted}
\caption{A simple example of a dynamically switching pattern}
\label{lst:dynamic_pattern_simple}
\end{listing}

Our design and implementation allow for a lot of flexibility in how this feature is used; we explore this in Section \ref{sec:dynamic_pattern_examples}. We also expand the test suite of the original library to include tests for this feature.

\subsubsection{Design}

An actor contains a matcher, and a matcher contains one matching tree for each join pattern. These matching trees store all partial matches that are made on messages. In order to switch join definition, the matching trees need to be changed so that they reflect the partial matches that can be made with the new join definition.

Our switching process operates using the following simple algorithm. When the right-hand side of a join pattern requests a pattern switch,

\begin{enumerate}
    \item The actor extracts from the current matcher all messages that have already been matched.

    \item The actor \textit{prepends} to its own mailbox these messages that have already been matched.

    \item The actor replaces its matcher.

    \item The actor runs its new matcher on its mailbox.
\end{enumerate}

In essence, we simply construct the new matching trees from scratch. By prepending to the mailbox of the actor, we ensure that the new matcher is able to make use of all messages that the old matcher has taken but not consumed. Notably, we do not make use of the structure of the old matching trees to construct the new trees. Since the join definitions may be completely different, the information in the old trees is not relevant to the new ones.

When the new matcher receives the mailbox with prepended messages, the matcher is free to perform whatever actions it requires based on the prepended messages: this may include stopping the actor, or even performing another switch. If another switch is performed, the process is simply repeated, and a new layer of prepended messages is added to the mailbox. This would have been much more difficult with a more complex algorithm, such as one making use of a finite state machine to store the current join definition being used. By using a simple algorithm, we allow for much flexibility in its use.

\subsubsection{Implementation}

The \texttt{JoinActors} library uses a type called \texttt{Result} as the ``return" type of the right-hand side of every join definition. This is a union type, and the old library has two cases for it: \texttt{Continue}, to tell the actor to continue matching, and \texttt{Stop}, to tell the actor to return a result. We extend the \texttt{Result} type with a third case, \texttt{Switch}. This case is parameterized with the instance of the \texttt{Matcher} type that the user would like the actor to use. A small detail worth mentioning is that we changed \texttt{Result} from a discriminated union type to an ``ad-hoc" union, introduced in Scala 3.\footnote{\url{https://docs.scala-lang.org/scala3/book/types-union.html}} We did this to avoid problems related to type variance, since the \texttt{Switch} case introduces a new invariant type variable.

The old library uses the Java collection type \texttt{LinkedTransferQueue} to implement actor mailboxes. This is an efficient concurrent data structure that is well suited to this use case, but unfortunately it does not support the prepend operation that we need in our algorithm. We therefore implement a wrapper over this collection type called \texttt{PrependableLinkedTransferQueue}. Listing \ref{lst:prependable_queue} shows the full code of this class. This wrapper class stores a \texttt{LinkedTransferQueue} in a field called \texttt{delegate} (line 2), as well as a Scala \texttt{ArrayDeque} in a field called \texttt{prepends} (line 3). We define a \texttt{put} method (lines 5-6) which simply forwards to the \texttt{put} method of \texttt{delegate}. We also define a method called \texttt{prependAll} (lines 8-9) which takes a collection and adds all of the collection's elements to \texttt{prepends}. Finally, we define a \texttt{take} method (lines 11-13): if \texttt{prepends} is not empty, we take from \texttt{prepends}, and otherwise we take from \texttt{delegate}.

\begin{listing}[h!]
\begin{minted}[bgcolor=bg, linenos]{scala}
class PrependableLinkedTransferQueue[M]:
  private val delegate = LinkedTransferQueue[M]()
  private val prepends = mutable.ArrayDeque[M]()

  def put(m: M): Unit =
    delegate.put(m)

  def prependAll(msgs: IterableOnce[M]): Unit =
    prepends.prependAll(msgs)

  def take(): M =
    if prepends.nonEmpty then prepends.removeHead(false)
    else delegate.take()
\end{minted}
\caption{The code of the \texttt{PrependableLinkedTransferQueue} class}
\label{lst:prependable_queue}
\end{listing}

The accesses to the \texttt{delegate} field in the \texttt{put} and \texttt{take} methods are thread-safe, since the \texttt{LinkedTransferQueue} class is thread-safe. On the other hand, the accesses to \texttt{prepends} in \texttt{take} and \texttt{prependAll} are not thread-safe. Nevertheless, this does not cause any issues, as all calls to \texttt{take} and \texttt{prependAll} are done in the thread running the actor.

We add code to the \texttt{run} method of the \texttt{Actor} class to handle the new \texttt{Switch} result type, implementing our switching algorithm. Listing \ref{lst:actor_dynamic_pattern_run} shows the new \texttt{run} method, with the new case on lines 6-11. On line 7 we make use of a new method we add to the \texttt{Matcher} trait called \texttt{storedMessages}, which provides the messages that the matcher has taken from the mailbox but not consumed.

\begin{listing}[h!]
\begin{minted}[bgcolor=bg, linenos]{scala}
@tailrec
private def run(promise: Promise[T]): Unit =
  matcher(mailbox)(self) match
    case Continue    => run(promise)
    case Stop(value) => promise.success(value)
    case Switch(newMatcher) =>
      val storedMessages = this.matcher.storedMessages
      mailbox.prependAll(storedMessages)
    
      this.matcher = newMatcher
      run(promise)
\end{minted}
\caption{The new \texttt{run} method in the \texttt{Actor} class}
\label{lst:actor_dynamic_pattern_run}
\end{listing}

\subsubsection{Example uses}
\label{sec:dynamic_pattern_examples}

We now demonstrate various interesting things that can be done with this dynamic pattern switching feature. All of these examples are taken from the test suite of the repository, with slight modifications.

As mentioned previously, when a matcher switch is performed, it is possible to perform a second switch even while processing the backlog of messages created by the first switch. Listing \ref{lst:dynamic_pre_double_switch} shows such an example. Here, three messages are sent to the actor: \texttt{M1}, \texttt{M2}, and \texttt{M3}, in that order. The first matcher (lines 11-14) waits for the \textit{third} message \texttt{M3} and switches to the second matcher. The second matcher (lines 6-9) takes the second message \texttt{M2} \textit{from the backlog} and switches to the third matcher. The third matcher (lines 1-4) then takes the first message \texttt{M1} from the backlog and stops execution.

\begin{listing}[h!]
\begin{minted}[bgcolor=bg, linenos]{scala}
val matcher3 =
  receive[MsgPlain, Unit] { (_) => {
    case M1() => Stop(())
  }}(algorithm)

val matcher2 =
  receive[MsgPlain, Unit] { (_) => {
    case M2() => Switch(matcher3)
  }}(algorithm)

val matcher1 =
  receive[MsgPlain, Unit] { (_) => {
    case M3() => Switch(matcher2)
  }}(algorithm)

val actor = Actor(matcher1)
val (futureResult, actorRef) = actor.start()
actorRef ! M1()
actorRef ! M2()
actorRef ! M3()
\end{minted}
\caption{An example showcasing nested pattern switches}
\label{lst:dynamic_pre_double_switch}
\end{listing}

Listing \ref{lst:dynamic_pre_and_post_switch} shows an example similar to the last, but now with a join pattern in the second matcher, showing how backlog messages can be joined together with new messages. The first matcher (lines 6-9) waits for \texttt{M2} and switches to the second matcher. The second matcher (lines 1-4) then waits for both \texttt{M1}, which it receives from the backlog, and \texttt{M3}, which it receives after the switch happens.

\begin{listing}[h!]
\begin{minted}[bgcolor=bg, linenos]{scala}
val matcher2 =
  receive[MsgPlain, Unit] { (_) => {
    case M1() &:& M3() => Stop(())
  }}(algorithm)

val matcher1 =
  receive[MsgPlain, Unit] { (_) => {
    case M2() => Switch(matcher2)
  }}(algorithm)

val actor = Actor(matcher1)
val (futureResult, actorRef) = actor.start()
actorRef ! M1()
actorRef ! M2()
actorRef ! M3()
\end{minted}
\caption{An example showcasing a nested pattern switch with a join pattern}
\label{lst:dynamic_pre_and_post_switch}
\end{listing}

It is also possible to switch to another matcher, and subsequently switch back to the first. However, this requires both matchers to refer to each other, and so the matcher variables need to be defined in a way that allows for forward references. The \texttt{lazy val} construct in Scala is suitable for this purpose. Listing \ref{lst:dynamic_switch_and_switch_back} shows such an example. The messages sent to the actor are, as before, \texttt{M1}, \texttt{M2}, and \texttt{M3}. The first matcher (lines 1-5) consumes \texttt{M1} and switches to the second matcher. The second matcher (lines 7-10) then consumes \texttt{M2} and switches back to the first matcher. Finally, the first matcher consumes \texttt{M3} and stops execution.

\begin{listing}[h!]
\begin{minted}[bgcolor=bg, linenos]{scala}
lazy val matcher1: Matcher[MsgPlain, Result[MsgPlain, Unit]] =
  receive[MsgPlain, Unit] { (_) => {
    case M1() => Switch(matcher2)
    case M3() => Stop(())
  }}(algorithm)

lazy val matcher2: Matcher[MsgPlain, Result[MsgPlain, Unit]] =
  receive[MsgPlain, Unit] { (_) => {
    case M2() => Switch(matcher1)
  }}(algorithm)

val actor = Actor(matcher1)
val (futureResult, actorRef) = actor.start()
actorRef ! M1()
actorRef ! M2()
actorRef ! M3()
\end{minted}
\caption{An example showcasing an actor that switches between two matchers in both directions}
\label{lst:dynamic_switch_and_switch_back}
\end{listing}

As a last example, we show how it is even possible to switch to a matcher received \textit{as the payload} of a message. Listing \ref{lst:dynamic_matcher_in_payload} demonstrates this. We define a matcher called \texttt{matcherInPayload} on lines 1-4. We send two messages to the actor (lines 13-14): one containing \texttt{matcherInPayload}, and one called \texttt{MM1}. The first matcher used by the actor (lines 6-9) waits for a message containing a matcher in its payload. When it receives this, it switches to this matcher in the payload. The matcher which we provide in the payload, \texttt{matcherInPayload}, then waits for the message \texttt{MM1} and stops execution.

\begin{listing}[h!]
\begin{minted}[bgcolor=bg, linenos]{scala}
val matcherInPayload =
  receive[MsgWithMatcher, Unit] { (_) => {
     case MM1() => Stop(())
  }}(algorithm)

val matcher1 =
  receive[MsgWithMatcher, Unit] { (_) => {
    case MMatcher(payloadMatcher) => Switch(payloadMatcher)
  }}(algorithm)

val actor = Actor(matcher1)
val (futureResult, actorRef) = actor.start()
actorRef ! MMatcher(matcherInPayload)
actorRef ! MM1()
\end{minted}
\caption{An example showcasing an actor that switches to a matcher contained in a message payload}
\label{lst:dynamic_matcher_in_payload}
\end{listing}
\section{Advanced example: Payment microservices}
\label{sec:payment}

In addition to the aforementioned  use cases provided in the benchmark suite, we also implement a more advanced model use case of a payment system with a microservice infrastructure. This model payment system was originally created as part of a course project at DTU; \cite{dtu_pay} here we reuse elements of its architecture in the context of join patterns.

We model the backend of a payment service similar to MobilePay,\footnote{\url{https://en.wikipedia.org/wiki/MobilePay}} where customers can make requests to transfer money to a merchant. We implement each microservice as an actor, and all communication between the microservices occurs through message passing. The microservices match on these messages using join patterns to perform the appropriate tasks. Join patterns are especially suited to the purpose of modeling microservices, as it is common for a microservice to await multiple events before it performs an action. We have two instances of this in our implementation.

We model four microservices:

\begin{itemize}
    \item \textbf{Core service:} This microservice receives payment requests from the outside and forwards them appropriately.

    \item \textbf{Account service:} This microservice stores the account information of the payment system. Given an account, it is able to validate whether the account exists.

    \item \textbf{Token service:} This microservice handles single-use tokens used to make payments. A customer can request tokens to be generated, which the customer is then tasked with storing. When a customer wants to make a payment, they need to provide a previously generated token, and this is validated by the token service.

    \item \textbf{Payment service:} This microservice communicates with a bank to effectuate payment requests.
\end{itemize}

We implement two idealized event flows through these microservices: one for customers requesting a token, and one for customers requesting a payment.

When a customer requests a token, the following process occurs in the system:

\begin{itemize}
    \item An \texttt{ExternalTokenGenerationRequest} is sent to the core service. The core service generates a \textit{correlation ID}, which will be attached to every event associated with this request. It then sends a \texttt{TokenGenerationRequested} event with this correlation ID to the account service and token service.

    \item When the account service receives \texttt{TokenGenerationRequested}, it sends \texttt{CustomerValidated} to the token service.

    \item When the token service has \textit{both} a \texttt{TokenGenerationRequested} and a \texttt{CustomerValidated} in its mailbox with the same correlation ID, it sends \texttt{TokenGenerated} back to the core service. This consumption of two events is implemented using a join pattern on the two event types, with a guard testing for equal correlation IDs.
\end{itemize}

In a real implementation, an account ID would be associated with the token request, and a token would be included in the \texttt{TokenGenerated} event and sent back to the customer. Moreover, the account may be fake, in which case the account service would \textit{invalidate} the account and the user would receive an error message. We choose to omit these error cases for simplicity.

When a customer requests a payment, the following process occurs in the system:

\begin{itemize}
    \item An \texttt{ExternalPaymentRequest} is sent to the core service. The core service again generates a correlation ID so that the flow of this request can be tracked. It then sends a \texttt{PaymentRequested} event to the account service, the token service, and the payment service.

    \item When the account service receives \texttt{PaymentRequested}, it sends \texttt{MerchantValidated} to the payment service.

    \item When the token service receives \texttt{PaymentRequested}, it sends \texttt{TokenConsumed} to the account service.

    \item When the account service receives \texttt{TokenConsumed}, it sends \texttt{CustomerValidated} to the payment service.

    \item When the payment service receives \textit{all three of} \texttt{PaymentRequested}, \texttt{MerchantValidated}, and \texttt{CustomerValidated}, all with the same correlation ID, it sends \texttt{PaymentSucceeded} back to the core service. Again, this consumption of multiple events with the same correlation ID is implemented using a join pattern with a guard.
\end{itemize}

In a real implementation, the payment request would include e.g. a sending account ID, a receiving account ID, an amount, and authentication information. In addition, any of the validation steps could fail; we again omit these failure cases for simplicity.

\begin{figure}[h!]
    \centering
    \includegraphics[width=1\linewidth]{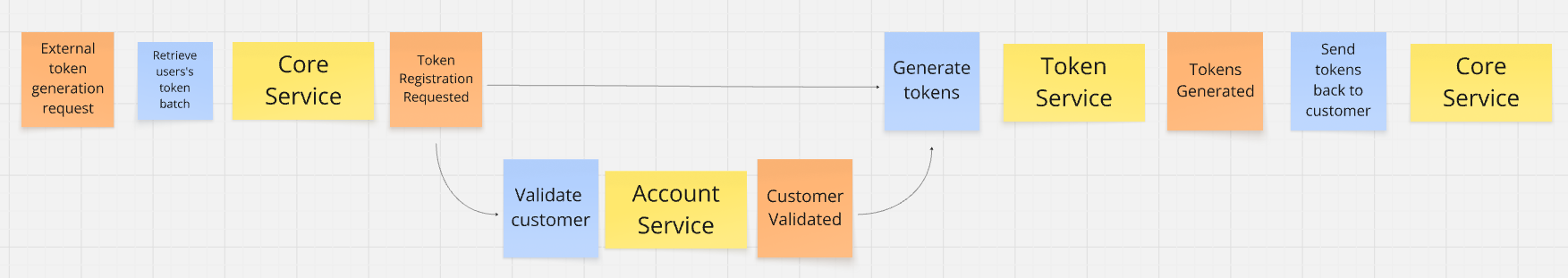}
    \caption{An event storming diagram for the token generation process}
    \label{fig:token_generation_event_storming}
\end{figure}

\begin{figure}[h!]
    \centering
    \includegraphics[width=1\linewidth]{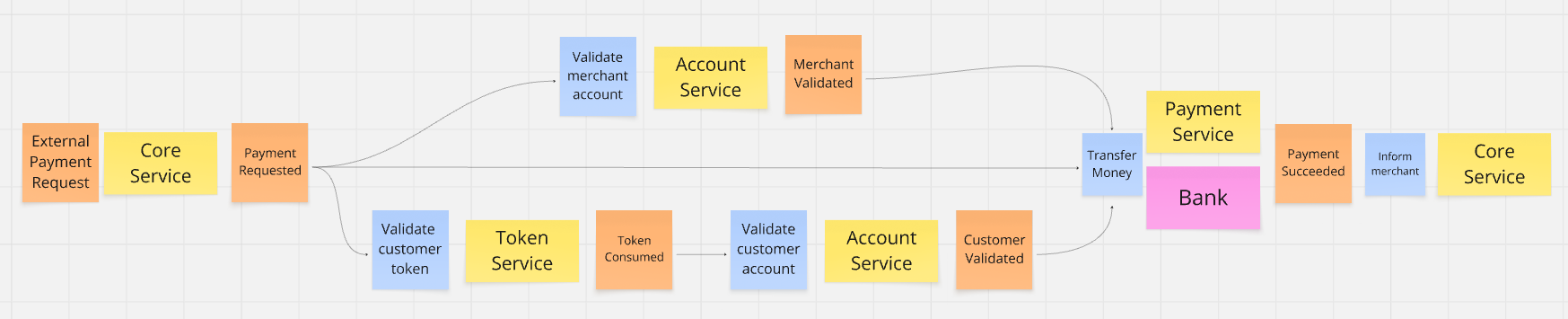}
    \caption{An event storming diagram for the payment process}
    \label{fig:payment_request_event_storming}
\end{figure}

To visualize these processes, Figure \ref{fig:token_generation_event_storming} shows a simplified event storming diagram\footnote{\url{https://en.wikipedia.org/wiki/Event_storming}} for the token request process, and Figure \ref{fig:payment_request_event_storming} shows one for the payment process.

Listing \ref{lst:payment_actor} shows the most complex join pattern in our implementation of this model, namely the one used by the payment service to wait for events from all three of the other microservices. As mentioned previously, this is a join pattern with three constructor patterns, as well as a guard that checks for equality of the correlation IDs.

\begin{listing}[h!]
\begin{minted}[bgcolor=bg, linenos]{scala}
val matcher = receive { selfRef => {
  case PaymentRequested(id1)
    &:& MerchantValidated(id2)
    &:& CustomerValidated(id3)
    if (id1 == id2) && (id2 == id3) =>
    coreService.get ! PaymentSuceeded(id1)

    Continue
...
\end{minted}
\caption{The main join pattern in the payment service actor}
\label{lst:payment_actor}
\end{listing}

We include our implementation of this payment service model as a runnable example in the \texttt{core} package of the repository.

\section{Conclusion}

We investigated various avenues for optimizing the fair join pattern matching algorithm in the \texttt{JoinActors} library. In doing so, we established a significantly faster new baseline in \texttt{WhileLazyAlgorithm}, as well as various more specific optimizations that lead to performance improvements in certain situations. We overhauled the benchmark suite for fair join pattern matching, making it more flexible and user-friendly, while also facilitating future extensions. We improved the syntax for join patterns, and added the powerful new feature of dynamic pattern switching. As a final complement, we also presented a new use case for join patterns, as a proof of concept for how they can be applied to the common real-world use case of microservice web architectures.

The work in this thesis and its predecessor paper raises multiple opportunities for future work:

\begin{enumerate}
    \item We have focused on optimizing the stateful tree-based matching algorithm from the predecessor paper. However, that paper also implements a brute-force matching algorithm in the \texttt{BruteForceMatcher} class. This naive algorithm works better in certain scenarios, so it may be beneficial to apply optimizations to this as well. In particular, the use of a mutable coding style and the guard filtering optimization could easily be applied.

    \item We have tackled the problem of parallelizing the stateful matching algorithm; while the results show noticeable improvements in performance, we are still far from the optimal speedup that could be achieved from parallelism. One could investigate whether this performance could be further improved by reducing the synchronization overhead.

    \item We tried to explain the performance differences between our algorithms on our benchmarks, but due to time and scope constraints, we have studied these differences elementarily. A further study could attempt to more systematically explain the performance characteristics of these algorithms. This could be done either by more closely inspecting the flame graphs, or by using other debugging tools such as Valgrind.\footnote{\url{https://valgrind.org/}}

    \item Both this thesis and the predecessor paper use simple plots to compare the performance of different matching algorithms. For a more informed comparison, one could also evaluate the statistical significance of the difference in performance between algorithms.

    \item Our guard filtering optimization (Sections \ref{sec:message_filtering}, \ref{sec:filtering_matchers}) achieves impressive results, but it could still benefit from further development. For example, in the current implementation, if a message is not discarded by the filter, the filtering clause is evaluated twice: once for the filter, and once for the final guard check. This duplicate work could be avoided. Furthermore, the concept of guard filters could be extended to overhaul the entire system for checking guards. All clauses in a guard could be assigned to a subset of constructor patterns in the join pattern, and evaluated whenever a partial match containing this subset is made. This would significantly advance filtering, as clauses that refer to payloads from multiple messages would also be used to remove messages from the matching tree early.

    \item We tried to change the internal representation of the matching tree structure from a red-black tree to an array in Sections \ref{sec:replacing_matching_trees} and \ref{sec:array_matchers}, albeit with limited success. With further focus on how to best represent the matching tree, it may be possible to identify a data structure that performs better across all benchmarks.

    \item In Section \ref{sec:failed_optimizations}, we attempted certain optimizations that had theoretical backing, but did not lead to any empirical performance improvement. These could be reevaluated, and perhaps a different implementation strategy could achieve the expected performance improvements.

    \item The implementation of pattern matching in the \texttt{JoinActors} library is quite limited, as it does not support recursive matching of algebraic data types such as lists, tuples, unions, etc. This has previously been explored in \cite{recursive_matching}. With a more advanced macro, it would be possible to have e.g. a join pattern of the form

    \begin{center}
        \texttt{M1(x::xs) \&:\& M2((a, b, c)) if x == b}
    \end{center}

    where the pattern within \texttt{M1} deconstructs a list and that within \texttt{M2} deconstructs a tuple.

    \item The ability to specify a bound on the mailboxes of actors would increase the practical usability of the library, as it would prevent matching trees from growing indefinitely.

    \item Our guard filtering optimization was partially inspired by the Rete algorithm~\cite{rete_original}. It may be possible to develop further optimizations based on Rete or similar algorithms such as TREAT~\cite{treat} and LEAPS~\cite{leaps}.

    \item The predecessor paper proposed automatically switching the matching algorithm to suit the input traffic. Now that we have developed a wealth of new matching algorithms, some suited to highly specific settings, this idea appears even more appealing.
\end{enumerate}

\printbibliography

\begin{appendices}

\section{Commands used to run benchmarks}
\label{sec:benchmark_commands}

\subsection{Simple Smart House}
The Simple Smart House benchmark in Figures \ref{fig:baseline_vs_stateful}, \ref{fig:immutable_vs_mutable}, \ref{fig:mutable_vs_lazy}, \ref{fig:lazy_vs_while}, \ref{fig:while_eager_vs_eager_parallel}, \ref{fig:while_lazy_vs_lazy_parallel}, 
\ref{fig:array_parallel_vs_array_while}, \ref{fig:simple_smart_house_all}, \ref{fig:simple_smart_house_long}, and \ref{fig:evrete_comparison} was run with the following command:

\sloppy \texttt{benchmarks/run simple-smart-house --min-param 0 --max-param 20 --matches 25 --algorithms "stateful, mutable, lazy-mutable, while-lazy, while-eager, eager-parallel, lazy-parallel, array-while, array-parallel" --repetitions 20 --warmup 5 --smoothen}

The corresponding \texttt{evrete-smarthouse} command for the Evrete data was the following:

\sloppy \texttt{evrete-smarthouse --min-param 0 --max-param 20 --matches 25 --repetitions 20 --warmup 5}

The shorter benchmark with heavy guards in Figure \ref{fig:evrete_comparison_heavy_guards} was run with the following command:

\sloppy \texttt{benchmarks/run simple-smart-house --min-param 0 --param-step 1 --max-param 5 --matches 25 --algorithms "while-lazy, lazy-parallel" --repetitions 20 --warmup 2 --heavy-guard --smoothen}

The corresponding \texttt{evrete-smarthouse} command for the Evrete data was the following:

\sloppy \texttt{evrete-smarthouse --min-param 0 --max-param 5 --matches 25 --repetitions 20 --warmup 2 --heavy-guard}

\subsection{Complex Smart House}
The short Complex Smart House benchmark in Figures \ref{fig:filtering_matchers} and \ref{fig:complex_smart_house_all} was run with the following command:

\sloppy \texttt{benchmarks/run complex-smart-house --min-param 0 --max-param 25 --matches 10 --algorithms "stateful, mutable, lazy-mutable, while-lazy, while-eager, eager-parallel, lazy-parallel, array-while, array-parallel, filtering-while, filtering-parallel" --repetitions 20 --warmup 5 --smoothen}

The long Complex Smart House benchmark in Figure \ref{fig:filtering_matchers_long} was run with the following command:

\sloppy \texttt{benchmarks/run complex-smart-house --min-param 0 --max-param 100 --matches 10 --algorithms "filtering-while, filtering-parallel" --repetitions 20 --warmup 5 --smoothen"}

\subsection{Bounded Buffer}
The short Bounded Buffer benchmark in Figure \ref{fig:bounded_buffer_all} was run with the following command:

\sloppy \texttt{benchmarks/run bounded-buffer --min-param 0 --max-param 20 --bufferBound 100 --count 100 --algorithms "stateful, mutable, lazy-mutable, while-lazy, while-eager, eager-parallel, lazy-parallel, array-while, array-parallel" --repetitions 20 --warmup 15 --smoothen}

The long Bounded Buffer benchmark in Figures \ref{fig:while_vs_array} and \ref{fig:bounded_buffer_long} was run with the following command:

\sloppy \texttt{benchmarks/run bounded-buffer --min-param 0 --max-param 50 --bufferBound 100 --count 100 --algorithms "stateful, mutable, lazy-mutable, while-lazy, while-eager, array-while" --repetitions 20 --warmup 15 --smoothen}

\subsection{Size and Size with guards}
The ``Size" and ``Size with guards" benchmarks in Figure \ref{fig:size_benchmarks} were run with the following commands:

\sloppy \texttt{benchmarks/run size --min-param 1 --max-param 6 --matches 5000 --algorithms "stateful, mutable, lazy-mutable, while-lazy, while-eager, eager-parallel, lazy-parallel, array-while, array-parallel" --repetitions 20 --warmup 4 --smoothen}

\sloppy \texttt{benchmarks/run size --min-param 1 --max-param 6 --matches 5000 --noise --algorithms "stateful, mutable, lazy-mutable, while-lazy, while-eager, eager-parallel, lazy-parallel, array-while, array-parallel" --repetitions 20 --warmup 4 --smoothen}

\sloppy \texttt{benchmarks/run size-with-guards --min-param 1 --max-param 6 --matches 5000 --variant normal --algorithms "stateful, mutable, lazy-mutable, while-lazy, while-eager, eager-parallel, lazy-parallel, array-while, array-parallel" --repetitions 20 --warmup 4 --smoothen}

\sloppy \texttt{benchmarks/run size-with-guards --min-param 1 --max-param 6 --matches 5000 --variant noisy --algorithms "stateful, mutable, lazy-mutable, while-lazy, while-eager, eager-parallel, lazy-parallel, array-while, array-parallel" --repetitions 20 --warmup 4 --smoothen}

\sloppy \texttt{benchmarks/run size-with-guards --min-param 1 --max-param 6 --matches 4 --variant non-satisfying --algorithms "stateful, mutable, lazy-mutable, while-lazy, while-eager, eager-parallel, lazy-parallel, array-while, array-parallel" --repetitions 20 --warmup 4 --smoothen}

\subsection{Baseline measurements}
All the baseline measurements were made on the \texttt{dev-baseline} branch of the repository using the same commands, but with the \texttt{--algorithms} parameter set to \texttt{stateful}.

\subsection{``Other attempted optimizations" measurements}

The measurements for the ``Other attempted optimizations" section (Section \ref{sec:failed_optimizations}) were made with similar commands on the \texttt{dev-growing-array-ref}, \texttt{dev-buffer}, and \texttt{dev-hash-map} branches of the repository. On \texttt{dev-growing-array-ref} and \texttt{dev-hash-map}, the \texttt{--algorithms} parameter was set to \texttt{while-lazy}, and the results were combined with the results from the main benchmarks. On \texttt{dev-buffer}, the \texttt{--algorithms} parameter was set to \texttt{array-while, buffer-while}, and only results from these runs were used to make the plots.

\end{appendices}

\end{document}